\pdfoutput=1 
\documentclass[twocolumn,tighten]{aastex63}

\usepackage{amsmath}
\usepackage{xspace} 
\usepackage{array}

\received{February 2020}
\submitjournal{ApJ}

\shorttitle{Low-mass interstellar probe downlink}
\shortauthors{Messerschmitt et al.}

\newcommand	{\incfig}[3]	
	{
	\begin{figure}[!t]
    \includegraphics[#2]
    {fig-#1}
    \caption{#3}
    \label{fig:#1}
	\end{figure}
	}

\renewcommand{\arraystretch}{1}
    
\newcommand {\doctable}
	[5] {
	\begin{table}[!t]
	\footnotesize
	\caption{#2}
	\label{tbl:#1}
	\centering
	\begin{tabular}{#3}
	#4
	\end{tabular}
	\end{table}
	}

\newcommand{\tc}[1]{\multicolumn{2}{c|}{$#1$}}
\newcommand{\tct}[1]{\multicolumn{2}{l}{#1}}

\newcommand
	{\comment}
	[1]
	{\textcolor
    	{red}
        {\textbf{#1}}
        }
\renewcommand
	{\comment}
	[1]
	{}

\newcommand{\eqnref}[1] {\eqref{eq:#1}}
\newcommand{\secref}[1] {\S\ref{sec:#1}}
\newcommand{\tblref}[1] {Tbl.\ref{tbl:#1}}
\newcommand{\figref}[1] {Fig.\ref{fig:#1}}

\newcommand{\parameters}{
	Tbls.\ref{tbl:missionParameters},
	\ref{tbl:transmitterParameters}, 
	\ref{tbl:receiverParameters},
	and \ref{tbl:bppmParameters}\xspace
	}

\newcommand {\citeref}
	[1] {\citep{RefNumber#1}}

\newcommand {\citereftwo}
	[2] {\citep{RefNumber#1, RefNumber#2}}

\newcommand {\citereffour}
	[4] {\citep{RefNumber#1, RefNumber#2, RefNumber#3, RefNumber#4}}

\newcommand {\citereffive}
	[5] {\citep{RefNumber#1, RefNumber#2, RefNumber#3, RefNumber#4,RefNumber#5}
	}

\newcommand{\ile}[1]{\mbox{$#1$}} \newcommand{\diff}{\,\mathrm d} \newcommand{\prob}[1]{\text{Pr} \left\{ #1 \right\}}

\newcommand{\rate}{\mathcal R}

\newcommand{\capacity}{\mathcal C}

\newcommand{\backpwr}{\Lambda_B}
\newcommand{\peakpwr}{\Lambda_P^R}
\newcommand{\avgpwr}{\Lambda_A^R}

\newcommand{\timeslot}{T_s}

\newcommand{\PAR}{\text{PAR}}
\newcommand{\SBR}{\text{SBR}}

\newcommand{\SNR}{\text{SNR}}
\newcommand{\SDR}{\text{SDR}}

\newcommand{\BPP}{\text{BPP}}
\newcommand{\PPD}{\text{PPD}}

\newcommand{\volume}{\mathcal V}

\newcommand{\massratio}{\zeta}

\newcommand{\pE}{P_\text{E}}
\newcommand{\pW}{P_\text{W}}
\newcommand{\pD}{P_\text{D}}

\newcommand{\blue}[1]{\textcolor{blue}{\bfseries #1}}
\renewcommand{\blue}[1]{#1}

\newcommand{\levelone}{\blue{collector}\xspace}

\newcommand{\leveltwo}{\blue{aperture}\xspace}
\newcommand{\levelthree}{\blue{element}\xspace}
\newcommand{\Leveltwo}{\blue{Aperture}\xspace}

\begin{document}

\title{
Challenges in Scientific Data Communication from\\
Low-Mass Interstellar Probes\footnote{
Copyright\copyright 2020.
}
 	}


\author[0000-0002-5889-7346]{David G Messerschmitt}
\affiliation{University of Calfornia at Berkeley,
Department of Electrical Engineering and Computer Sciences, USA}

\author[0000-0002-4973-8896]{Philip Lubin}
\affil{University of California at Santa Barbara,
Department of Physics, USA}

\author[0000-0003-0833-0541]{Ian Morrison}
\affiliation{Swinburne University of Technology,
Centre for Astrophysics and Supercomputing, Australia}
\affiliation{Now with Curtin University, International Centre for Radio Astronomy Research, Australia}

\begin{abstract}
A downlink for the return of scientific data from space probes at interstellar distances is studied.  
The context is probes moving at relativistic speed using a terrestrial directed-energy beam for propulsion, necessitating very-low mass probes.
Achieving simultaneous communication from a swarm of probes launched at regular intervals to a target at the distance of Proxima Centauri is addressed.
The analysis focuses on fundamental physical and statistical communication limitations on downlink performance
rather than a concrete implementation.
Transmission time/distance and probe mass are chosen to achieve the best data latency vs volume tradeoff.
Challenges in targeting multiple probe trajectories with a single receiver are addressed, including multiplexing, parallax,
and target star proper motion.
Relevant sources of background radiation, including cosmic, atmospheric,
and receiver dark count are identified and estimated. 
Direct detection enables high photon efficiency and incoherent aperture combining.
A novel burst pulse-position modulation (BPPM) beneficially expands the 
optical bandwidth and ameliorates receiver dark counts.
A canonical receive optical \levelone combines minimum transmit power with constrained swarm-probe coverage.
Theoretical limits on reliable data recovery and sensitivity to the
various BPPM model parameters are applied,
including a wide range of total collector areas.
Significant near-term technological obstacles are identified.
Enabling innovations include a high peak-to-average power ratio,
a large source extinguishing factor, the shortest atmosphere-transparent wavelength
to minimize target star interference, 
adaptive optics for atmospheric turbulence, very selective bandpass filtering (possibly with multiple passbands), 
very low dark-count single-photon superconducting detectors,
and very accurate attitude control and pointing mechanisms.
\end{abstract}

 


%
\section*{Nomenclature}

\begin{center}
\footnotesize
\begin{tabular}{ l p{7cm} }
$X^{\{T,S,R,TR\}}$ &
     Design values of parameter ``$X$'' at
     $T{=}$transmitter,
      $S{=}$receiver \leveltwo, 
      $R{=}$entire receive \levelone,
      or $TR{=}$end-to-end.
      $S$ and $R$ correspond to the canonical receive \levelone
      defined in \secref{collection}.
       \\[5pt]
        $Y_{\{A,P,N,I,D\}}$ &
       Values of performance metric ``$Y$'' corresponding to
      $A{=}$average power, 
      $P{=}$peak power,
      $N{=}$noise, 
      $I{=}$interference, 
      or $D{=}$dark counts.
\\[2pt]
      Others & See \parameters\  and \tblref{performanceMetrics}
\end{tabular}
\end{center}

\section{Introduction}

Since exoplanets are commonplace,
interest in and concrete efforts toward the
scientific exploration of exoplanets are growing.
This paper addresses the technology development challenges
behind the goal of launching low-mass interstellar space probes to
the vicinity of the nearest star system (Alpha Centauri)
and returning the acquired scientific data before the end of the 21st century.
Prominent among the myriad challenges
are propulsion and communication.
Here we address the communication of
scientific data from a probe at interstellar
distances and traveling at a relativistic speed.

First proposed by Robert Forward in 1962 \citeref{1014},
a low-mass probe with a sail propelled by directed energy from earth
is the only known technological option for close exploration of nearby stars
that may be accessible with available or near-term technology,
and which completes its mission in a matter of decades.
This has led to concrete efforts to validate the concept and its
supporting technology in the context of the NASA StarLight
and Breakthrough StarShot programs
\citereffour{811}{851}{1017}{867}.

Such a probe achieves minimum mass (and hence highest velocity) by
carrying only a sail for propulsion, 
scientific instrumentation, a downlink communications system,
photon thrusters for attitude control, and an electrical power generator.
Coupled with very high power directed energy beaming from earth
and a sail that captures the momentum from that beam,
it can reach 10-20\% of light speed.
Following a flyby (likely of Proxima Centauri
or other targets within our stellar neighborhood), the probe would transmit
the scientific data collected back to a receiver on the earth's surface,
which would consist of a large set of optical \leveltwo{s} with incoherent combining.
Location on the earth's surface (rather than on a
space-based platform) is motivated by the large dimensions
of the directed-energy and receive infrastructures,
and the large energy requirements for the former.

The focus here is on the probe-to-earth communication downlink design with the
expectation that any earth-to-probe uplink communication 
is short-lived and benefits from relative proximity to earth.
Relying heavily on a reusable fixed 
propulsion and communication infrastructure 
confined to the earth's surface,
a swarm of multiple probes can be launched
with a diversity of scientific instruments capturing
images, magnetic field strengths, spectroscopy, etc.

We quantify the tradeoff between probe mass, speed and data rate,
with the goal of minimizing the data latency, defined as
the time from probe launch to the completion of the data download.
As the total data volume is increased,
the probe mass and latency necessarily increase, and speed decreases.
Later numerical results focus on
the lowest-mass "wafer scale" version.

Today most interest in interstellar exploration
by space probes resides in the astronomy and
astrophysics
communities, but some essential 
knowledge and experience lies
with the communications sciences.
With the goal of informing all these communities 
and uniting them around this challenge,
we are careful to define terminology and
emphasize intuitive explanation rather than detailed derivation
of theoretical results.

\subsection{Goals}
\label{sec:goals}

As of this writing it is early in the conception
of the interstellar mission based on low-mass
probes.
Our goal \emph{is} to address the physical and
statistical communication limitations on a scientific data downlink,
as well as their implications for the requisite technological capabilities.
Our goal is \emph{not} to propose
a concrete and fully specified
design for such a communication downlink,
nor to address the numerous engineering challenges that will be encountered,
as there are too many
uncertainties, interactions between launch and downlink communication,
and questions about the technologies that may be
available in the timeframe of the first
operational downlink.
An initial conclusion is that wavelengths in the
visible optical regime are appropriate due to the
severe probe electrical power and mass restrictions.
We propose an architecture that
meets operational requirements.
Within that architecture we
study and quantity the inevitable dependencies
and tradeoffs
among the different components, 
and in general attempt to gauge the
feasibility of the endeavor.

Because we can use higher power or larger
apertures,
the laws of physics do not limit the
achievable distances or data rates.
Rather, practical limitations are due to low probe mass and available
technologies and their performance
characteristics.
Because probes will be launched in a couple
decades at the earliest, and
the beginning of downlink operation will be
delayed for at least five decades (see \secref{timeline}),
we don't consider
currently available technology to be limiting.
Rather we identify the
areas of greatest technological challenge,
and identify new capabilities that are needed.
This mission can leverage
general technology advances over the coming decades, but may
also require specific technological innovation and development.
We also identify and discuss tradeoffs among technologies,
and assume that greater advancement in some areas can
offset slower advancement in others.

Thus we can identify the specific goals of this paper
as:
\begin{itemize}
    \item
    Identify the physical limitations on such a
    mission, especially those related to the
    space environment (such as unwanted sources
    of noise and interference).
    \item
    Propose a system architecture, within which
    we identify the system components and quantify
    needed capabilities.
    \item
    Understand the interactions and tradeoffs among those
    components.
    \item
    Identify components where currently available technology
    is inadequate and estimate the necessary future capabilities.
    A more concise summary of these technology inadequacies is
    available \citeref{1011}.
    \item
    Generally assess the feasibility of the endeavor
    in the envisioned timeframe.
\end{itemize}
Implementation and practice always brings
some bad news.
While we attempt to anticipate the greatest challenges that
will be encountered, we do not attempt to accurately predict an actual deployed system
performance.
Thus, the results here always fall on the idealistic side.
One of the goals of any subsequent technology development will
be to come as close to these ideals as possible.

\subsection{Related work}

Most effort and publication devoted to low-mass probes
has emphasized the challenges of
propulsion by directed energy and light-sail
\citereffive{852}{853}{855}{856}{857}
and \citereffour{858}{859}{860}{861}.

In some ways this challenge mirrors the extensively studied requirements
for interplanetary spacecraft communication within our
solar system \citeref{796}.
The substantial differences include the severe limitations on the probe's
available electrical power, processing, and
transmit optics,
as well as the much greater propagation loss,
different sources of background radiation,
ground-based reception with many issues related to
atmospheric turbulence, scattering and weather,
and the expectation of a swarm of probe
downlinks operating concurrently.

Space communication applications have stimulated extensive
research and engineering into long-distance free space
optical communication.
Results most relevant to the present challenge come from
JPL's interplanetary network (IPN) \citeref{802}.
This includes an encouraging laboratory demonstration of high
photon efficiency in a communication link
with insignificant background radiation \citeref{834}.

Interstellar communication differs from terrestrial
communication in its emphasis on energy efficiency
rather than spectral efficiency.
At radio wavelengths,
energy-efficient interstellar communication
has been studied in the
context of METI/SETI \citereftwo{656}{708}.
What is different about space probes is the feasibility of transmitter-receiver coordination in the
interest of achieving high photon efficiencies to help overcome
power/size/mass limitations.

The germane sources of 
optical-wavelength background radiation for
interstellar communication are a current area
of research \citeref{1017}.
These background calculations have been independently
quantified in \citeref{803},
which quantifies the ultimate quantum
limits on communication.
The present paper addresses what is
achievable in practice with existing or foreseeable
technologies.

Here we explore a probe-swarm or single-probe receiver that achieves the
minimum transmit power-area product in the interest of
minimizing probe mass by operating near the physical limits of the
received-signal noise floor.
We also explore the implications of varying \levelone area
over a wide range.
A recent paper \citeref{1015} addresses the same application
but addresses only the single probe case
(requiring complete receiver duplication for each probe in a swarm)
and focuses on minimizing receive \levelone area rather than probe mass (see \secref{reducingArea}).

\section{Scientific mission}
\label{sec:mission}

The low-mass probe mission comprises the science objectives
and how they are achieved by a swarm of low-mass probes, each supporting
an individual downlink to return scientific data by that probe to earth.
\subsection{Numerical parameters}
\label{sec:numerical}

The set of parameters directly related to the science mission are
listed in \tblref{missionParameters}.
Assumed numerical parameters in this and the subsequent parameter and
performance metric tables are merely reference points and
do \emph{not} carry the weight of
design objectives.
This would be premature, as there are many design
and technology issues and interactions among
various system components that will weigh heavily on the
final design.
Consonant with our goals
(see \secref{goals})
these choices serve to gauge the
realism (or lack thereof) of the system design
using an idealized model that ignores many practical challenges
that will undoubtedly arise.
More importantly, scaling relationships
that follow from varying these parameters
are studied (both theoretically and numerically) to appreciate 
the implications of changing assumptions (see \secref{model}).

\doctable
	{missionParameters}
    {Scientific mission parameters}
    {|lp{6cm} | c |}
    {
    \hline
    & \textbf{Description} & \textbf{Value}
     \\ \hline  \\[-2ex]
$\rate_0$ & Nominal data rate for scientific data \newline immediately following 
    encounter \newline during non-outage periods (b/s) & $1.$
     \\
    $R_a$ & $R_0$ reduced to account for 
    outages (b/s) & $0.432$
    \\
    $\volume$ & Volume of total data (Mb) & $28.9$
    \\
    $L$ & Volume of data segment (Mb) & 1.0
     \\
    $P_c$ & Probability of correct segment 
    recovery & 0.99
    \\
    $T_L$ & Latency of total data return (yr) & $28.$
     \\
    $J_p$ & Probes transmitting concurrently & $26$
    \\ \hline
}

\subsection{Scientific objective}
\label{sec:scientific}

Each individual low-mass probe carries one or more scientific instruments to gather relevant
scientific information in the vicinity of the target star.
Given the severe mass-power objective, each probe may carry only a single instrument.

Individual probes may carry different types of scientific instrumentation, subject to severe mass limitations.
The primary relevance of the scientific objective to the
downlink design is the total data volume $\volume$
to be downloaded\footnote{
We measure $\volume$ in bits (b) rather than bytes (B). 
For example, $10^6$ bits ($1$ Mb)
equals $125$ kilobytes ($125$ kB).}
and the reliability
with which that data is recovered.
These two parameters are certainly related to the
type of instrumentation carried by a probe.
Early missions are likely to emphasize imaging, so
this is the application addressed in 
our examples.
An additional parameter of importance to scientists is the data latency $T_L$,
which is the time elapsed from probe launch to the return of data volume
$\volume$ in its entirety.

For purposes of reliability, scientific data can
be divided into \emph{segments}, 
each with volume $L$.
It is the reliability of recovery of each segment 
at the receiver that
is of greatest interest.
For example, for an imaging mission a segment
would be one image (2-D representation in
terms of matrix of pixels),
and we assume \ile{L=1\ \text{Mb}}.\footnote{
This will accommodate a $1000{\times}1000$
set of pixels at one bit per pixel following compression.}
A segment is either recovered correctly in its entirety or
corrupted in some fashion due to 
one or more bit errors.\footnote{
After compression, even a single bit
in error often propagates across the image
and thus has serious consequences.}
$P_c$ is the probability that each individual
segment is recovered correctly in its entirety.

\subsection{Scientific data rate}

The volume $\volume$ and latency $T_L$ in \tblref{missionParameters}
are indirectly related to the scientific data rate $\rate$.
$\rate$ actually decreases as the square of distance (see \secref{increasingDistance}),
so $\rate_0$ is the initial (highest) value of $\rate$ immediately following encounter.
In addition, outages due to atmospheric effects reduce the effective data rate to
\ile{\rate_a < \rate_0} (see \secref{outages}), and $\rate_a$ is used in calculating $\volume$.

\subsection{A swarm of probes}
\label{sec:swarm}

The project budget will be heavily concentrated
in the terrestrial launch and communication infrastructure.
The incremental cost of each probe launch
is insignificant in comparison \citeref{867},
suggesting the repeated launching of probes.
Such a swarm of probes can increase the scientific return by following
different trajectories near the target star,
providing redundancy to account for navigational errors or other sources of failure on
individual probes,
and enabling a diversity of scientific instruments while accommodating a tight
power/mass budget.

Because the cost of the receiver system
is likely to be large,
our primary analysis focuses on
a single receiver that is shared
among all the concurrent downlinks.
Thus the relevant parameter is not the total number of
probes launched, but rather the number of probes $J_p$ for which
downlink data transmission is concurrent from the receiver perspective.
The specific values of $J_p$ and $T_L$ illustrated are based on the minimization of $T_L$
from each probe for a given $\volume$ and a specific launch schedule (see \secref{masstradeoffs}).

Concurrent data reception causes serious
complications in the design and operation of
the terrestrial receiver infrastructure
and is a major concern addressed in the
following (see \secref{trajectories}).
To minimize this challenge, we don't 
expect the downlinks to operate during the transit period.
Dedication of a receiver to a single downlink
(or multiple receivers, one for each downlink as in \citeref{1015})
is a special case that falls within the scope of our analysis.

\subsection{Timeline}
\label{sec:timeline}

Some significant technology advances
are needed to approach the performance metrics envisioned in \secref{model}.
It is fortunate that there is considerable time
available to benefit from the ongoing evolution
of technology.
There is also the opportunity to target certain
technologies for faster advances targeted on the
needs of this application and interstellar
exploration more generally.
Transmitter technologies
must be finalized before the first launch
(perhaps 1-3 decades).
While an operational receiver deployed not long thereafter
could have benefits, delaying the download of
scientific data until 25+ years following the
first launch is an option.
Thus for the receiver we have available at least 4-5
decades of technology advancement and development,
although there should be a high confidence in the technical feasibility
and budget availability prior to the first probe launch.

\section{Probe transmitter}
\label{sec:transmitter}

The communications transmitter is integrally related to
other functional units in the probe, such as 
electric power generation and navigation and pointing control.
We assume transmitter parameters as listed in \tblref{transmitterParameters}.

\doctable
	{transmitterParameters}
    {Probe transmitter parameters}
    {| l p{4.5cm} | c |}
    {
    \hline
     	& \textbf{Description} & \textbf{Value}
     \\ \hline  \\[-2ex]
     	$\massratio$ & Probe mass ratio & 1.
	\\
    	$u_0$ & Probe speed (c) & 0.2
    \\
    	$D_{\{0,1\}}$ & \{Start,end\} of downlink
    	\newline
   	propagation distance (ly) & \{4.24,4.66\}
    \\
 	$A_e^T$ 
    	& Effective transmit aperture 
    	\newline
    	area (cm\textsuperscript{2})
    	& 100
    \\
    	$F_x$ & Optical source extinction ratio & $10^{-7}$
    \\ \hline
}
    
\subsection{Launch and propulsion}

The data latency $T_L$ is the sum of the transit time to the target star,
the duration of the downlink transmission, and the signal propagation time
from the farthest reaches of transmission (see \secref{latencyVolume}).
Minimizing $T_L$ subject to a constraint on $\volume$
is a principled approach to determining the best choice of probe mass
(see \secref{masstradeoffs}).
The smallest mass option (\ile{\massratio{=}1}) is assumed,
with increases in transmit power and transmit aperture area
possible when $\massratio$ is increased (see \secref{masstradeoffs}).

The goal of acquiring scientific data from instrumentation
in proximity to even nearby stars
within the 21st century requires a probe speed $u_0$ that is
a significant fraction of the speed
of light $c$ \citereftwo{865}{867}.
For example, the distance to the nearest star
(Proxima Centauri) is 4.24 ly, and the probe transit time
at \ile{u_0{=}0.2 c} is 21.2 yr.
If a probe transmits data beginning immediately after
encounter with that star and its exoplanets,
the data would begin to reach earth 25.44 years after launch.

Other transmit parameters include the propagation distance $D_{\{0,1\}}$,
which is determined by the probe speed $u_0$, the data volume $\volume$,
and the initial scientific data rate $\rate_0$.
The transmit aperture effective area $A_e$ is a measure of the transmission 
power-to-flux efficiency (see \secref{antennaTheory}), and equals the geometric area in the diffraction limit.
The transmit aperture is assumed to be standalone and relatively small in deference to the low mass requirement.
Other proposals assume the much larger probe sail area is exploited as a transmit aperture \citeref{1015}.
Finally the extinction ratio $F_x$ is a measure of the spurious power from the transmit laser
in its ``off'' state relative to its ``on'' state (see \secref{extinction}).

\subsection{Optical modulation}
\label{sec:tx}

Scientific payload data is embedded in the transmitted optical signal power vs. time by
the modulation.  
We propose
\emph{burst pulse-position modulation} (BPPM), which can achieve the required
efficiency while compatible with the imposed constraints
(see \secref{BPPM}).

In BPPM a semiconductor laser in the transmitter generates short
pulses of light intensity with a duration on the order of 0.1 to \ile{1\ \mu\text{s}}
with a repetition rate of about 1-2 Hz.
Scientific data is supplemented by error-correction redundancy
and then imposed on the pulses by adjusting their timing (see \secref{BPPM} and \secref{highBPP2}).
The average power is $P_A^T$, on the order of \ile{1{-}100\ \text{mW}}, with a peak transmitted power $P_P^T$
that is about $10^6$ larger.\footnote{
This description is oversimplfied,
as achieving a $P_P^T$ this large will require
a more complicated optical arrangement
(see \secref{laser}).
}
The intensity-modulated light is emitted in the direction of earth by a transmit aperture with
effective area $A_e^T$
(see \secref{antennaTheory}).
Accurate attitude control based on photon thrusters is necessary to ensure that the earth-based
receive \levelone is illuminated with the maximum power (see \secref{attitudeControl}).

The probe incorporates processing to compress scientific data
and perform the coding side of error-correction coding (ECC, see \secref{ECC}).
In addition, it performs data interleaving and adds additional redundancy
to counter outages resulting from daylight and weather events (and possibly other sources)
at the earth-side receiver (see \secref{outagemitigation}).

The probe-to-earth distance increases up to 10\% during the
operation of the downlink, which may last from \ile{2{-}9\ \text{yr}}
(see \secref{masstradeoffs}).
Accommodation of increasing propagation distance requires the
data rate $\rate$ to be reduced by slowing the rate of transmitted
pulses and increasing the peak transmitted power 
(by a maximum of about 20\%) to maintain
fixed average power.

\section{Terrestrial receiver}
\label{sec:receiver}

A terrestrial receiver is assumed, with
the parameters chosen for receiver operation as listed in \tblref{receiverParameters}
for two cases: a swarm of probes and a single probe.
The choice of a visible optical wavelength $\lambda_0^R$ is based on aperture area considerations (see \secref{wavelength}),
and is chosen close to the shortest wavelength for which the atmosphere is transparent because this minimizes
radiation from the target star (see \secref{bnumerical}).

\doctable
	{receiverParameters}
    {Terrestrial receiver parameters}
     {| l p{4.5cm} | c | c |}
    {
    \hline
    	& \textbf{Description} & \textbf{Swarm} & \textbf{Single}
     \\ \hline  \\[-2ex]
    	$\lambda_0^R$ 
    	& Received optical wavelength (nm) & \tc{400}
     \\ \cline{3-4}
    	$\Omega_A$ & Coverage solid angle (arcsec\textsuperscript{2}) & 10. & 0.01
    \\
 	$A_e^S$ 
    	& Effective \leveltwo 
    	area (cm\textsuperscript{2}) & $6.8$ & $6807.$
     \\ \cline{3-4}
    	$P_W$ & Probability of weather outage & \tc{0.1}
    \\
    	$P_D$ & Probability of daylight outage & \tc{0.52}
    \\
    	$\eta$ & Optical detector efficiency & \tc{1.0}
    \\
    	BPP & Non-outage photon efficiency (b/ph) & \tc{10.9}
     \\
        SBR & Average signal to average background
    	photons during non-outage PPM frames
    	& \tc{$4$}
     \\
    	$F_c$ & Coronagraph rejection & \tc{0.01}
    \\ \hline
}

\subsection{Trajectories and coverage}
\label{sec:trajectories}

The role of a receive optical \levelone is to convert
incident optical power to photon detection events.
It can be thought of as a single-pixel
optical telescope
which does not attempt to
image the swarm of probes (since their
trajectories are similar) but rather
captures a superposition of their signals as they reach the earth.

The coverage of a collector
is defined as the solid angle $\Omega_A$
over which transmitted probe signals
are recovered with nearly equal sensitivity.\footnote{
Coverage is similar to the field-of-view of
a single-pixel optical telescope,
or the resolution of a single pixel
in a multiple-pixel imaging telescope.
In contrast to optical telescopes, the
coverage that results from the beamforming
of an optical phased array
can be non-circular in shape.
}
To accommodate a swarm of probes with
a single receiver,
$\Omega_A$ has to be considerably larger than it would be for a single probe.
The \levelone is composed of a large number of individual
\leveltwo{s} whose photon detections are combined incoherently.
The coverage of the \levelone equals the coverage of the
\leveltwo{s} and is related to the effective area $A_e^S$
of the \levelone's constituent \leveltwo{s}
 (see \secref{collection}).
An increase in $\Omega_A$ results in
an inevitable decrease in receiver
sensitivity, 
which is quantified by a decrease in $A_e^S$
 (see \secref{collection}).

The trajectories of different probes in a swarm are illustrated in a simplifed schematic 2-D form in \figref{Trajectories}.
There are five phases to each 
individual probe mission:
launch of the probe, transit to the neighborhood
of the target star, encounter
with that star
(observations and scientific data collection),
    downlink operation
    (transmission of that scientific data to
    earth), and finally permanent probe silence.

\incfig
    {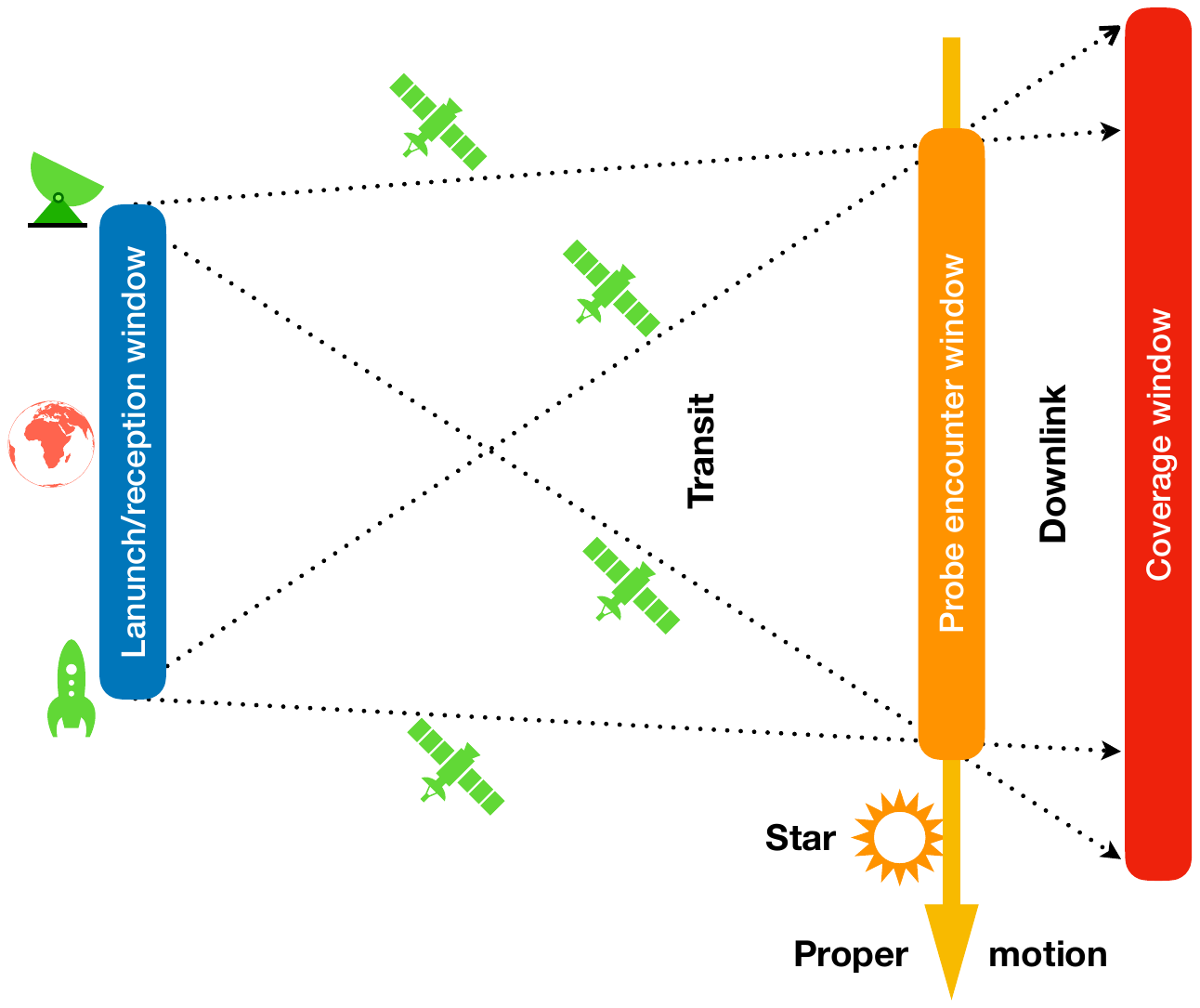}
   {
  trim=100 00 30 30,
  clip,
   width=1\linewidth
   }
    {
    A 2-D schematic representation of the
    4 most extreme probe trajectories as viewed from earth.
    The launch/reception window captures
    the seasonal variation in earth's position, and the probe
    encounter window captures the proper motion of the target star.
    As shown all encounters
    are assumed to fall on the same side 
    of the target star, which moves
    away from the encounter positions.
    Downlink operation follows encounter.
    Receiver coverage is assumed to
    cover all concurrently-transmitting probes, and a coronagraph
    function takes advantage of spatial
    separation to reject a portion of the
    target star's radiation.
    }

Shown specifically are
four probe trajectories at the most extreme angles, taking into account orbital
differences between launch and reception
coordinates as
well as the proper motion of the target star
(both of which are exaggerated in scale).
Of greatest concern for receiver design 
are the relative variations in
the positions of the different probes and the
target star, which
determine the
required spatial coverage of the receiver
and the coronagraph rejection
(see \secref{subarrayArchitecture}).
There are two primary
influences:
\begin{itemize}
\item
    A single
    probe's apparent position varies due to the
    parallax effect of the earth's orbital
    motion.
\item    
    The target star's proper motion
    (movement in relation to the
    galactic background)
    implies that different probe
    encounters are spatially separated \citeref{1015}.
    These probe encounter positions
    trail behind the 
    moving star, resembling the tail of
    a comet.
\end{itemize}

\subsection{Source of electric power}
\label{sec:powergeneration}

A probe launched with directed energy and cruising at constant
speed following launch requires no propulsion.
Electrical power is nevertheless necessary for probe attitude control,
scientific instruments,
and downlink communications \citeref{1012}.
Examples of power sources include
a radioisotope thermoelectric generator (RTG) or forward-edge ISM proton-impact conversion during the cruise phase
(before and after encounter) with the possible addition of
photovoltaic power from the target star during the encounter.
At relativistic speeds the latter will be short-lived (on the order of hours),
but may be valuable for scientific instruments,
as well as the processing for data compression,
outage mitigation (see \secref{outagemitigation}) and
error-control coding (see \secref{ECC}).

The available electrical power limits the available transmit optical power.
Here we make no prior assumption about transmit power, but rather characterize the
minimum transmit power necessary subject to the other constraints, and emphasize
the tradeoff between transmit power and total area of the transmit aperture and receive collector.

\subsection{Heterodyne vs direct detection}
\label{sec:directheterodyne}

There are two common ways to detect the optical
signal from a probe in the receiver.
\emph{Heterodyne} mixes that optical signal with a local oscillator (LO),
and optical square-law detection results in
components of sum and difference frequencies as well as amplification.
The difference-frequency component can fall at a microwave wavelength.
Here we adopt (for reasons explained in \secref{BPP}) the alternative of
\emph{direct detection}, which eliminates the LO and 
relies on single-photon counting detectors.
A significant advantage is the reliance of direct detection on intensity (and not phase)
modulation at the source, obviating
any need for coherence across the \leveltwo{s} comprising the \levelone
(see \secref{apertureArchitecture}).

Heterodyne would offer some compelling advantages.
In addition to signal amplification as a part of optical detection,
channel separation and bandpass filtering can be performed 
using available microwave technologies.
It also opens up a wider class of modulation schemes based
on the phase as well as magnitude of the incident wavefront.

At optical wavelengths (where the energy per photon is significantly higher)
individual photons can be directly detected, although due to quantum effects the
number of photons detected per unit time is only stochastically related to the receive power.
The result is signal self-noise
(sometimes called \emph{shot noise}
or \emph{quantum noise}).

\subsection{Background radiation}
\label{sec:backgroundSources}

Photon counts originating from sources other than shot noise are
lumped into the category of background radiation.
For a given transmit power and transmit/receive apertures,
the initial data rate $\rate_0$ that can be achieved
consonant with a sufficiently high signal-to-background ratio (SBR) is limited by
the sources of background radiation.
No matter how large SBR, signal shot noise remains as
a limitation on the reliability with which the scientific
data can be recovered (see \secref{highBPP2}).

Four distinct types of background
radiation can be identified:\footnote{
These are terms commonly used in the communications
literature. We adopt them here because (a)
they are descriptive of these types of impairments
and (b) because 
design techniques are adopted from the communication
literature.}
\begin{itemize}
\item
Imperfect \emph{laser extinction} at the transmitter
during intervals where zero power is sought will
result in unwanted photon counts at the receiver
(see \secref{laser}).
\item 
\emph{Noise} is radiation that cannot be separated
from the data signal because of its
time, wavelength, \emph{and} spatial overlap.
In the current application such radiation is broadband, and must be reduced by limiting the optical bandwidth.
Significant sources of noise are the
cosmic background radiation, the deep star field,
zodiacal radiation,
and scattered sunlight and moonlight from the earth's atmosphere.
\item
\emph{Interference} is unwanted radiation which
is distinguishable from the data signal in one
or more physical parameters (time, wavelength, or spatial)
and can therefore be partially or
wholly rejected by technological means.
The major interference considered here
is radiation from the target star, which is spatially
separated from the probes,
but which may be challenging to reject because of
its close proximity to the probe trajectories.
The coronagraph function in the receiver can partially reject
this interference due to its spatial separation.
We assume the rejection factor is $F_c$ relative to the signal
originating from the probe.
\item
\emph{Dark counts} mimic detection of
photons originating from sky or cosmic sources,
but originate within the receiver 
and would be present even if all incident light radiation
were blocked from the receiver.
\end{itemize}
Incomplete extinguishment
and dark counts generally cannot be limited by bandpass filtering (with the exception of black body radiation in the optics).
Numerical calculations suggest that operation during periods of sunlight is not feasible,
and with that assumption interference, scattered moonlight, and
dark counts are the limiting factors placing a lower bound on the probe's transmit power (see \secref{background}).

\subsection{High photon efficiency}
\label{sec:BPP}

One motivation for adopting direct detection is the higher photon efficiency that
can be achieved, which results in a reduction in the total \levelone area.
This is of practical significance because a \levelone will likely be kilometer-scale.
If the rate at which data is reliably recovered is $\rate_0\ \text{b/s}$ and the signal
average photon rate is $\Lambda_A^R\ \text{ph/s}$, photon efficiency BPP (in b/ph) is defined by
\begin{equation}
\label{eq:rateFromBPP}
\rate_0 = \BPP \cdot \Lambda_A^R
\,.
\end{equation}

Due to excess shot-noise introduced by the high-power LO,
 heterodyne cannot
achieve a comparable BPP to direct detection.
The theoretical limit is
\ile{
	\BPP < \eta/\log 2 = 1.44 \eta\ \text{b/ph}
	} 
for detector quantum efficiency $\eta$ consistent with reliable data
recovery \citeref{815}, while direct detection imposes no theoretical limit.
The assumption in \tblref{receiverParameters} is \ile{\BPP{=}10.9\ \text{b/ph}}.
This is consistent with a laboratory demonstration of \ile{\BPP = 13\ \text{b/ph}} \citeref{834}.
The significant photon efficiency advantage of direct detection makes it worthwhile to
overcome the technological challenges it introduces, which is our focus here.

\subsection{Signal-to-background ratio (SBR)}
\label{sec:SBR}

SBR at the receive \levelone output is \ile{\Lambda_A^R/\Lambda_B^R},
where $\Lambda_A^R$ is the average rate of signal photons and $\Lambda_B^R$ is the accumulated average photon rate for all
sources of background radiation.
Our goal is to achieve an SBR sufficiently large that background
radiation has limited impact on data reliability.
The impact of SBR is measured by the theoretical effect on
BPP (see \secref{photonCountingCapacity}).
With the goal of limiting background to a minor impact
we choose \ile{\SBR{=}4} in
\tblref{receiverParameters}.
This results in a 5\% reduction in the theoretically obtainable BPP as
compared to \ile{\SBR{=}\infty}.
Achieving a specific SBR places a lower limit \ile{\Lambda_A^R > \Lambda_B^R \cdot \SBR}
on the received signal average photon rate
$\Lambda_A^R$, and
and hence a lower bound on average transmit power $P_A^T$.

\section{Receiver photon collection}
\label{sec:collection}

The major element of the receiver is the \levelone,
which is similar to a single-pixel optical telescope.
It consists of a large number of \leveltwo{s}, each with a dedicated optical detector
that responds to individual photons
from the incident radiation (background as well as signal)
and introduces spurious dark counts as well 
(see \secref{apertureArchitecture}).
Thus no coordination or optical phasing of \leveltwo{s} is necessary.
It would be advantageous to share optical detectors over multiple
{\leveltwo}s if optical interference can be avoided (see \secref{detectorShare}).
In contrast to an optical telescope, the optical path
includes a highly selective
optical bandpass filter that
eliminates most out-of-band background radiation
(see \secref{bandpassFilter}), and
may serve to separate the signals from the individual probes as
well (see \secref{WDM}).
Photon detection events are logged at the individual \leveltwo{s},
accumulated over a network,
and a post-processing stage interprets the
totality of photon events
to reliably recover the scientific data.
This canonical receive \levelone is now described in greater detail.

\subsection{Canonical receive \levelone}
\label{sec:apertureArchitecture}

The receive \levelone in a direct-detection receiver includes
the optics and photonics necessary to capture the power of
an incoming electromagnetic wave and convert it to a digital representation
of individual photon detection events.
The receive \levelone will be quite large, and is 
thus a major capital and operational cost.

The receive \levelone must meet
some challenging requirements:
\begin{itemize}
\item
Accurate pointing toward the target star and swarm of probes
 has to be maintained
in the face of earth orbital dynamics and atmospheric
refraction.
\item
Both the total geometric size
(physical dimensions) and total collection area
are large as required to achieve a receiver sensitivity that accommodates
severe limitations on the probe due to its
mass restriction; namely, transmit power and
transmit aperture size.
A distinction between size and area is
needed because the receive \levelone is not
likely to be a single filled optical aperture.
\item

Achieve a coverage solid angle $\Omega_A$
which is relatively large
(but no larger than necessary) and
has an atypical elongated shape
in the direction of the target star's proper motion
(see \secref{trajectories}).

\item
It will likely incorporate adaptive optics to
compensate for atmospheric seeing conditions (see \secref{atmosphere}).
\item
Separation of signals from
different probes in the optical domain
is required for
some approaches to multiplexing
signals from different probes, such as WDM
(see \secref{WDM}).
\item
Whatever means possible should limit the background radiation.
This includes optical bandpass filtering
(see \secref{bandpassFilter}),
rejection of interference with coronagraph functionality
(see \secref{coronagraph}),
and modulation design to limit dark counts (see \secref{BPPM}).
\end{itemize}

\subsection{\Leveltwo{s}}
\label{sec:subarrayArchitecture}

A large size (kilometer scale) of the receive optical \levelone as a whole
is required to achieve low transmit power with adequate signal level to counter shot noise.
If such a \levelone were single mode (a single optical detector) and diffraction-limited, not only would its
coverage solid angle $\Omega_A$ be too small to cover a swarm of probes, but adequate
pointing accuracy to track a single probe would be practically unrealizable.

These challenges and the principles of antenna design 
suggest
a  canonical receive optical \levelone architecture illustrated in
\figref{receiveApertureTop}.
The entire \levelone must be decomposed into a large number $N^S$ of smaller {\leveltwo}s.
Each \leveltwo is required to achieve
all the objectives of the receive array
as a whole except a sufficient photon
rate to support the desired data rate.
In particular, the \leveltwo
accurately points at the 
coverage window as pictured in
\figref{Trajectories},
achieves the desired $\Omega_A$,
and achieves a required SBR.
Alternatives to this architecture can be explored,
such as abandoning the single-mode diffraction-limited constraint on an aperture.
For example, with a conventional optical telescope,
photons in individual pixels containing probe trajectories can be counted.
Photon counts accumulate incoherently across \leveltwo{s}.
Since photon counts for both the probe signal and the various sources
of background accumulate by a factor of $N^S$, SBR remains fixed
independent of $N^S$.
The value of $N^S$ is chosen to achieve the desired signal photon rate
commensurate with the desired rate $\rate_0$ commensurate with \eqnref{rateFromBPP}.
\Leveltwo replication by $N^S$ is called a \emph{scale out} because the photon counts increase linearly with $N^S$.\footnote{`Scale out' is an
	engineering term referring to an $N^S{\times}$ replication with a performance that
	increases proportionally to $N^S$. In this case `performance' is measured by the overall sensitivity
	of the receiver while maintaining fixed coverage and SBR.}

Each \leveltwo (which may or may not be further decomposed into a phased array of {\levelthree}s as described shortly)
may have a dedicated optical detector responsible for generating a single
stream of photon detection events in response to
the incident electromagnetic radiation.

\incfig
	{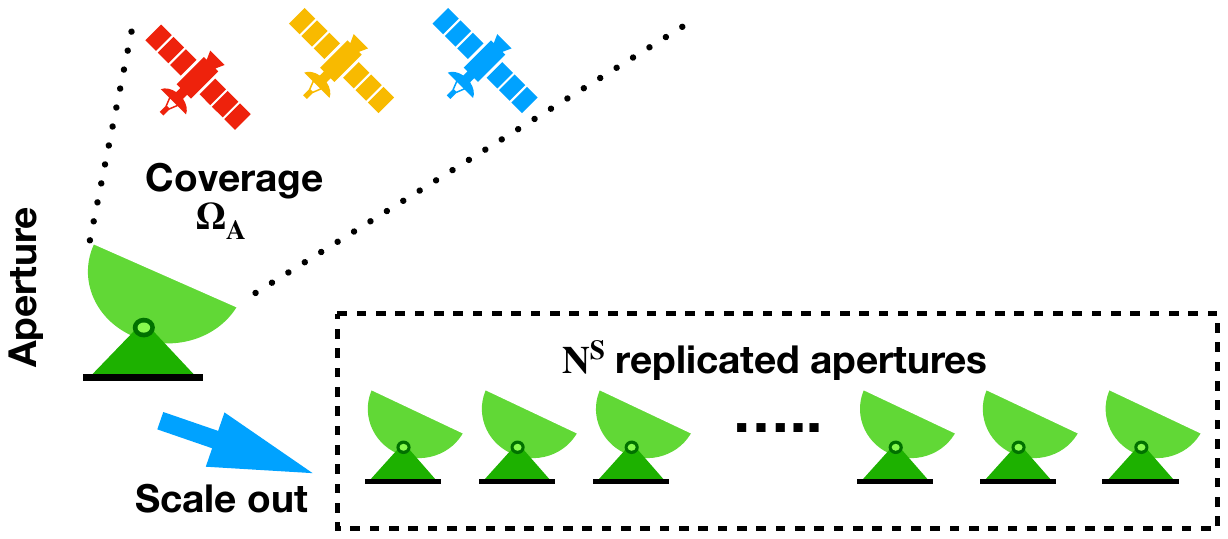}
	{
	 trim=130 310 20 0,
    clip,
    width=1\linewidth
	}
	{
	The receive optical \levelone is composed as an $N^S$-way
	replication (called scaling out) of the \leveltwo.
	An individual \leveltwo focuses on achieving the maximum
	and uniform sensitivity to the signals from all probes in a swarm
	across a coverage solid angle $\Omega_A$, and
	has effective area $A_e^S$.
	It is also responsible for achieving a sufficiently
	large SBR to achieve the target BPP, 
	and may also be responsible for separating the signals
	from different probes.
	The $N^S$ {\leveltwo}s independently capture and log photon detection events,
	with their only coordination a common clock for time stamps.
	The resulting incoherent accumulation of the incident optical power
	across the entire \levelone
	achieves a sufficient average photon detection rate $\Lambda_A^R$ 
	to recover the scientific data at rate $\rate$,
	while influencing neither the coverage nor SBR.
	}

\Leveltwo sensitivity 
is quantified by its
	equivalent area $A_e^S$
	(defined as the ratio of the intensity of
	an incident plane wave to detected power).
	 This determines the required
transmit power in the probe and the
order $N^S$ of scale-out.
	Although we like a larger $A_e^S$,
	in fact it is predetermined
	by the relationship between $A_e^S$ and coverage $\Omega_A$.\footnote{
	Each \leveltwo qualifies as an `antenna' that is subject to the theory of
	\secref{antennaTheory} assuming it is single mode (has a single optical detector)
	and diffraction-limited.
	This sensitivity-coverage relationship is based on
	the simplifying assumption (which can be approximated in practice) 
	that the sensitivity is uniform over solid angle
	$\Omega_A$ and zero elsewhere.
	}
In \secref{antennaTheory} this relationship is determined in
\eqnref{areaSolidAngle} to be \ile{A_e^S \propto \Omega_A^{-1}}.
Larger coverage, as dictated by our
requirement to service a swarm of probes
with a common receiver, implies reduced sensitivity.

	Each \leveltwo is responsible for limiting noise
	and interference via an optical bandpass filter (see \secref{bandpassFilter}) and coronagraph functionality (see \secref{coronagraph}).
Accounting for a non-ideal filter, define
the \emph{effective bandwidth} $W_e$ of a
bandpass filter as the ratio of output power
to input power spectral density (assumed to be
constant).
If the transfer function is $H(f)$, then
\begin{equation}
\label{eq:WeDef}
W_e = \int_0^\infty \big| H(f) \big|^2 \,\text d f
\,,
\end{equation}
in which case \ile{W_e = W} for an ideal
bandpass filter with bandwidth $W$.
More generally $P \cdot W_e$ is the filter output power for a constant spectral density input $P$.
The dark count rate for its optical detector should be as small as technology allows (see \secref{opticalDetector}).
	The coronagraph partially rejects interference from the target
	star by exploiting its small angular separation from the
	probe trajectories.

Within the context of a specific \leveltwo design,
including particularly sensitivity,
coverage, coronagraph rejection, and dark count rates,
the probes are responsible for a sufficiently large power-area metric \ile{\xi_A^T = P_A^T A_e^T}
and pointing accuracy to achieve the desired SBR
(see \secref{methodology}).
There is thus a tradeoff between $P_A^T$ and transmit aperture effective area $A_e^T$.

\subsubsection{\Leveltwo scale out}
\label{sec:scaleout}

Since the receive optical \levelone 
as a whole has \ile{N^S{\gg}1} \leveltwo{s}
that are incoherently combined,
it is not an antenna in the sense of \secref{antennaTheory} and thus
the concept of effective area $A_e^R$ does not apply.
Instead we use \emph{total effective area} metric $N^S A_e^S$.
This metric gives us a sense of the total geometric area of the entire \levelone.

While the numerical values of $N^S$ are large, the total collector area
$A_e^S N^S$ is a more significant indicator of capital costs.
For example it would be attractive to combine apertures into larger assemblages with joint fabrication.
Optics similar to that used in smart phone technology can be fabricated in substantial
assemblies, and it may also be possible to share a single optical detector over
multiple {\leveltwo}s (if optical interference, which would modify the coverage,
can be avoided, see \secref{detectorShare}).
\newpage
\subsection{Post processing}

A more detailed picture of scale out as well as \leveltwo decomposition into
smaller {\levelthree}s is shown in \figref{apertureArchitecture}.
Each \leveltwo not only detects individual photons, but it logs those
photon-detection events as a time-stamp 
and, if the \leveltwo performs de-multiplexing, a probe identifier.
There being no need for real-time processing of these events, they can be
shared in a common post-processing stage in which
timing recovery is performed
followed by scientific data extraction from the photon events.
A permanent stored log of photon events
(probe-number and time-stamps)
also constitutes
the permanent scientific record of a swarm of probe missions.

\incfig
    {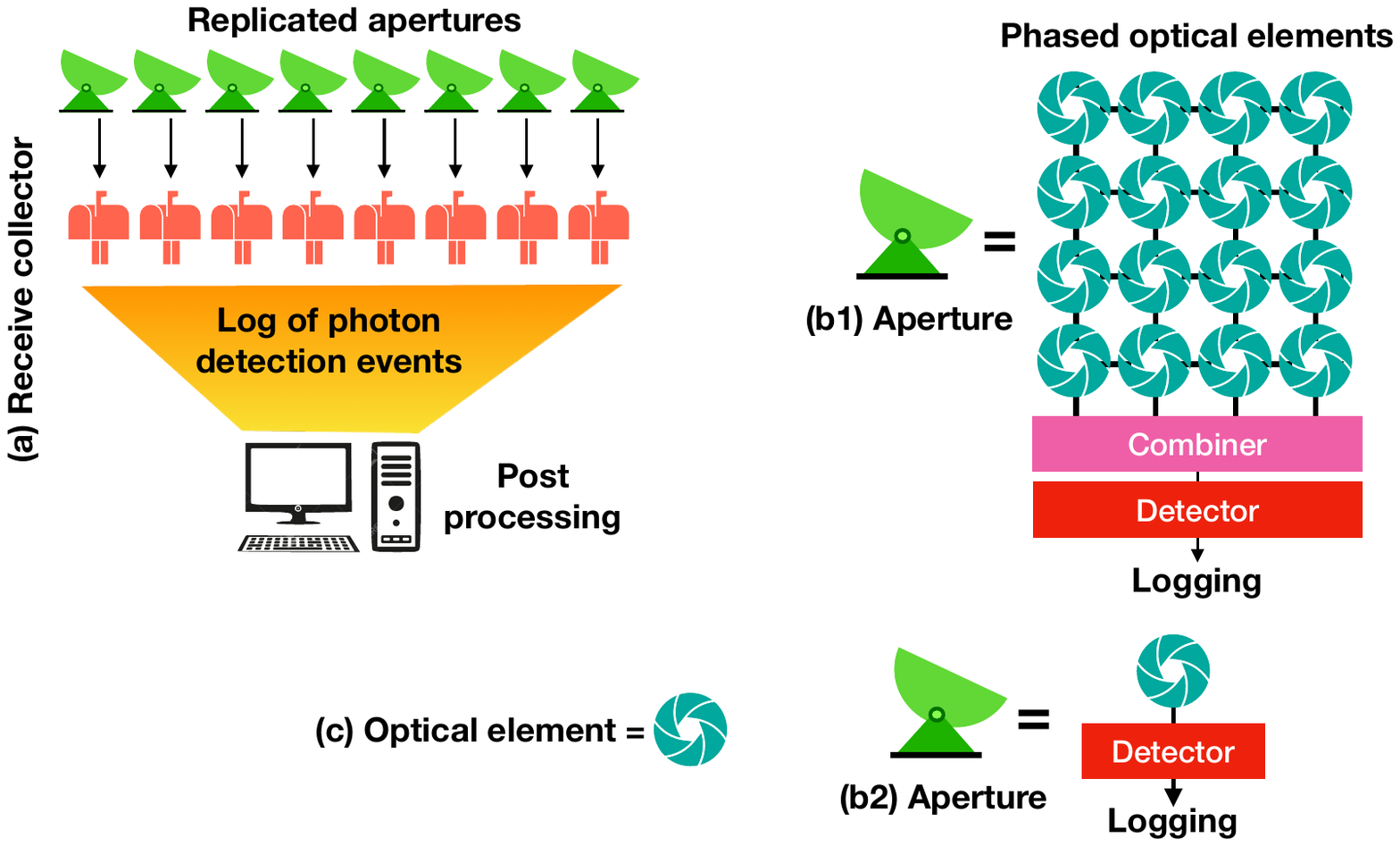}
    {
    trim=0 100 0 0,
    clip,
    width=1\linewidth
    }
    {
    A schematic of a
    two-level receive optical \levelone decomposition.
    (a) The replicated {\leveltwo}s independently log
    photon detection events, and in a post-processing
    stage the aggregate events are processed to recover
    the scientific data originating from multiple probes.
    (b1) Each \leveltwo is composed of a set of phased {\levelthree}s
    with a combiner including precision phase shifts feeding a single optical detector.
(b2) A degenerate case is an \leveltwo comprised of a single \levelthree.
    (c) Each \levelthree is an optical element (composed of lenses, reflectors or thin films).
    }

The collection of {\leveltwo}s operate independently.
There is no attempt to align the phases among individual
{\leveltwo}s; indeed, the {\leveltwo}s each have their
own optical detector which captures the power
of the incident radiation to that \leveltwo, 
so all phase information 
has been deliberately discarded.
This arrangement is colloquially called
 `photon buckets'.
The relative placement of \leveltwo{s} is somewhat flexible,
allowing for geographic distribution and relative
placement for the convenience of construction or maintenance.
Atmospheric turbulence is not an issue except at the \leveltwo level
(see \secref{turbulence}).
Placement of \leveltwo{s} divided among a set of one or more independent space platforms
is also an intriguing possibility.

The {\leveltwo}s all connect to a communication network
for distribution of target coordinates
(which change at a slow rate),
consolidation of photon events for processing,
and status updates.
The total number of events to be logged across all {\leveltwo}s is
of order
0.1 per second per probe for the parameters of \tblref{transmitterParameters}
and \tblref{receiverParameters},
so the storage and communication
requirements are readily achieved.

The synchronization of time stamps across {\leveltwo}s must be maintained,
and this requires a common time reference \citeref{806}.
Synchrony with the multiple probe clocks must also be maintained,
all this with an accuracy related to the time-slot $T_s$ (see \secref{clocks}).

\subsection{Shared optical detectors}
\label{sec:detectorShare}

One troublesome feature of \figref{apertureArchitecture}
is the $N^S$-way duplication of optical detectors,
which multiplies the dark count rate by the same factor and
imposes a very strict requirement
on individual detector dark counts.
Sharing of a single optical detector over multiple
\leveltwo{s} would reduce the overall dark count rate, as well as
reduce the cost and complexity of the scale-out function.
However to avoid any impact on the coverage pattern of the overall
receive array, any interference in the optical domain among 
\leveltwo signals would have to be avoided
(see \secref{opticalDetector}).

\subsection{Summary}

The two-level hierarchy of {\leveltwo}s
combined with scale out to a full aperture
is required to \emph{simultaneously}
achieve a desired (non-zero) coverage $\Omega_A$,
a desired SBR, and a desired
sensitivity (signal photon rate $\Lambda_A^R$).\footnote{
If as in \citeref{1015} there were a single probe and pointing accuracy is not a consideration,
the coverage could be as small as needed to
achieve sufficient sensitivity.
}
The consequence of covering multiple probes
with a single receiver
is inevitably a larger $\Omega_A$ and hence a higher $P_A^T A_e^T$ product
for each probe.
A desirable feature of the canonical receive aperture
is that the
difficult technological challenges are confined to the
\leveltwo realization, while the large number
$N^S$ of
{\leveltwo}s is primarily a budgetary and operational issue.

\subsection{Phased \levelthree{s}}
\label{sec:subapertureDecomposition}

The canonical receive aperture of \figref{apertureArchitecture}
anticipates that the \leveltwo{}s are themselves
composed of phased \levelthree{}s.
Each \levelthree is
envisioned as an (ideally) diffraction-limited
monolithic optical element, which may be constructed
from a lens or reflector.
Following a combiner,
these \levelthree{}s share
an optical detector.
Within the combiner
the magnitude and phase
for each \levelthree optical signal as it arrives
at the detector must
be precisely controlled.
The relatively small effective area $A_e^S$ of the \leveltwo
 make this technically feasible,
 and indeed for the values in \tblref{receiverParameters} it may not even be necessary.
Depending on the \leveltwo dimensions
an adaptive-optics component to compensate for
atmospheric turbulence may be needed
(see \secref{atmosphere}).

There are several requirements that suggest
the use of a phased \levelthree:
A coverage pattern with solid angle $\Omega_A$ that may be
non-circular in shape, a coronagraph function
which realizes a null in the overall 
sensitivity pattern at the angle of the
interfering target star, and adaptive optics.

\section{Transmit modulation and receive optical processing}

The optical modulation and associated receive processing are now
described.
A specific modulation coding technique called burst-mode
pulse-position (BPPM) modulation technique is adopted.
The parameters of this technique are listed in \tblref{bppmParameters}.

\doctable
	{bppmParameters}
    {Burst-mode pulse-position modulation parameters}
     {| l p{4.5cm} | c |}
    {
    \hline
    	& \textbf{Description} & \textbf{Value}
     \\ \hline  \\[-2ex]
    	$m$ & Bits per PPM frame & $12$
    \\
    	\ile{M=2^m} & Slots per PPM frame & $4096$
    \\
    	$T_s$ & PPM slot duration ($\mu$s) & $0.1$
    \\
    	$W_e$ & Effective optical bandwidth (MHz) & 10.
    \\
    	$T_F$ & PPM frame duration (ms) & $0.41$
    \\
    	$T_I$ & Inter-PPM-frame interval (s) & $2.2$
    \\
    	$\delta$ & BPPM duty cycle & \ile{1.9{\cdot}10^{-4}}
    \\
    	PAR & Peak-to-average ratio & 4096.
   \\ \hline
    }

\subsection{High photon efficiency and peak power}
\label{sec:highBPP}

There are two impairments that limit the 
photon efficiency BPP that can be achieved.
The first is background radiation, the impact
of which is limited by placing a
theoretical lower bound on SBR (see \secref{photonCountingCapacity}).
The quantum efficiency $\eta$ of the entire optical
path reduces signal, noise, and interference
equally, but does not affect the rate of dark counts.
Thus its effect on SBR is only through the dark count
component of background, and
\ile{\eta < 1} renders dark counts relatively more important.
Because the $\eta$ achievable decades from now is uncertain,
our nominal assumption is \ile{\eta{=}1} together with a sensitivity analysis.

Once background radiation is rendered insignificant,
signal shot noise determines a 
theoretical upper bound on BPP.
High BPP with reliable data recovery is achieved by judicious choice
of the mapping from scientific data to
light intensity vs time in the transmitter (see \secref{BPPM})
and error-correction coding (see \secref{ECC}).
Regardless of how these are structured, there is a theoretical
limit on the shot-noise-limited BPP that can be achieved
in conjunction with reliable recovery of the
scientific data.
For large SBR this limit is
\begin{equation}
\label{eq:logPAR}
\BPP < \log_2 \PAR \,,
\end{equation}
where PAR
is the peak-to-average power ratio (see \secref{photonCountingCapacity}).
Thus to achieve \ile{\BPP{=}10.9} in \tblref{receiverParameters} we must have, at minimum,
\ile{\PAR{>}2^{10.9}{=}1911}.

 High PAR is beneficial to photon efficiency, but it does have implications 
 for the peak transmit power $P_P^T$ required in the transmitter (see \secref{laser}).
     A high peak power also requires short-term energy storage
     and possibly voltage conversion
    in the probe, since electric
    power is likely to be generated continuously
    (see \secref{powergeneration}).

PAR and optical bandwidth $W_e$ are related in a subtle
but significant way.
A larger PAR is inevitably associated with a larger $W_e$, since it requires relatively high-energy
pulses with short duration and low duty cycle.
A larger $W_e$ is beneficial in reducing the required selectivity of the receive optical bandpass filters.
A larger $W_e$ would also appear to admit more background radiation to the
optical detector, thereby decreasing SBR.
This is not entirely the case as now shown.

\subsection{Burst pulse-position modulation (BPPM)}
\label{sec:BPPM}

The modulation
has a major influence on photon efficiency.
For the purposes of our design exploration we have chosen
a novel burst pulse-position modulation (BPPM).
The structure of the intensity-modulated signal is
defined in \figref{BPPM}.
The signal is divided into \emph{PPM frames}, each with
\emph{frame duration} $T_F$.
Frames are typically interspersed with
\emph{blank intervals}, 
during which no optical power is transmitted.
The total time interval between
the beginning of successive frames is
the \emph{frame interval time} $T_I$.

Within each frame, the signal is structured
as exactly one pulse of light, which may occur
in one of \ile{M=2^m} short time intervals called
\emph{slots}, each with \emph{slot duration} $T_s$, 
so that \ile{T_F = M T_s}.
This internal frame structure
is called \emph{pulse-position modulation} (PPM).
PPM is commonly used in optical communications
because it trades greater bandwidth
(often an ample resource)
for improved energy efficiency.
However, it is typically used in the conventional PPM
configuration in which blank intervals are omitted.
Slot time $T_s$ is substantially reduced in BPPM without
changing $M$ or $T_I$ and thus without any effect on data rate.

\incfig
	{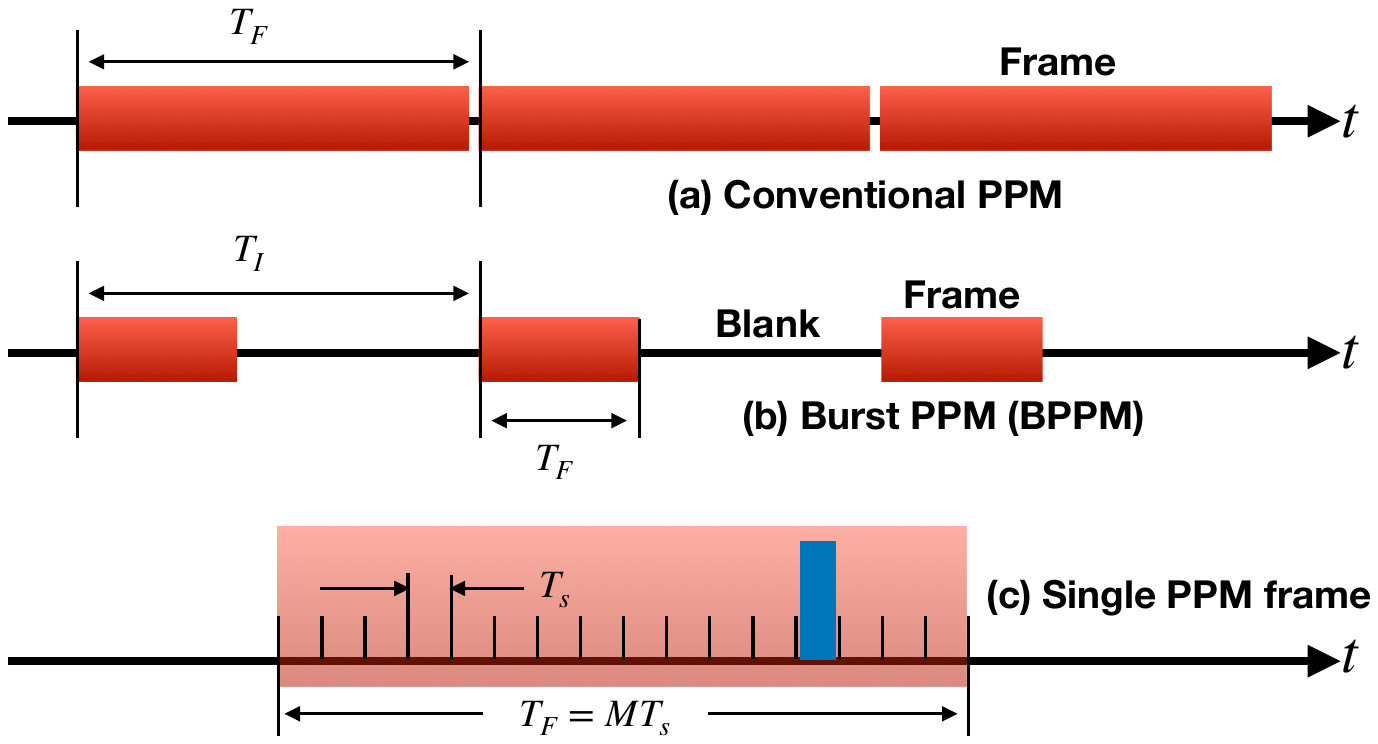}
	{
	trim=80 200 20 20,
    	clip,
    	width=1\linewidth
	}
	{
	An illustration of two forms of pulse-position modulation (PPM).
	(a) In conventional PPM the time axis is filled with PPM frames, or \ile{T_F = T_I}.
	(b) In burst-PPM (BPPM) the PPM frame duration $T_F$ is decoupled from
	the inter-frame interval $T_I$ by choosing a very small value for $T_s$, allowing for blank intervals between frames.
	(c) The internal structure of each PPM frame consists of $M$ time slots, each
	with duration equal to the slot time $T_s$.
	}

The disadvantage of BPPM over conventional PPM 
is an increase in peak transmit power 
(due to the reduction in $T_s$)
$P_P^T$.\footnote{In the dark-count dominated regime BPPM
doesn't actually increase $P_P^T$,
because it increases PAR by 
reducing average power rather than increasing peak power
(see \figref{powerVsSlotVsDarkPeak}).
}
However, this is offset by some significant advantages:
\begin{itemize}
\item
Optical dark counts
originating as blackbody radiation in the optical path and in the optical detector
are a major issue due to the potentially large number $N^S$ of optical subsystems
and optical detectors.
The impact of dark counts in the receiver can be reduced to any degree desired
by reducing slot time $T_s$.
 (see \secref{opportunities}).
\item
As the propagation distance $D$ from the probe increases,
the data rate $\rate$ has to decrease
(starting at $\rate_0$ at the beginning
of transmission following the completion of scientific data collection)
to accommodate the decreasing signal power reaching the receiver.
Similarly probes with different masses or more advanced technology 
may have greater electrical power available, allowing for an increase in $\rate$.
In PPM it is cumbersome to modify $\rate$, but with BPPM changes in rate are
trivially accomplished by changing the blank interval time 
while keeping frame interval $T_F$ constant.
This is a major theoretical and practical simplification of the design,
because no adjustment in optical bandwidth, ECC or other aspect of the design is required
(see \secref{increasingDistance}).
\item
It is difficult to achieve the desired selectivity in a receiver optical
bandpass filter with current technology, and adjustment of bandwidth to a changing
data rate is also a challenge.
With BPPM the optical bandwidth can be increased to any degree desired,
relaxing the selectivity requirement, and can be held fixed across time, distance,
and probes with different masses.
This accrues without penalty in background radiation
since the increased bandwidth is compensated by the longer blank intervals
 (see \secref{opportunities}).
\end{itemize}

The receiver, as part of the post-processing, must synchronize its timing
with frames and slots, a functionality called
\emph{timing recovery}.\footnote{
Timing recovery is a standard function of
all digital communication systems, and will
not be addressed further.
This is one of the post-processing functions
included in \figref{apertureArchitecture}.
}
With knowledge of the frame and slot timing,
there are three possibilities for how many photons are detected:
\begin{itemize}
\item
One or more photons are detected within
the slot populated at the transmitter,
but none in the other slots nor the blanking interval.
\item
No photons are detected within the entire frame,
which is declared a \emph{frame erasure}.
\item
One or more photons are detected within
two or more slots, which is declared a \emph{recognized frame error}.
\item
One or more photons are detected within exactly
one slot not populated at the transmitter.
This is an \emph{unrecognized frame error}.
\end{itemize}
Erasures are a manifestation of
signal shot noise and
frame errors are attributable to background radiation.

Knowledge of erasures and recognized frame errors
is preserved for subsequent processing.
We find that to achieve high photon efficiency BPP the
rate of erasures must be 80-90\%, but the
inclusion of substantial error-correction redundancy
can overcome these erasures and yield acceptably low
error rates in the recovery of the scientific data
(see \secref{highBPP2}).

In the conventional PPM shown in \figref{BPPM}a
the blank intervals are eliminated, but in the BPPM shown
in \figref{BPPM}b blank intervals between frames
are allowed (conventional PPM is a special case).
In this case the \emph{duty cycle} \ile{\delta = T_F/T_I} is defined.
We can think of BPPM as a higher-data-rate transmission
scheme operating with a low duty cycle \ile{\delta{\ll}1}.
It is important to note that \ile{m/T_I \gg \rate} where $m/T_I$ is the ``raw'' 
data rate carried by BPPM frames and $\rate$ is the rate of reliable
recovery of scientific data.
This disparity is attributable to the added redundant data
associated with error-correction coding (see \secref{ECC}).

\section{Probe mass tradeoffs}
\label{sec:masstradeoffs}

A feature of a directed-energy launcher is that, for a fixed launcher infrastructure,
a variety of probe versions with different masses can be launched.\footnote{
The many uses of a launcher, its deployment,
as well as the benefits and issues of various mass missions
are discussed in \citeref{1017}.
}
The probe speed $u_0$ and initial
data rate $\rate_0$ in \tblref{transmitterParameters}
are used in the numerical examples here.
These are estimates for a lowest-mass ``wafer scale'' version.
The question is, on what basis do we choose a probe mass?

In terms of scientific outcomes the  $\rate_0$ is of
only indirect interest.
In terms of the scientific mission
the performance metrics of direct interest are
the total data volume $\volume$ returned,
the data latency $T_L$, and the reliability with which the data is
recovered (see \secref{scientific}).
We now show that there is an optimum mass $m$
such that $T_L$ is minimized for a given
$\volume$, or $\volume$ is maximized for
a given $T_L$.
Thus a principled way to choose the mass
is to incorporate any constraints on
$T_L$ or $\volume$.
For comparison, estimates of these metrics
for two outer-planetary missions are listed
in \tblref{planetaryData}.

\doctable
	{planetaryData}
    {
    Data latency/volume for planetary missions\footnote{
    9000 images were returned from Neptune
    \citeref{831} at
    2 Mb per image \citeref{832}.
    The data rate was programmable and variable.
    The Pluto scientific data return was 50 Gb
    \citeref{828}. 
    }
    }
    {lp{2.5cm}p{2.5cm}}
    {
     \textbf{Target} & \textbf{Neptune} & \textbf{Pluto}
     \\
	\textbf{Mission} & \textbf{Voyager 2} & \textbf{New Horizons}
     \\[3pt] \hline \\ [-2ex]
     Data rate $\rate$ & $\sim 1$ kbps & $\sim 1$ kbps
     \\
     Data latency $T_L$ & 12 yr & 10.5 yr
     \\
     Data volume $\volume$ & 18 Gb & 50 Gb
    }

\subsection{Mass scaling approximations}
\label{sec:massScaling}

The mass $m$ strongly influences the probe speed $u_0$ and 
the initial scientific data rate $\rate_0$, 
which in turn determine $T_L$ and $\volume$.
We define a set of scaling laws,
\begin{align}
\notag
&m(\massratio) = \massratio \cdot m(1)
\\
\label{eq:massScaling}
&u_0 (\massratio) = \massratio^{-1/4} \cdot u_0 (1) \,,\quad
\rate_0 (\massratio) = \massratio^2 \cdot \rate_0 (1) \,.
\end{align}
The mass ratio \ile{\massratio \ge 1} is the factor by which
mass is increased. 
Transmission starts at the target star distance \ile{D = D_0}
and ends at distance $D_1{>}D_0$.
The data rate starts at $\rate_0$ and decreases
with $D$ in accordance with \eqnref{rateVsDistance}.
The scaling of $\rate_0$ in \eqnref{massScaling}  assumes that \ile{\{P_A^T, P_P^T, A_e^T\} \propto \massratio}
(the detailed scaling depends on the specifics of the probe design).
It is shown later (see \eqnref{linkbudget} and \eqnref{rateFromBPP} )
that in this case $\rate_0$
follows \eqnref{massScaling}, with one factor of
$\massratio$ due to increased $P_A^T$ and another factor of $\massratio$
due to increased $A_e^T$.
The scaling ratio for $u_0$ follows from the launch dynamics,
assuming a fixed launch infrastructure \citeref{1017}.

\subsection{Latency-volume tradeoff}
\label{sec:latencyVolumeTradeoff}

The relationship between $\volume$ and $T_L$ parameterized
by $\massratio$ is determined
in \secref{latencyVolume} and plotted in \figref{latencyVsVolume}
for five values of $\massratio$.
The volume scales as \ile{\volume \propto \rate_0 (1)}, so a
change in $\rate_0 (1)$ would shift all curves horizontally.
There are three components of $T_L$: transit time, transmission time, and propagation time.
The minimum $T_L$ increases with $\massratio$ due to the
greater transit time.
Eventually $T_L$ increases rapidly with $\volume$
due to the decreasing $\rate$ with $D$.

\incfig
	{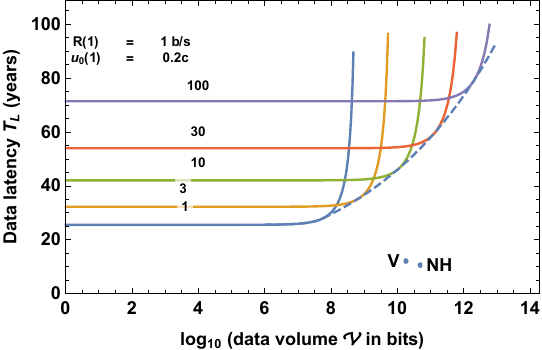}
    {width=1\linewidth}
    {
    Plot of the data latency $T_L$ in yr
    vs the logarithm of the total data volume
    $\volume$ in bits for different values of the 
    mass ratio $\massratio$
    (the curves are labeled with $\massratio$).
    For example \ile{\log_{10} \volume = 9} corresponds to
    $10^9$ bits or \ile{1\ \text{Gb}}.
    The dashed line is, for each value of $\volume$,
     the minimum $T_L$ achievable
    by the optimum choice of $\massratio$.
    It is assumed that for \ile{\massratio = 1} 
    (the lowest-mass probe),
    the speed is  \ile{u_0 (1) = 0.2 c}, and the 
    initial transmit data rate is \ile{\rate_0 (1) = 1\ \text{b/s}}. 
    The Voyager (V) and New Horizons (NH)
    points are obtained from \tblref{planetaryData}.
    }

For each $\volume$ there is an optimum choice of $\massratio$
that minimizes $T_L$.
The resulting minimum $T_L$ vs $\volume$ is plotted as the dashed curve
in  \figref{latencyVsVolume} and is
also plotted in \figref{dataLatencyDecomposition}.
The latter plot shows the decomposition of $T_L$ into its
three components.
The
latency-minimizing value of $\massratio$,
the resulting transmission time,
and the distance $D_1$ at the end of transmission
are plotted in \figref{transmissionAndMassRatio}.

\incfig
	{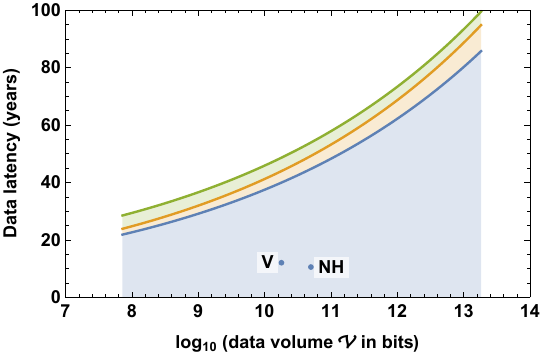}
	{width=1\linewidth}
	{
	A plot of the data latency $T_L$ vs data volume $\volume$,
	where for each value of $\volume$ the latency-minimizing
	mass ratio $\massratio$ is chosen.
	This ranges about 28 to 100 years, where larger $\volume$
	is associated with a larger $\massratio$ and a larger $T_L$.
	$T_L$ is decomposed into its three components, from bottom to top: 
	Transit time to the star, the transmission time, and the
	return signal propagation time, all measured by an earth clock.
	Latency is dominated by the transit time.
	}

\incfig
	{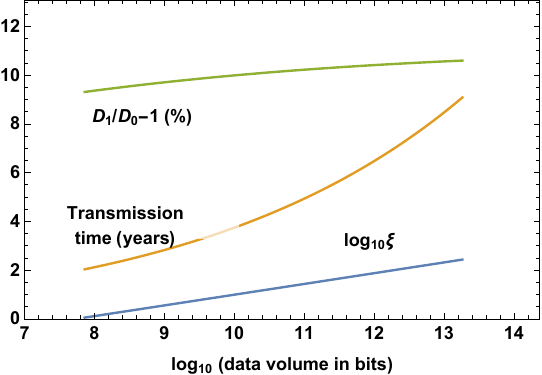}
	{width=1\linewidth}
	{
	Three quantities of particular interest are plotted as a function of
	data volume $\volume$.
	First is the log of the optimum mass ratio $\massratio$ that
	minimizes the data latency $T_L$.
	A plot of $T_L$ vs $\volume$ for this optimum $\massratio$ is
	shown in \figref{dataLatencyDecomposition}.
	Second is the transmission time component of $T_L$,
	which varies in the range of two to nine years.
	Third is the distance increase \ile{(D_1/D_0-1)} expressed as a percentage.
	The space probe travels a fairly consistent 9\% to 10\% farther than the target star
	during downlink transmission.
	}

If $\volume$ is increased, both the optimum $\massratio$ and $T_L$ increase.
For the optimum $\massratio$ the transmission time is always much
shorter than the transit time.
Obtaining a larger $\volume$ requires increased transmission time, but
the distance traveled during transmission remains relatively constant
since the probe's speed is lower.
For the range of parameters shown,
the minimum $T_L$ falls in the 30 to 80 year range for
a data volume in the 100 Mb to Tb range.
Volumes $\volume$ comparable to the planetary missions are feasible with the assumed $\rate_0 (1)$, 
but only with \ile{\massratio \sim 20} and \ile{T_L \sim 50\ \text{yr}}.
The $\volume$ and $T_L$ parameters of \tblref{transmitterParameters} assume that downlink operation
is 10\% of the probe transit time.

\section{Choice of wavelength}
\label{sec:wavelength}

The two spectrum ranges offering atmospheric transparency
for a terrestrial receiver are millimeter wave radio and visible optical.
This choice affects nearly every aspect
of the design.
We study the optical case in this paper, but the radio possibility
should also be considered.

\subsection{Propagation loss}
\label{sec:LinkBudget}

As compared to the outer
planets, the propagation losses from the
nearest stars approach a factor of ${>}10^8$.
This obstacle is overcome by an
adjustment to other parameters in \parameters{,}
including prominently wavelength, data rate,
and receive aperture size.

The appropriate signal level at the receiver
is measured differently at radio and optical
wavelengths.
For radio wavelengths it is the average power
\ile{P_A^R = N^S P_A^S}, 
while for optical (assuming
direct detection) it is
the average rate of detected photons 
\ile{\Lambda_A^R = N^S \Lambda_A^S}.
These metrics are affected by the
propagation loss and
atmospheric effects,
with the latter more significant at optical
wavelengths (see \secref{atmosphere}).
Neglecting
atmospheric absorption and turbulence,
these metrics are determined by the average 
transmitted power $P_A^T$ through a link budget,
\begin{align}
\label{eq:FriisPower}
&P_A^S =
\eta \cdot \frac{A_e^T A_e^S}{\lambda_0^2 D_0^2}
\cdot P_A^T
\\
\label{eq:linkbudget}
&\Lambda_A^S =  \frac{\lambda_0}{h c} \cdot P_A^S =
\eta \cdot 
\frac{A_e^T A_e^S}{h c \lambda_0 D_0^2}
\cdot P_A^T
\end{align}
(see \secref{signalResponse} and \eqnref{Friis}).
The factor $\lambda_0/hc$ in \eqnref{linkbudget} converts from power to photons per second.

The \emph{efficiency factor} \ile{0 < \eta \le 1}
in \eqnref{linkbudget} accounts
for non-propagation losses, such as 
aperture pointing error (see \secref{attitudeControl}),
as well as attenuation and 
the quantum efficiency in the receiver.
A Doppler red shift in wavelength 
from transmitter to receiver has also
been neglected in \eqnref{linkbudget} by setting \ile{\lambda_0^R = \lambda_0^R = \lambda_0}
(for \ile{u_0 = 0.2 c} this shift is about 20\%).
Relativistic aberration has also been neglected.

\subsection{Radio}
\label{sec:radioWavelength}

Although radio is used for deep space
missions today, the primary objection to radio
at interstellar distances is the large
aperture sizes (or alternatively large transmit power).
As an order-of-magnitude estimate, factors affecting the link budget for
the New Horizons (NH) mission are scaled to
accomodate interstellar distances in \tblref{interstellaradio}.
Although it would be desirable if transmit power and aperture size
could be increased to compensate for the greater distance,
they will likely be
substantially \emph{decreased} for low-mass probes.
In principle these adverse elements can be compensated by a lower data rate,
a substantially larger receive aperture area,
and/or shorter wavelength.
The shortest radio wavelength that consistently penetrates the atmosphere is
chosen.

\doctable
	{interstellaradio}
    {For equivalent SNR, scaling of the NH radio downlink to a low mass probe illustrating the impractically large receive aperture size
    that may be needed.}
    {lcc}
    {
     \textbf{Parameter} & \textbf{NH} & \textbf{Low mass}
\\[3pt] \hline \\ [-2ex]
     Distance $D_0$ & 4.5 lh & 4.24 ly 
\\
     Wavelength $\lambda_0$ &  3.6 cm & 10 mm 
\\
     Transmit power $P_A^T$ & 24 W & 0.1 W 
\\
     Transmit aperture diameter $\sqrt{4 A_e^T / \pi}$ & 2.1 m & 10 cm 
\\
     Receive aperture diameter  $\sqrt{4 A_e^R / \pi}$ & 70 m & 1652 km 
\\
     Data rate $\rate$ & 1 kb/s & 1 b/s 
}

An appropriate link budget for radio is \eqnref{FriisPower}.
Assuming that there is one single-mode diffraction-limited \leveltwo which receives signal power $P_A^S$,
the parameters contributing to this received power $P_A^S$ are listed
in  \tblref{interstellaradio}, with reasonable assumptions for the transmit power
and aperture size.
When substituted into \eqnref{FriisPower} the received power is found to be a factor of $10^3$
smaller for the low-mass probe.
This results in an equivalent signal-to-noise ratio SNR at the receiver for two reasons.
First the major noise impairments are isotropic cosmic background radiation
and receiver thermal noise,
which have the same white power spectral density independent of receive aperture area
(see \secref{noiseResponse})
and assuming equivalent receiver noise factor.
Second for an equivalent modulation and coding scheme the receiver bandwidth can be
reduced by a factor of $10^3$ in the low-mass case due to the lower data rate,
thereby reducing the noise component of SNR by the same factor.

The receive aperture is impractically large, especially considering that it
is assumed to be diffraction limited.
Such an aperture would also require an impractical pointing accuracy,
although this requirement could in principle be relaxed by implementing a phased array
with equivalent total collection area.\footnote{
The incoherent accumulation of power as in \figref{apertureArchitecture} is not acceptable
since the NH modulation and coding scheme strongly exploits both signal amplitude and phase.
Accurate phasing across constituent apertures is necessary.
}
A larger transmit aperture (with associated greater pointing accuracy), higher transmit power, or
lower data rate would allow the receive aperture area to be reduced.

\subsection{Optical}
\label{sec:opticalWavelength}

Relative to radio,
choosing an optical wavelength yields a factor of
$10^4$ to $10^5$ in the link budget of \eqnref{linkbudget}
(depending on the two wavelengths being compared).
This is a considerable advantage, but
optical wavelengths also pose substantial challenges.
One is atmospheric effects (see \secref{atmosphere}), with the resulting
scintillation (fading in communication terms) and 
weather-based outages.

One challenge in optical communication
with direct detection (see \secref{directheterodyne}) is a
`mode shift' in photon efficiency BPP
vs SBR at low values of SBR.
This is observed in the theoretical limit
on feasible photon efficiency BPP,
which is essentially independent of SBR
(for fixed peak-to-average ratio PAR)
for large values of SBR,
but deteriorates rapidly for small SBR
(see \figref{bitsPerPhotonVsABlargeSBR}).
Intuitively this is due to the effect of modulation and
detection based on signal power (photon count),
and the effects of square-law detection on background radiation.
The case is made in \citereftwo{887}{868}
that a probe receding in distance from earth
will suffer a deterioration in SBR due
in part to a reduction in data rate and
signal power in the face of a fixed
(or even growing) background radiation,
and as a result the link budget switches dependency
from $D^2$ to $D^4$.
However, these papers make specific technological
assumptions as to modulation and parameter scaling relations with $D$
(such as holding optical bandwidth fixed even as the signal bandwidth decreases).\footnote{
There is also one error in \citeref{868},
with the assumption that isotropic sources of noise increase
as the effective area and directivity 
of the receive aperture is increased 
to compensate for greater distances.
Actually noise due to unresolved
sources is independent
of directivity because the greater
gain of a larger aperture
is exactly offset by its smaller
coverage (see \secref{noiseResponse}).
}

A desirable $D^{-2}$ behavior can be achieved at deteriorating SBR
by replacing PPM by a more nuanced modulation and coding \citeref{880},
although this is unproven in practice.
Here we use a simpler approach of simply constraining SBR to be relatively large.
Fortunately for a space mission and its
presumed maximum propagation distance \ile{D = D_1}
this challenge is circumvented by ensuring
that SBR is relatively large at distance $D_1$.
We constrain
SBR to a relatively large value at this distance (in our case \ile{\SBR{=}4}),
with a correspondingly large average transmit power $P_A^T$.
This is assisted by BPPM (see \secref{BPPM}), which allows us
to hold optical bandwidth $W_e$
fixed (subject to any technological limitations on bandpass filtering)
without penalty in SBR as $D$ increases, although a price is paid in
a correspondingly growing peak transmit power $P_P^T$.

Our numerical results in \secref{background} 
subject to these assumptions do strongly suggest that
communication with low-mass probes at
optical wavelengths is not
feasible given the current state of technology.
We identify here some key technologies that
must advance to empower optical communication
in this application.
These are: achieving relatively high peak powers
at the optical detector through pulse compression or other means
(see \secref{laser});
more selective optical bandpass filtering
(see \secref{bandpassFilter});
attenuation of interference by a coronagraph
function (see \secref{coronagraph});
receiver optics (including superconducting optical detectors) with very low dark counts
(see \secref{opticalDetector});
and highly accurate probe attitude control and aperture pointing.

Since the relative advances in these technologies
cannot be anticipated with any degree of
certainty,
BPPM is chosen as a modulation technique
that offers greater freedom in trading off advances
in some of these technologies (see \secref{BPPM}).
Achieving a relatively large and constrained
SBR at the greatest distance $D_1$
is aided by the minimal increase
in distance (about 10\%) during transmission
(see \figref{transmissionAndMassRatio}).
The requirements for
these four technologies and related tradeoffs
are studied by numerical modeling in
\secref{background}.

\subsection{Space-based receiver}

Although not considered further here, a
space-based receiver would be advantageous for 
the elimination of atmospheric effects
and also the possibility of going to ultraviolet (UV)
wavelengths (10 to 400 nm). 
UV would
further moderate aperture sizes and
substantially reduce the star interference 
(see \secref{background} and \citeref{1017}).
While UV laser communications is not currently technologically advanced
(due primarily to its inability to penetrate the atmosphere)
it is an option to consider for the future.

\section{Trajectories and coverage}
\label{sec:coverage}

The coverage $\Omega_A$ of each \leveltwo
determines the effective area $A_e^S$, which in turn influences the transmit power-area product $P_A^T A_e^T$
necessary to achieve the desired SBR at the receiver.
The parameters affecting coverage
are (beyond our control) parallax and the star's proper motion
and (under our control) the swarm launch schedule and
sophistication of the \leveltwo beamforming.

\subsection{Concurrent swarm probe communication}

The probe trajectories as viewed from Earth are
shown in \figref{parallaxTraj} for a specific launch schedule.\footnote{
The variation in angle due to
parallax does not depend on the chosen coordinate system.
To readily capture the earth's orbit,
a heliocentric ecliptic XYZ coordinate system
is used.
The earth orbit
and a fixed star position were
obtained from
Mathematica\textsuperscript{\textregistered}
\texttt{StarData["Heliocoordinates"]}
and
\texttt{PlanetData["Heliocoordinates"]}
followed by transformation to
spherical coordinates.
}
The chosen parameters, including duration of transmission
and interval between launches, results in 26 probes transmitting
concurrently \tblref{missionParameters}, each one operating a downlink for 2.21 yr
in \tblref{transmitterParameters} (see \secref{latencyVolumeTradeoff}).
For the position of Proxima Centauri
relative to the earth's orbit,
the overall longitudinal parallax angular variation
(\ile{{\sim}2\text{ as}})
is about twice as great as the latitudinal 
(\ile{{\sim}1\text{ as}}).

\incfig
    {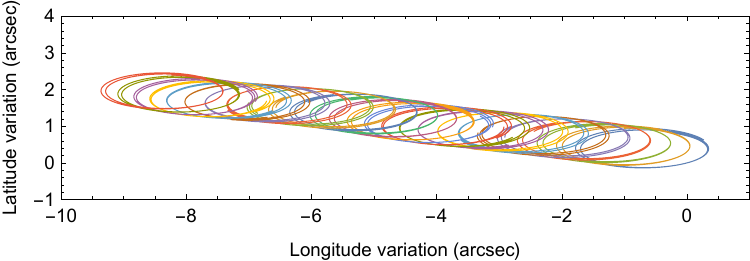}
    {width=1\linewidth}
    {
    The variation in probe coordinates
    as viewed by a common terrestrial receive
    aperture.
    The variation in the longitude
    and latitude (in arcsec
    in an ecliptic
    coordinate system)
   for 26 probes launched at 30-day intervals
   is shown.
    The origin corresponds to the angle of
    Proxima Centauri when viewed from the initial launch site.
    Each elliptical trajectory is due to the parallax
    effect of the receiver orbit about the sun over
    a period of 2.21 yr, which corresponds to a probe's
    travel 10\% farther than the star.
    The general drift of the probe positions is
    due to the star's proper motion, presenting
    a moving target for the various probes.
    }

The star's proper motion determines the
relative position of probes launched at different
times.
For Proxima Centauri this motion is
\ile{368\ \text{mas/yr}} and
\ile{- 3775\ \text{mas/yr}}
in declination and right ascension respectively.\footnote{
Obtained from
Mathematica\textsuperscript{\textregistered}
\texttt{StarData["ProperMotion"]}.
}
Fortuitously, the largest component of proper
motion is aligned with the largest overall parallax
(longitude and right ascension).
The trajectories of consecutively launched probes
will sometimes coincide
(unless the inter-launch interval is greater than about eight months)
and thus their signals could not be consistently separated spatially.

The coverage implications depend on the refinement
in beamforming in each \leveltwo.
Minimizing coverage (maximizing sensitivity)
requires beamforming
the precise location of each probe,
while the simplest beamforming simply
captures the full
variation in parallax shown in \figref{parallaxTraj}
for each probe and also across the relative positions
of successive probe launches.
The numerical results in \secref{background}
address these disparate cases by considering a range of
$\Omega_A$ values.

At longitude $l$ and latitude $b$
the solid angle of coverage is,
for small $\Delta l$ and $\Delta b$,
\begin{equation}
\label{eq:OmegaAvsBL}
\Omega_A \sim \cos (b) \cdot
\Delta b \, \Delta l
\end{equation}
Based on \figref{parallaxTraj} for the simplest
beamforming we have \ile{\Delta b \approx 1\ \text{as}},
and for Proxima Centauri \ile{\cos (b) = 0.71}.
$\Delta l$ is dominated by proper motion and is
\ile{\Delta l{\sim}10\text{ as}} for the assumptions in \figref{parallaxTraj},
and thus 
\ile{\Omega_A \approx 7.1\ \text{as}^2}.
To account for the practicalities of sidelobes we assume that
\ile{\Omega_A{=}10\ \text{as}^2} in \tblref{receiverParameters}.

The coverage can be manipulated by the launch schedule.
For example $\Omega_A$ could be reduced by 
moving toward a \emph{burst-mode launch schedule}, in which a
sub-swarm of probes are launched at closely spaced times interspersed
with launch-inactive time periods greater than the individual-downlink operation time
(on the order of two years for the parameters of 
\tblref{transmitterParameters} and \figref{transmissionAndMassRatio}).

\subsection{Single-probe communication}
\label{sec:singleProbe}

At the other extreme, if a receive aperture is dedicated to a single probe
and can dynamically compensate for parallax and atmospheric refraction of the signal path, 
the smallest coverage solid angle $\Omega_A$
will be determined by the \leveltwo pointing accuracy.
If we assume that with adaptive optics the pointing accuracy of the \leveltwo{s}
is \ile{{\approx}0.1\text{ as}}, then the coverage can be no smaller than \ile{\Omega_A{\approx}0.01\ \text{as}^2}
(this the value assumed in \tblref{receiverParameters}).\footnote{
The \ile{A_e^S{=}707.\ \text{m}^2} assumption in \citeref{1015} implies
\ile{\Omega_A{=}9.4\cdot 10^{-5}} and hence an \leveltwo pointing accuracy on the order of \ile{0.01\ \text{arcsec}} or better.
The \leveltwo sensitivity and the target star interference calculations are strongly dependent on this assumption.
}
Our goal in the numerical results is to consider this entire range of possibilities.

\section{Model and numerical results}
\label{sec:model}

Before discussing the key enabling technologies,
it is helpful to appreciate the
trade-offs available through the choice
of parameters of BPPM.
BPPM is
specifically designed to permit
a greater advancement of one technology
to offset a lesser advancement of another.
This is reflected in the modeling equations
for BPPM in \tblref{designEquations},
which are now described
and then used in an extended numerical example.

\doctable
	{designEquations}
    {Receive aperture design equations with BPPM}
    {ll}
    {
\hline \\ [-2ex]
\textbf{Parameters:} 
    \\
    \textbf{  Data reliability:} & \ile{\SBR, \BPP, \rate_0, W_e, K_s^R, m}
    \\
   \textbf{  Geometic:} & \ile{b, \Delta b, \Delta l, \Omega_A, \lambda^R, D_0}
    \\
    \textbf{  Physical:} & 
    \ile{\eta, F_c, F_x, T_s, \Lambda_{N,I,D}^S}
\\[3pt] \hline \\ [-2ex]
\textbf{Transmitter:}
    \\[3pt]
     \ile{T_F = M T_s}
    &
     \ile{W_e T_s > 1}
\\[3pt]
\ile{P_P^T T_s = P_A^T T_I}
&
 \ile{T_I \rate_0 = K_s^R \cdot \BPP}
 \\[2pt]
 \ile{\displaystyle{\BPP = \frac{1 - e^{-K_s^R}}{K_s^R} \cdot m}} & \ile{M = 2^m}
    \\[8pt] \hline \\[-2ex]
    \textbf{Receive \leveltwo:}
 	&
     \ile{\displaystyle{
		\Lambda_P^S = \frac{P_P^T A_e^T A_e^S}{h c \lambda^R D_0^2}
     }}
\\[8pt]
     \ile{K_s^S = \eta \Lambda_P^S T_s} & 
     \ile{K_X^S = \eta F_x \Lambda_P^S T_F}
\\[3pt]
    \ile{K_N^S = \eta \Lambda_N^S W_e T_F}  &
    \ile{K_I^S = \eta F_c \Lambda_I^S W_e A_e^S T_F}
\\[1pt]
     \ile{K_D^S = \Lambda_D^S T_F} &
    \ile{\displaystyle{ 
    	\SBR = 
   	 \frac{K_s^S}{K_X^S + K_N^S + K_I^S + K_D^S }
	 }}
\\[8pt]
     \ile{A_e^S \cdot \Omega_A = \big( \lambda^R \big)^2}
     &
     \ile{\Omega_A = \cos(b)\cdot \Delta b \Delta l}
\\[3pt] \hline \\ [-2ex]
\textbf{Receive \levelone:} & \ile{N^S = K_s^R/ K_s^S}
\\[3pt] \hline \\ [-2ex]
\textbf{Metrics:} & \ile{A_e^S, P_A^T, P_P^T, N^S}
 \\[3pt]
 \hline
     }

from the choice of parameters listed in \parameters
\ are listed in \tblref{performanceMetrics}, along with a few additional
chosen parameters.
and Excel\textsuperscript{\textregistered}
implementations of \tblref{designEquations}
are available for self-exploration of the parameter space \citeref{1021}.
%
This is our nominal design, and in the following the scaling of
performance with changes to chosen parameters is described.
%

Although the values of $N^S$ are quite large, the possibility of
multiple-\leveltwo assemblages addresses some concerns
(see \secref{scaleout}).
For example \ile{N^S{=}5.9{\cdot}10^7} can be accommodated by 5900 assemblages,
each accommodating a $100{\times}100$ array of \leveltwo{s}.
The total area metric $A_e^S N^S$ may be a more accurate indicator of \levelone capital cost.

\doctable
	{performanceMetrics}
    {Transmitter and receiver performance metrics}
    {| l p{4.5cm} | c | c |}
     {
     \hline
     & \textbf{Description} & \textbf{Swarm} & \textbf{Single}
     \\ \hline  \\[-2.5ex]
 	$P_{A}^{T}$ & Avg. transmit power (mW)
	 & $29.4$ & $0.8$
     \\[3pt]
	 $P_{P}^{T}$ & Peak transmit power (kW)
	 & $638.$ & $17.4$
    \\ \cline{3-4}
    $K_s$ & Average detected photons per slot & \tc{0.2}
    \\ 
       	$\rho_M$ & Reduction in moonlight \newline irradiance
   	relative to full moon
    	& \tc{1}
      \\
     	$\Lambda_{D}^S$ & Avg. rate of dark counts 
      	\newline referenced to each receive \leveltwo (ph/yr)
     	&
     	\tc{32.0}
     \\
     	SXR & Signal-to-extinction ratio & \tc{2441.}
    \\ \cline{3-4}
    	SIR & Signal-to-interference ratio & $152.$ & $4.15$
    \\
    	SNR & Signal-to-noise ratio & $8.3$ & $226.$
    \\
    	SDR & Signal-to-dark count ratio & $8.2$ & $223.$
    \\[2pt]
    	$N^S$ & Number of receive {\leveltwo}s & $5.9{\cdot}10^7$ & $2.2{\cdot}10^6$
    \\[2pt]
  	$A_e^S N^S$ 
    	& Total effective aperture area (km\textsuperscript{2}) & $0.04$ & $1.47$
    \\ \hline
    }

\subsection{Metrics and methodology}
\label{sec:methodology}

Achieving a gram-scale probe mass limit is challenging.
For this purpose a useful metric for a design is the product of the transmit 
\{average,peak\} power and aperture effective area \ile{\xi_{\{A,P\}}^T = P_{\{A,P\}}^T A_e^T},
which should be monotonically related to probe mass
due to their impacts on electrical power generation and the transmit aperture size (see \secref{massScaling}).
Fixing a value for $\xi_{A,P}^T$ admits a tradeoff between transmit power and aperture area without
affecting the signal level as seen by the receiver.

Our design methodology is to minimize the average-power-area metric $\xi_A^T$ subject to 
reliable recovery of scientific data,
constraints on receiver coverage accommodating probe-swarm trajectories,
and the physical limitations imposed by background radiation and self-noise in the receiver.
This minimization is achieved by operating at the lowest SBR consistent with
 reliable data recovery, operating near the theoretical limits of photon efficiency,
 and subject to the fundamental physical constraints on a receive aperture imposed by the
 desired coverage.
 A major component of receiver cost is the receive aperture.
 A metric relating to that cost is the receive aperture total effective area $A_e^S N^S$.
 We do not choose or constrain $N^S$, but
 rather ascertain the value required to achieve reliable data recovery
 subject to that minimum $\xi_A^T$.
 Subsequently we consider the implications of varying $N^S$ about this unconstrained choice.
 
 Recognizing a tradeoff between $\xi_{\{A,P\}}^T$ and receive aperture area,
 another useful metric for a design is the \emph{end-to-end} power-area-area metric
 \begin{equation*}
 \ile{\xi_{\{A,P\}}^{TR} = P_{\{A,P\}}^T A_e^T \cdot A_e^S N^S}
 =  \xi_{\{A,P\}}^T \cdot A_e^S N^S
  \,.
 \end{equation*}
 We now display the dependence of $\xi_{\{A,P\}}^{TR}$ on other parameters,
 which offers considerable insight.

\subsection{Transmitter/receiver tradeoffs}
\label{sec:reliableData}

There is a practically significant tradeoff between the transmitter power-area metric and
the required number of receive \leveltwo{s}.
As now established, increased transmitter resources (power and aperture area)
can be traded for a reduced receiver burden (number of \leveltwo{s}).

\newpage
\subsubsection{End-to-end metrics}

The design equations in \tblref{designEquations} determine
the end-to-end metrics $\xi_{\{A,P\}}^{TR}$ defined in \secref{methodology} as
 \begin{align}
 \label{eq:averagePowerAreaMetric}
 \xi_A^{TR} &= \xi_A^T \cdot A_e^S N^S =
 \frac{h c \lambda_0^R D^2}{\eta} \cdot \frac{\rate}{\BPP}
 \\
 \label{eq:peakPowerAreaMetric}
 \xi_P^{TR} &= \xi_P^T  \cdot A_e^S N^S =
 \frac{h c \lambda_0^R D^2}{\eta} \cdot \frac{K_s^R}{T_s}
 \,.
 \end{align}
$\xi_A^{TR}$ in \eqnref{averagePowerAreaMetric} is a 
reformulation of the single-\leveltwo link budget \eqnref{linkbudget}
 for the canonical receive \levelone of \figref{apertureArchitecture}.
 Expressed in terms of coverage solid angle $\Omega_A$, 
 \eqnref{averagePowerAreaMetric} and  \eqnref{peakPowerAreaMetric} become
\begin{equation}
\label{eq:xiTscaling}
 \xi_{\{A,P\}}^{T} =
 \frac{h c D^2}{\lambda_0^R \eta} \cdot \frac{\Omega_A}{N^S}  
 \cdot \left\{ \frac{\rate}{\BPP}, \frac{K_s^R}{T_s} \right\}
\,.
\end{equation}
Thus for fixed coverage we have that \ile{ \xi_{\{A,P\}}^{T} \propto \left( N^S \right)^{-1}},
so that if desired $N^S$ can be reduced at the expense of added transmit power or aperture area or both
(see \secref{reducingArea}).
This tradeoff is actually exercised by manipulating the hidden parameter SBR.
Increasing $\xi_{\{A,P\}}^{T}$ results in a higher SBR at each receiver \leveltwo
(because of a larger signal rather than lower background radiation),
and this in turn allows an adequate signal level at the optical detector with a smaller
number of \leveltwo{s}.

With respect to average power,
$\xi_A^{T}$  in \eqnref{xiTscaling}
 quantifies the adverse effect
 of achieving a smaller BPP  and/or smaller quantum efficiency $\eta$,
 and establishes a linear dependence of average power on scientific data rate $\rate$.
 It also shows that
 for fixed coverage $\Omega_A$, a longer wavelength is advantageous,
 although this fails to account for an important
 effect of wavelength on background radiation.
 Because it is blind to SBR,
 \eqnref{xiTscaling} also fails to capture a reduction in background radiation
 that follows from reducing BPP (see \secref{reducingPower}).

 With respect to peak power, \eqnref{xiTscaling}
shows an invariance of $\xi_P^{TR}$ to $\rate$ and BPP, 
 but rather a dependence on parameters $T_s$ and $K_s^R$.

\subsubsection{Parameter consistency issues}

For any set of \emph{consistent} parameter choices, \eqnref{xiTscaling} is satisfied.
However, \eqnref{xiTscaling} is not satisfied for any \emph{arbitrary} choice of the
data reliability parameters \ile{\{\SBR, m, K_s\}}, because BPP depends strongly on
these parameters and \eqnref{xiTscaling} assumes that a photon efficiency BPP is actually achieved.
We must chose \ile{\{\SBR, m, K_s\}} and then determine the resulting BPP before invoking  \eqnref{xiTscaling}.
The average power-area tradeoff is strongly influenced by BPP, and this imparts a strong motivation to increase BPP.

Parameter $K_s^R$,
the average number of detected signal photons for each PPM slot,
has a major influence on data reliability since it determines
the shot-noise induced rate of PPM frame erasures.
To achieve a large BPP, which beneficially reduces $\xi_A^{TR}$, it is perhaps
surprising that $K_s^R$ must be very small, resulting in a high probability of frame erasure
(see \secref{ECC}).
A nearly optimum choice across a range of conditions is \ile{K_s^R{\approx}0.2}
(see \secref{theoreticalPPMvsPhotonCounting}).
This is termed \emph{photon starvation} of the PPM frames, and small $K_s^R$ beneficially
reduces $\xi_P^{TR}$ as well.
On the other hand, the equivalent parameter $K_s^S$ at an \leveltwo output is determined
by SBR, and is not directly related to BPP or data reliability.
The purpose of the scale-out of \leveltwo{s} by the factor \ile{N^S = K_s^R/K_s^S} 
is to multiply the value of
$K_s$ by incoherently accumulating received photons.

The signal-to-background ratio SBR must also be large enough to support the assumed
photon efficiency BPP (see \secref{photonCountingCapacity}).
Since the background radiation is determined by the \leveltwo, the minimum SBR requirement
imposes a lower bound on the average signal power at each \leveltwo,
and thus places a lower bound on $\xi_{\{A,P\}}^{T}$ (see \secref{powerArea}).

\subsection{Changes with increasing propagation distance}
\label{sec:increasingDistance}

The downlink propagation
distance $D$ ranges over \ile{D_0{<}D{<}D_1},
where $D_0$
is the distance to the target star
(corresponding to scientific data acquisition)
and $D_1$ is the distance at which
transmission is completed.
As $D$ increases in this range,
a goal is to hold constant both the average
transmit power (which is determined by the
electrical generation capacity) and the
reliability of 
scientific data recovery.
It follows that the only parameters in the end-to-end
metrics that can change are data rate $\rate$ and
peak transmit power $P_P^T$.
Assuming \eqnref{averagePowerAreaMetric}  remains fixed,
we must have \ile{\rate \propto D^{-2}}.
If the initial data rate
at \ile{D = D_0} is $\rate_0$ then
\begin{equation}
\label{eq:rateVsDistance}
    \rate (D) = \rate_0 \cdot
    \left( \frac{D_0}{D} \right)^2 \,.
\end{equation}

For fixed data reliability we choose to
keep \ile{\{M, T_s, T_F, K_s \}} fixed with $D$.
Based on \tblref{designEquations}, $\rate$
can be changed simply by adjusting the frame
interval \ile{T_I \propto \rate^{-1}}.
In addition, based on \eqnref{peakPowerAreaMetric}
the peak transmit power has to be increased
as \ile{P_P^T \propto D^2}.

In summary, as $D$ increases $\rate$ is
adjusted downward by increasing $T_I$,
and $P_P^T$ is increased to maintain
a fixed reliability and also a fixed $P_A^T$.
All ECC parameters and algorithms remain fixed,
which is one of the desirable features of BPPM.

\subsection{Coverage}

The sensitivity of the \leveltwo is
determined by $A_e^S$, as seen in the
relation for $\Lambda_P^S$,
and thus coverage
and sensitivity are inversely related.
That is, a swarm of probes received concurrently
by a single receive aperture, to the extent
that this necessitates a larger $\Omega_A$,
increases the required $P_{\{A,P\}}^T$ for each probe
(see \tblref{performanceMetrics} for a specific example).

\subsection{Power-area tradeoff}
\label{sec:powerArea}

It is useful to numerically characterize the transmitter/receiver tradeoffs
described analytically in \secref{reliableData}.
The tradeoff between receive \levelone total area and transmit average and peak transmit powers
(for a fixed transmit aperture area $A_e^T$) is shown in \figref{PapVsAreaSwarm} and \figref{PapVsAreaSingle} for the swarm and
single probe coverage cases respectively, holding the transmit aperture $A_e^T$ fixed.
The range in powers and areas is very large, with impractically large powers on the left 
(corresponding to a single \levelone \leveltwo) and impractically large receive \levelone 
areas on the right (corresponding to the minimum photon efficiency).
As expected, the transmit powers are smaller in the single probe case of \figref{PapVsAreaSingle},
and as a result the received \levelone areas are generally larger.

\subsubsection{Reducing receive \levelone area}
\label{sec:reducingArea}

Relative to the nominal design with parameters defined in \parameters,
the receive \levelone area can be reduced by increasing SBR.
Since background radiation is not affected when
$A_e^S$ is fixed, an increase in $\xi_{A}^{T}$ is reflected by an increase in photon rates
at the \leveltwo output, an increase in SBR, and a commensurate reduction in $N^S$ for a fixed photon efficiency BPP.
The largest SBR and transmit power corresponds to a single \levelone \leveltwo (\ile{N^S{=}1}).\footnote{
The design assumption made in \citeref{1015} is \ile{N^S{=}1}, with the goal of minimizing the received \levelone area.
As a result the
required transmit area-power metric $\xi_{A}^{T}$ is more than eight orders of magnitude larger
than the nominal design here, but of course the receive collector area is much smaller.
}

\subsubsection{Reducing transmit power}
\label{sec:reducingPower}

Again relative to the nominal design parameters,
the transmit power can be \emph{reduced} in the specific context of PPM or BPPM by reducing the photon efficiency BPP
through a reduction in the number of slots $M$ per frame.
The primary effect of reducing BPP is to increase the $N^S$
needed to achieve a large enough photon rate to support the assumed scientific data rate $\rate$.
This larger receive \levelone area results in lower transmit powers.

Specifically this trend is due to a reduction in background radiation, and permitting the transmit powers
to be decreased while maintaining the desired SBR.
In PPM, BPP is controlled by adjusting the bits per PPM slot $m$, with \ile{\BPP \approx 0.9 m}
for the nominal value of $K_s$.
For conventional PPM, reducing $m$ also beneficially increases time-slot duration $T_s$,
and thus reduces optical bandwidth $W_e$ and thus $\Lambda_{I,N}^S$ in \tblref{SBR}.
For BPPM, holding $T_s$ and $W_e$ fixed imply that $\Lambda_{I,N}^S$ remains fixed,
but the frame duration $T_F$ is proportional to $2^m$ and thus reducing $m$ has the effect of
reducing background photon counts $K_{X,I,N,D}^S$ in \tblref{designEquations}.
Reducing BPP to increase aperture area is less efficient for average power than it is for peak power
due to the increase in area needed to accommodate poorer photon efficiency, which begins to offset
the beneficial reduction in background radiation at small BPP.

\incfig
	{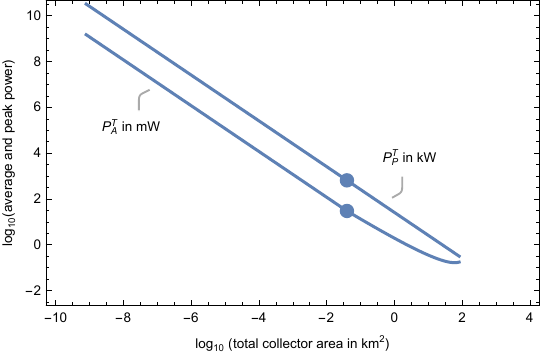}
	{
	trim=0 0 0 0,
    	clip,
    	width=1.\linewidth
    	}
	{
	A log-log plot of the average transmit power $P_A^T$ and peak transmit power $P_P^T$
	vs the total receive aperture effective area $N^S A_e^S$ with fixed coverage (\ile{\Omega_A = 10\ \text{arcsec}^2})
	corresponding to the swarm probe case.
	The units are milliwatts for average power and kilowatts for peak power, so there is a six order of magnitude difference
	between the two curves.
	The dots represent the nominal case for the parameters listed in \parameters.
	The larger powers to the left of the dots reflect a larger choice for SBR, culminating with the SBR corresponding
	to a single \leveltwo (\ile{N^S{=}1}).
	The smaller powers to the right of the dots reflect a smaller choice for BPP, culminating in \ile{\BPP{=}1},
	corresponding to a PPM frame with just two slots (\ile{M{=}2}).
	}

\incfig
	{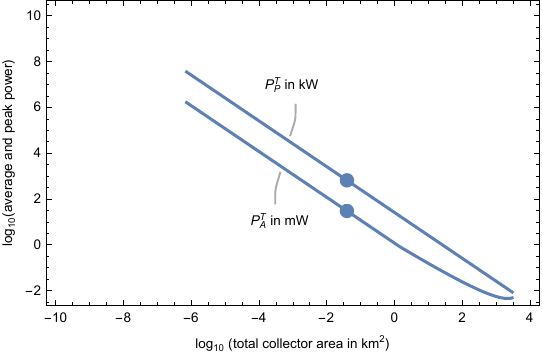}
	{
	trim=0 0 0 0,
    	clip,
    	width=1.\linewidth
    }
	{
	\figref{PapVsAreaSwarm} is repeated for fixed coverage (\ile{\Omega_A = 0.01\ \text{arcsec}^2})
	corresponding to the single probe case.
	}

\subsection{Background radiation}
\label{sec:backrad}

The relevant average detected photon rates
at each \leveltwo are listed in \tblref{SBR}
for each of the four
sources of background. 
These are related to the detected peak
photon rate $\Lambda_P^S$ and the
three physical parameters $\Lambda_{N,I,D}$,
the appropriate units for which
are listed in the third column.
Like the probe signal,
the three incident sources of radiation are
affected by detection efficiency $\eta$,
whereas the dark count rate is intrinsic to the
detector and thus is not affected by $\eta$.
Thus dark counts become a more significant
contributor to SBR as $\eta$ decreases.

\doctable
	{SBR}
    {Constituents of background radiation}
    {lll}
    {
    \textbf{Source} &  \textbf{Avg. ph/s} 
    & \textbf{$\mathbf{\Lambda_{P,N,I,D}}$ units}
     \\[1pt]
     \hline\\[-2ex]
Source extinction
    & \ile{\eta F_x \Lambda_P^S}
    & ph/s
    \\
     Unresolved noise 
     & \ile{\eta \Lambda_N W_e} 
     & ph/s-Hz
     \\
     Host star interference 
     & \ile{\eta F_c \Lambda_I A_e^S W_e} 
     & ph/s-Hz-m\textsuperscript{2}
     \\
     Detector dark counts & \ile{\Lambda_D^S}
     & ph/s
    }

Noise and interference are broadband
sources of radiation, and thus as they
reach the optical detector are
proportional to $W_e$, the bandwidth of the
receive optical bandpass filter which
precedes that detector.
The extinction ratio \ile{F_x{<}1} determines the
background following the
optical bandpass filter that is related to incomplete extinction
of the source.

Noise
(due to unresolved sources of radiation)
is considered to be
isotropic over the coverage solid angle $\Omega_A$,
and thus the captured photon rate is
independent of $A_e^S$ for a single-mode diffraction limited \leveltwo.
This is because any changes to collection area are
offset by the resulting change in coverage solid angle
(see \secref{noiseResponse}).

Interference (originating with the target
star) is essentially a point source
like the probe transmitter.
It can be approximated as
a plane wave when it reaches the \leveltwo,
and the captured photon
rate is thus proportional to the
\leveltwo effective area $A_e^S$.
Any rejection of this interference due to
it falling outside the coverage angle
or other coronagraph functionality is
modeled by the coronagraph rejection ratio
\ile{F_c{\le}1}.

Dark counts are distinctive in being
unaffected by either $W_e$ or $A_e^S$.\footnote{
This may not be true of dark counts originating as blackbody
radiation from the optics preceding the optical bandpass filtering
(see \secref{opticalDetector}).
}
This presents a unique challenge because
we cannot limit dark counts originating in the detector
(as opposed to receive optics) by optical bandpass filtering.

\subsection{Conditions on extinction ratio}
\label{sec:extinction}

Returning to \tblref{designEquations},
the average photon counts per frame
for signal ($K_s^S$) and for incomplete
extinction ($K_X^S$) are related to the
peak signal power $\Lambda_P^S$.
This implies an upper
limit on the permissible extinction ratio $F_x$
that is related to the desired SBR.
Eliminating $\Lambda_X^S$ from the equations,
\begin{equation}
\label{eq:extinctionMinimum}
K_s^S = \frac
    {\SBR \big( K_N^S + K_I^S + K_D^S \big)}
    {1- M F_x \cdot \SBR}
    \,.
\end{equation}
Thus incomplete extinction (\ile{F_x{>}0}) results in a compensating
increase in the required $K_s^S$, which
in turn requires an increase in $P_P^T$.
This increase is unrelated to other sources
of background, and since it is not a function
of $T_I$ is also unrelated to
the benefits of BPPM (see \secref{opportunities}).
Based on \eqnref{extinctionMinimum},
a feasible design requires
\begin{equation}
\label{eq:FxLimitation}
\ile{F_x < (M \cdot \SBR)^{-1}}
\,.
\end{equation}
This conclusion also follows directly from
\ile{\SBR < \text{SXR}} (since incomplete extinction is one component of SBR)
and \ile{\text{SXR} = 1/ M F_x}.

\subsection{Opportunities afforded by BPPM}
\label{sec:opportunities}

The slot time $T_s$ plays a central role
in \tblref{designEquations}.
First, choosing a smaller $T_s$ assists
in achieving the targeted SBR by reducing
all three of $K_{\{N,I,D\}}^S$.
This boost in SBR results from
ignoring all photon detection events that
occur during the blank intervals in \figref{BPPM}.
The price paid for smaller $T_s$ is an increase
in peak transmitted power $P_P^T$. 

The reduction in dark counts arising from smaller $T_s$
is especially advantageous, since dark counts originating in the detector 
(as opposed to receive optics) are
unabated by bandlimiting.
This is significant because the numerical
results in \secref{background} suggest that
dark counts are typically a very significant significant
contributor to overall background radiation for larger coverages $\Omega_A$.
This parallels radio telescopes, where thermal
noise introduced in the receiver electronics
typically dominates cosmic
sources of noise.

Another important parameter in 
\tblref{designEquations} that relates to
technology issues is the optical bandwidth
$W_e$ imposed at the receiver, since this determines
the level of noise and interference reaching the
optical detector.
The minimum bandwidth is determined by
\ile{W_e T_s{\sim}1}, and it is natural to keep
$T_s$ fixed as distance $D$ increases
as well as across different probes
(even if they have differing masses and data rates).
As a result,
the optical bandwidth does not
require agility or configurability.

The total noise and interference
is always proportional to $W_e T_s$, and thus
remains fixed if $W_e$ is coordinated with $T_s$.
Thus BPPM does not help (or hurt) with respect to total
noise and interference at the optical detector.
A reduction in $T_s$ to minimize dark counts also increases
the optical bandwidth $W_e$ without increasing background radiation.
In terms of optical bandpass filter technology
this is desirable since it relaxes the required selectivity.
In effect BPPM trades off background
suppression (except for dark counts) between the frequency and time domains.
Time-domain suppression is technologically preferable
because it is simpler to implement and more precise.
Thus all considerations (other than peak power)
argue in favor of making $T_s$ smaller.

\newpage
\subsection{Parameter-metric sensitivity}
\label{sec:background}

We now justify the choice of parameters in \parameters,
 followed by a study of the sensitivity of these
parameters to changing assumptions. 

\subsubsection{Choice of nominal parameters}
\label{sec:nominal}

The choice of \ile{\rate_0 = 1\ \text{b/s}} is a convenient 
(but rather arbitrary) choice.
This is the scientific data rate during periods of non-outage,
but for the chosen daylight and weather outage probabilities the net scientific data rate is
\ile{\rate_a = 0.432\ \text{b/s}} (see \secref{outagemitigation}).
For fixed coverage and also holding $\{K_s^R,T_s\}$ fixed,
based on \eqnref{xiTscaling} reducing $\rate_0$ is a way to either reduce
the transmitter average power-area metric $\xi_A^T$ (which has no effect on peak power $P_P^T$)
or alternatively reduce the number of receive \leveltwo{s} $N^S$
(which has the adverse effect of increasing the required peak power-area metric $\xi_P^T$).
Surprisingly reducing $\xi_P^T$ requires us to \emph{increase} $\rate_0$ and \ile{N^S \propto \rate_0}.
Alternatively $\xi_P^T$ can be reduced by increasing PPM slot time $T_s$, but this
increases susceptibility to dark counts.

The reliability parameters in \tblref{receiverParameters} must be
self-consistent based on available theoretical bounds on
the achievable BPP consistent with reliable recovery of scientific data.
First, neglecting background radiation (that is  \ile{\SBR{=}\infty}),
the choice \ile{m{=}12} places an upper bound \ile{\BPP{<}12\ \text{b/ph}}
based on \eqnref{logPAR}.
This is true of any possible modulation coding scheme.
For the specific case of PPM, the theoretical upper bound is
 \ile{\BPP{<}10.9\ \text{b/ph}} for the chosen $m$ and $K_s$
based on \eqnref{PPM_BPP}.
Second, taking into account background radiation,
for this value of \ile{\PAR = M}, the choice \ile{\SBR{=}4}
results in the theoretical bound \ile{\BPP{\le}11.1\ \text{b/ph}} for any photon-counting detector
(see \secref{photonCountingCapacity}).
Thus all available theoretical bounds are consistent with the assumption \ile{\BPP{=}10.9}
in \tblref{receiverParameters}.
Concrete ECC methods are not expected to quite achieve these theoretical bounds,
so this BPP can be considered mildly optimistic.

A rather arbitrary choice is the dark count rate \ile{\Lambda_D^S{=}32\ \text{yr}^{-1}},
because we don't know what value may be feasible in the future.
Assuming nighttime-only operation of the downlink (see \secref{nightime}),
the two most significant sources of background 
(for the shorter wavelength)
are moonlight and dark counts.
The value $\Lambda_D^S$ is chosen so that these two sources are about equal
(\ile{\SNR\approx\SDR}) at the largest coverage (\ile{\Omega_A{=}10\ \text{as}^2}).
Thus, a larger (or smaller) $\Lambda_D^S$ will render dark counts (or moonlight) the
dominant source of background radiation
under these specific conditions.

The choice of transmit aperture effective area $A_e^T$ is also rather arbitrary,
choosing an area that seems consistent with a low-mass probe.
Both the transmit powers $P_{\{A,P\}}^T$ are inversely proportional to $A_e^T$,
so increasing this area is directly beneficial in terms of power.

From \eqnref{FxLimitation}
\ile{F_x{<}6{\times}10^{-5}} for the parameters in \parameters.
A 77 dB extinction ratio
(\ile{F_x{=}2{\times}10^{-8}}) 
has been
achieved in bench testing \citeref{834}.
The \ile{F_x{=}10^{-7}} value  in \tblref{transmitterParameters} thus appears feasible today,
and renders incomplete extinction a relatively minor contributor to the background.

The effect of changes to wavelength $\lambda_0^R$ are
multifaceted.
If the coverage $\Omega_A$ is held fixed,
effective area $A_e^S$ is proportional to $\big( \lambda_0^R \big)^2$
and thus the signal power increases in proportion to $\lambda_0^R$
in \eqnref{linkbudget}, the interference increases in proportion
to $\big( \lambda_0^R \big)^2$ from \tblref{SBR}, and hence
overall there is increased interference.
For a red-dwarf star such as Proxima Centauri the interference
is magnified further because of the wavelength-dependence of the star
black body radiation (see \tblref{background}).
Increasing the wavelength to \ile{\lambda^R{=}1\ \mu\text{m}} in \tblref{receiverParameters}
causes interference to replace moonlight and dark counts
as the dominant component of background radiation,
and results in a substantial increase in transmit power
(to \ile{P_A^T{=}562\ \text{mW}} and  \ile{P_P^T{=}12.2\ \text{MW}})
for the swarm case in \tblref{performanceMetrics}.
This larger interference could be mitigated by a larger
coronagraph rejection (smaller $F_c$).

Reduction in $\Omega_A$ from the swarm to single-probe case
in \tblref{receiverParameters} beneficially reduces the transmit powers
due to the larger effective area $A_e^S$ of the \leveltwo,
but increases the total receive aperture area $A_e^S N_S$.
Thus, reducing $\Omega_A$ yields the expected tradeoff between
a larger receive aperture and reduced transmit powers.
Similarly a larger transmit aperture effective area $A_e^T$
can be directly traded for smaller transmit powers.

Unfortunately it appears that the required $P_P^T$
cannot be achieved by available semiconductor lasers.
Some ideas for potentially circumventing this
issue, none of them currently qualified, are
discussed \secref{laser}.

Each PPM frame in BPPM represents
$K_s^R \cdot \BPP$ bits of scientific data
on average, and since this equals the number
of scientific data bits $T_I \rate_0$
in each frame, it follows that
\ile{T_I \propto K_s^R}.
This implies that in the photon
starvation regime where \ile{K_s{\ll}1} the rate at which
frames are transmitted $T_I^{-1}$ is relatively
large.
For example, for the parameters in \parameters\  we have
\ile{T_I = 2.2\text{ s/frame}}, and only
2.2 scientific data bits are represented
by each frame on average.
When \ile{m{=}12}
each PPM frame actually represents
12 bits, which we infer represents 2.2 scientific data bits and 9.8 bits of redundancy.
This redundancy is exploited by ECC to
achieve reliability in the face of a high
rate of frame erasures (see \secref{ECC}).
ECC achieves reliable data recovery by simultaneously mapping a large number of
scientific data bits into a large number of
PPM frames, thereby statistically averaging
over the random frame erasures (see \secref{ECC}).

We now give an extended numerical
example based on the design relations in \tblref{designEquations}
with the goal of more fully appreciating the
tradeoffs inherent in a BPPM design.This numerical tradeoff is idealized, neglecting some material
degradations in performance from other factors such as transmit aperture and
\leveltwo pointing inaccuracy and atmospheric effects 
(refraction and turbulence).

\subsubsection{\Leveltwo effective area}

The area $\log_{10} A_e^S$ is plotted in \figref{areaVsCoverage}
for two wavelengths of interest
as a function of the coverage solid angle
$\Omega_A$ over the range bracketed by the
swarm and single-probe cases in \tblref{receiverParameters}.
This tradeoff is straightforward as it does not involve
other factors such as background radiation.
Choosing a shorter wavelength reduces $A_e^S$
(and therefore \leveltwo sensitivity) for fixed coverage.
Increased coverage also reduces \leveltwo sensitivity
at a fixed wavelength.

\incfig
    {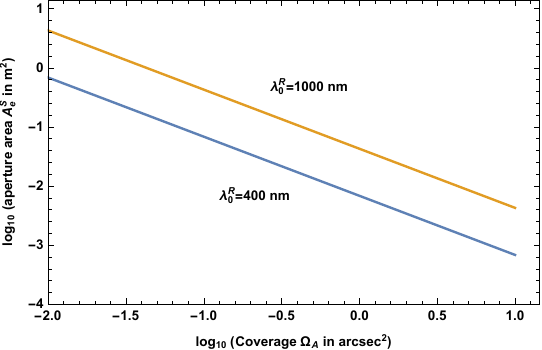}
    {width=1.\linewidth}
    {
    A log plot of the effective area $A_e^S$
    of an \leveltwo vs coverage
    solid angle $\Omega_A$.
    This assumes an idealized \leveltwo
    that achieves a uniform sensitivity over $\Omega_A$.
    }

\subsubsection{Background radiation}
\label{sec:bnumerical}

The physical parameters characterizing the noise
and interference used in the following numerical calculations
are listed in  \tblref{background}.\footnote{
Zodiacal noise is adopted from \citeref{1016} and
the sky noise and star interference are
adopted from \citeref{1017}.
The atmospheric scattering values are justified in \secref{skyRadiance}.
}
The values given apply to a \emph{single} \leveltwo
in the canonical receive \levelone
of \figref{apertureArchitecture},
since each \leveltwo has a single optical detector and thus
qualifies as a single-mode `antenna' in the sense of \secref{antennaTheory}.\footnote{
This applies even if
an optical detector is shared among
multiple {\leveltwo}s as described in
\secref{detectorShare} as long as optical interference among \leveltwo{s} is avoided.
}
The units in \tblref{background} are consistent with \tblref{SBR}.

\newcommand{\unitsa}{ph/s-m\textsuperscript{2}-Hz}
\newcommand{\unitsb}{ph/s-Hz}
\newcommand{\unitsc}{cm\textsuperscript{2}}
\newcommand{\unitsd}{counts/yr}

\doctable
	{background}
    {
    Assumed numerical values for background photon rates
    following a single-mode diffraction-limited \leveltwo.
    }
    {llll}
    {
    \hline
    \textbf{Radiation source} & Units & 400 \;nm & $1.0 \;\mu$m
    \\
    \hline
\tct{\textbf{Point-source interference}}
    \\
    \ \ Proxima Centauri $\Lambda_I$ & \unitsa
    & $8.0{\cdot}10^{-10}$ &  $2.0{\cdot}10^{-7}$
    \\
    \textbf{Unresolved sky noise}
    \\
    \ \ Zodiacal light $\Lambda_{N,zodi}$ & \unitsb
    \\
\ \ \ \ 90 degrees to ecliptic& & $2.0{\cdot}10^{-16}$ & $1.0{\cdot}10^{-14}$
    \\
\ \ Faint-star light $\Lambda_{N,star}$ & \unitsb
    & $2.0{\cdot}10^{-16}$ &  $1.0{\cdot}10^{-14}$
    \\
    \tct{\ \ Atmospheric scattering}
    \\
    \ \ \ \ Daylight  $\Lambda_{N,sun}$ & \unitsb
    & $4.1{\cdot}10^{-8}$ & $5.0{\cdot}10^{-7}$
    \\
    \ \ \ \ Fullmoon $\Lambda_{N,moon}$ & \unitsb
    & $1.0{\cdot}10^{-13}$ & $1.3{\cdot}10^{-12}$
}

Two wavelengths are compared in
\tblref{background},
\ile{1\ \mu\text{m}} (near infrared) and \ile{400\ \text{nm}} (blue visible).
The latter is of interest because of the substantially
smaller interference from Proxima
Centauri, which as a red dwarf star with a power spectrum
weighted toward the red.
The values in \tblref{background} are typical, but Proxima Centauri
is a flare star which exhibits short-term large increases in radiation.
These flares will likely contribute to outages (see \secref{outages}), unless the coronagraph
factor $F_c$  is considerably smaller than considered here.

The total noise contribution to background radiation during nighttime operation is 
\begin{equation*}
\Lambda_N = \Lambda_{N,zodi}  + \Lambda_{N,stars} + \rho_{moon} \cdot \Lambda_{N,moon} \,.
\end{equation*}
While $\Lambda_{N,moon}$ assumes a full moon, the factor
\ile{0{\le}\rho_{moon}{\le}1} can adjust for the lower radiance during other phases of the moon.
Generally $\Lambda_{N,moon}$ is the dominant sky noise source.
Note that assuming \ile{\rho_{moon}{<}1} implies that some nighttime periods (with a brighter moon)
are considered outages, and the actual scientific data rate $R_a$ will be correspondingly lower.\footnote{
Strictly speaking when \ile{\rho_{moon} = 0} there will be some black body atmospheric radiation,
but this is neglected since this irradiance is dominated by Zodiacal radiation 
$\Lambda_{N,zodi}$ from the solar system.
This case is also uninteresting since there would be 100\% outages.
}

\subsubsection{Daylight outage assumption}
\label{sec:nightime}

As is evident from \tblref{background}, daylight would have a
major impact on background radiation.
For the swarm case in \tblref{receiverParameters} the transmit
power would be impractically large to achieve daylight operation
(\ile{P_A^T{=}5.8\text{ kW}} and \ile{P_P^T{=}126\text{ GW}}).
For this reason, the metrics in \tblref{performanceMetrics} and the following calculations assume nighttime (fullmoon)
sky radiance.
The value \ile{\rho_{moon}{=}1} in \tblref{receiverParameters} and the following
numerical calculations assume the entire nighttime is a non-outage condition, barring weather events.
Since the receiver is aware of day-night and weather conditions, it can force
complete erasures during daytime and weather events, and these become outage conditions
in the sense of \secref{outagemitigation}.
Knowledge of the receiver's day-night or weather conditions is fortunately not necessary
at the probe transmitter.

\subsubsection{Average transmit power}

The average transmit power $P_A^T$ is plotted in \figref{powerVsAngle}
as a function of coverage solid angle $\Omega_A$.
This shows that the shorter wavelength is quite advantageous, especially
so at smaller coverages (where interference is more significant).

\incfig
	{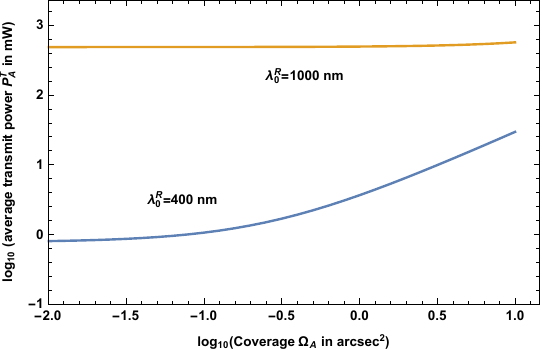}
	{
	trim=0 0 0 0,
    	clip,
    	width=1.\linewidth
	}
	{
	A log plot of the average transmit power $P_A^T$ in mW against the coverage solid angle
	$\Omega_A$, with all other parameters taken from \parameters.
	Two wavelengths are plotted, and for small
	$\Omega_A$ the required transmit power is much larger at the longer
	wavelength due to the greater relative \leveltwo sensitivity to interference
	and additionally the larger irradiance of a red-dwarf star.
	}

The differing behavior at the two wavelengths can be
better understood by comparing three components of SBR:
Interference (SIR), noise (SNR), and dark counts (SDR).
The smallest component is the dominant source of background radiation.
At the shorter wavelength (see \figref{sbrComponentsVsAngle400})
$\Lambda_D^S$ was chosen so that noise and dark counts are
equal contributors to background radiation, and hence \ile{\SNR \approx \SDR}
for all coverages (since neither noise nor dark counts are dependent on \leveltwo effective area).
Interference is an insignificant contributor compared to moonlight and dark counts at the larger coverages.

At the longer wavelength (see \figref{sbrComponentsVsAngle1000}) interference becomes the
dominant contributor to SBR at all coverages.
As a result, the dark count target could be considerably relaxed without a substantial impact on
overall performance metrics, but of course that performance would be considerably degraded
relative to the shorter wavelength.

\incfig
	{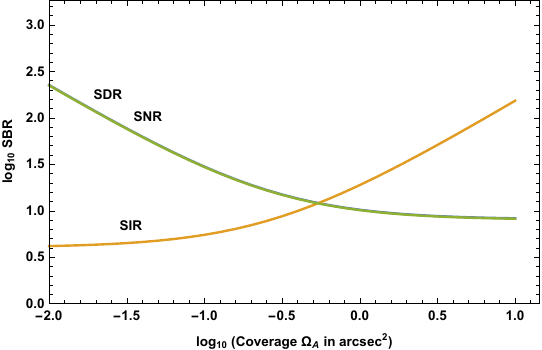}
	{
	trim=0 0 0 0,
    	clip,
    	width=1.\linewidth
	}
	{
	As an aid to understanding the small-coverage behavior 
	exhibited in \figref{powerVsAngle},
	the log of the three components of \ile{\SBR{=}4} are plotted
	vs coverage solid angle $\Omega_A$
	for \ile{\lambda=400\ \text{nm}}.
	For small coverages SIR is the smallest, and hence interference
	is the dominant source of background radiation.
	}
	
\incfig
	{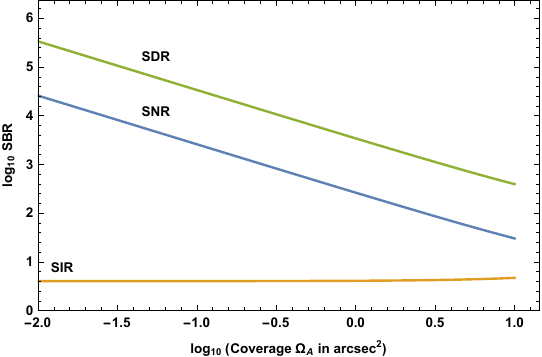}
	{
	trim=0 0 0 0,
    	clip,
    	width=1.\linewidth
	}
	{
	\figref{sbrComponentsVsAngle400} is repeated
	for  \ile{\lambda=1\ \mu\text{m}}.
	The interference
	becomes dominant for all coverages.
	}

\subsubsection{Number of {\leveltwo}s}

SBR is determined by the design of the \leveltwo and the choice of $P_A^T$,
and remains fixed during the scale-out to the entire receive \levelone.
The purpose of \leveltwo replication in scale-out is to increase the average detected photons
per slot $K_s^R$ to the value needed to achieve reliable recovery of scientific data
in the face of shot noise.
The value of $K_s^R$ and hence the number of \leveltwo{s} $N^S$ does not depend on the data rate $\rate_0$.
The resulting $N^S$ is very large, as plotted in \figref{numberSubArraysVsAngle}.

\incfig
	{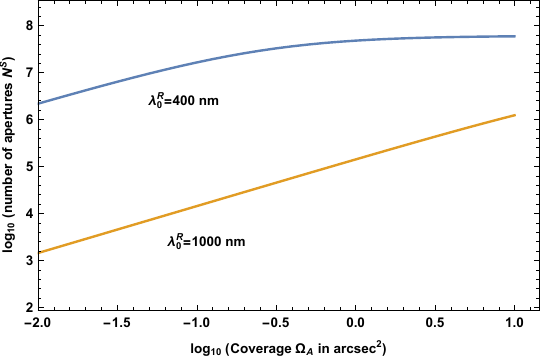}
	{
	trim=0 0 0 0,
    	clip,
    	width=0.95\linewidth
	}
	{
	A log plot of the number $N^S$ of {\leveltwo}s under the same conditions as \figref{powerVsAngle}.
	It is significantly larger at the shorter wavelength, which is due to (a) the smaller-area \leveltwo
and (b) the smaller peak transmit power $P_P^T$.
}

The metric $N^S A_e^S$, which is a measure of the total effective area across all \leveltwo{s},\footnote{
This is not the effective area of the receive aperture as a whole,
which is not defined since the receive aperture is not
single-mode and not diffraction-limited. 
}
is plotted in \figref{totalEffectiveArea}.
 It is significantly larger at the shorter wavelength, with the
    compensating benefit that the probe's average transmit power is lower
    (see \figref{powerVsAngle}).

\incfig
    {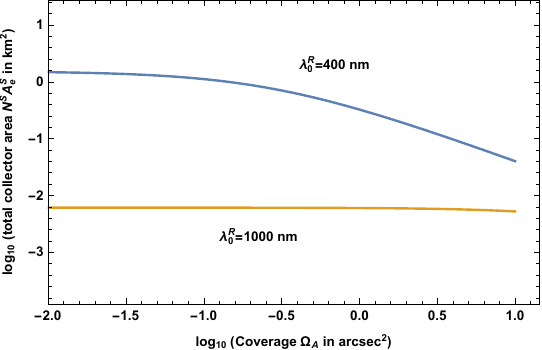}
    {
    trim=0 0 0 0,
    	clip,
    	width=0.95\linewidth
    }
    {Based on the \leveltwo effective area $A_e^S$
    in \figref{areaVsCoverage} and the number
    $N^S$ of \leveltwo{s} in \figref{numberSubArraysVsAngle},
    a log-plot of their product $N^S A_e^S$
    in km\textsuperscript{2}
    as a function of the coverage.
    Generally it is less than a 
    km\textsuperscript{2} under these
    ideal conditions, except for very small coverages.
    }

\subsubsection{Dark counts}
\label{sec:darkcounts}

To see the effect of changing dark-count rates,
$P_A^T$ is plotted in \figref{powerVsDarkVsFOV400}
and \figref{powerVsDarkVsFOV1000}
as a function of dark counts, coverage angle,
and wavelength.
This quantifies the obvious conclusion that the shorter
wavelength and a very low rate of dark counts are advantageous in achieving
low average transmit power.
Reducing the solid angle of coverage is also
advantageous.

\incfig
	{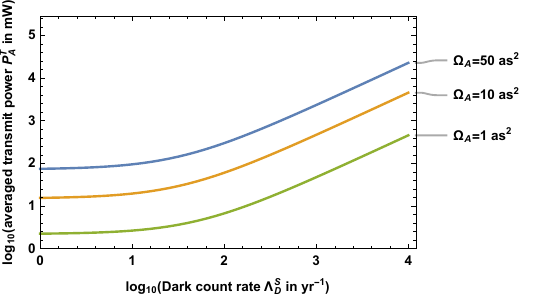}
	{
	trim=0 0 0 0,
    	clip,
    	width=1.\linewidth
    	}
	{
	A log-log plot of the average
	transmit power $P_A^T$
	for \ile{\lambda = 400\ \text{nm}}
	vs the average dark counts $\Lambda_D^S$ per year (when referenced to individual \leveltwo
	optical detectors).
	Sharing optical detectors across multiple
	\leveltwo{s} may be helpful in reducing
	$\Lambda_D^S$
	(see \secref{detectorShare}).
	Idealized performance is assumed for other physical characteristics
	like quantum efficiency, pointing accuracy, and atmospheric refraction and turbulence.
	The different curves illustrate the benefit of reducing the coverage angle.
	The highest dark count rate \ile{\Lambda_D = 10^4\ \text{yr}^{-1}} corresponds to
	\ile{1.14\ \text{hr}^{-1}}.
	}
	
\incfig
	{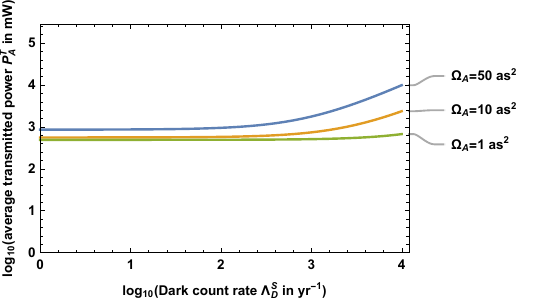}
	{
	trim=0 0 0 0,
    	clip,
    	width=1.\linewidth
    	}
	{
	\figref{powerVsDarkVsFOV400} is repeated
	for \ile{\lambda = 1\ \mu\text{m}}.
	A floor in transmit power appears
	because interference 
	becomes the dominant contributor to SBR
	for small dark count rate $\Lambda_D^S$.
	(see \figref{powerVsAngle} and \figref{sbrComponentsVsAngle1000}).
	}

\subsubsection{Coronagraph}
\label{sec:ceffect}

The effect of changing the coronagraph rejection is illustrated in
\figref{powerVsDarkVsCoronagraph}.
The value of $F_c$ makes a substantial difference at low dark count rates
because that is where interference dominates.

\incfig
	{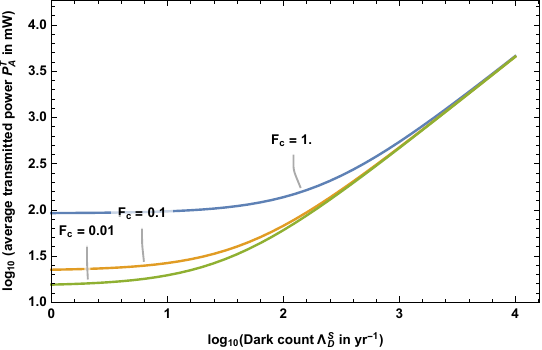}
	{
	trim=0 0 0 0,
    	clip,
    	width=1.\linewidth
    	}
	{
	\figref{powerVsDarkVsFOV400} is repeated
	with three different values for the
	coronagraph rejection $F_c$.
	At low dark count rates the interference
	becomes the dominant source of background
	radiation.
	}

\newpage
\subsubsection{Slot time and average transmit power}
\label{sec:slotpeak}

In BPPM reducing the slot duration parameter $T_s$
reduces the effective
dark count rate (by reducing the duty cycle $\delta$).
The resulting benefit of $T_s$ on reducing average transmit power 
$P_A^T$ is illustrated in
\figref{powerVsSlotVsDarkAvg}.
There is, however, a point of diminishing returns
as interference becomes the dominant source of background radiation.

\incfig
	{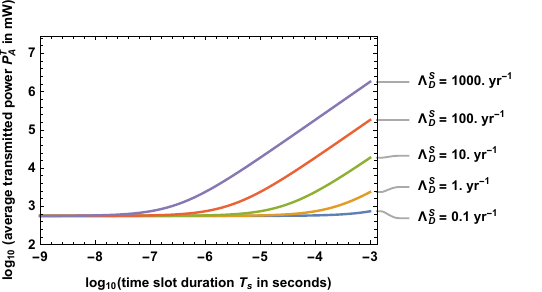}
	{
	trim=0 0 0 0,
    	clip,
    	width=1.\linewidth
    }
	{
	A log-log plot of the
	average transmit power $P_A^T$ vs slot time $T_s$ (between 1 ns and 1 ms)
	and different values of dark count rate
	$\Lambda_D^S$ in yr\textsuperscript{-1}.
	The largest feasible value
	(at \ile{\delta{=}1})
	is \ile{T_s{=}0.98\ \text{ms}}.
Reducing $T_s$ also reduces
    $P_A^T$ due to its beneficial limiting
    of dark counts.
    However, the benefit saturates at small $T_s$
	as interference dominates there.
	}

\subsubsection{Slot time and peak transmit power}

As seen in \eqnref{peakPowerAreaMetric}, the peak transmit power $P_P^T$
is not dependent on $\rate_0$ or BPP, but rather is determined by the
BPPM parameter $T_s$ and data reliability parameter $K_s^R$.
The effect of $T_s$ on peak transmit power
$P_P^T$ is shown in \figref{powerVsSlotVsDarkPeak}.
Increasing $T_s$ reduces $P_P^T$
because the pulse energy $P_P^T T_s$ remains relatively fixed.
This effect saturates at larger $T_s$
because the increase in $P_A^T$ seen in \figref{powerVsSlotVsDarkAvg} indirectly offsets
any reduction in $P_P^T$.
The peak power increases for smaller $T_s$ (where
dark counts have been largely suppressed)
because the average power is relatively constant in this interference-dominated regime.

\incfig
	{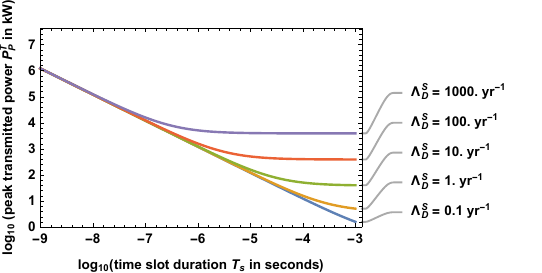}
	{
	trim=0 0 0 0,
    	clip,
    	width=1.\linewidth
    }
	{
	\figref{powerVsSlotVsDarkAvg} repeated for 
	peak transmitted power
	$P_P^T$.
	}

For the other parameters chosen in
\figref{powerVsSlotVsDarkPeak}, the best choice of 
slot time is \ile{T_s \sim 1\ \mu{s}}.
Any smaller $T_s$ than this and $P_P^T$
 increases,
and any larger $T_s$ results in an increase in $P_A^T$.
However this tradeoff is influenced
by the assumed dark count rate $\Lambda_D^S$.

\subsubsection{Moonlight variation}

The assumed outage probability assumes that the downlink
operates during all moon phases.
There is an opportunity to reduce the transmit powers
if operation is restricted to reduced moonlight by choosing
\ile{\rho_{moon} {<}1}, but of course the outage probability will also increase.
This effect on transmit power is quantified in \figref{powerVsDarkVsMoonlight}.

\incfig
    {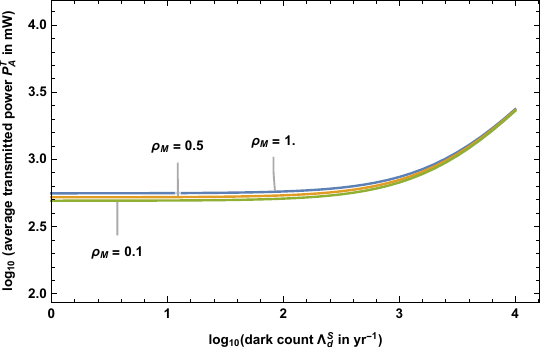}
    {
    trim=0 0 0 0,
    	clip,
    	width=1.\linewidth
    }
    {
    \figref{powerVsDarkVsFOV400} is repeated
    with all parameters held constant
    except the maximum moonlight irradiance parameter $\rho_{moon}$.
    Note that reducing $\rho_{moon}$ implies a larger outage probability
    and hence a lower post-outage data rate $\rate_a$.
    For larger $\Lambda_D^S$ a smaller $\rho_{moon}$ has little
    benefit since dark counts are the dominant source of background radiation.
    }

\subsubsection{Excess optical bandwidth}
\label{sec:bandpasstradeoff}

If it is necessary for some reason to chose
a $W_e$ that is significantly larger than the
minimum value $T_s^{-1}$,
the noise and interference
(but not dark count) contributions to
background at the optical detector are increased.
In this event probe transmit power has to
be increased to maintain the desired SBR,
and dark counts become relatively less
important.
The effect of increasing $W_e$
on average transmit power 
$P_A^T$ is plotted in
\figref{powerVsDarkVsBandwidth}.

\incfig
    {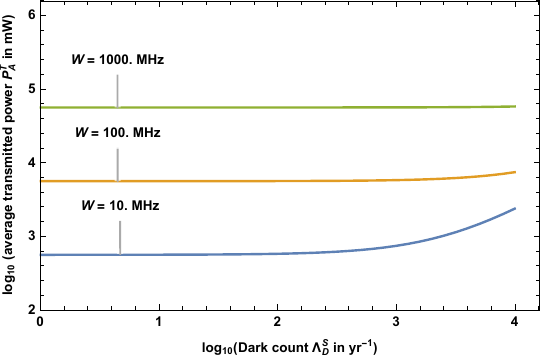}
    {
    trim=0 0 0 0,
    	clip,
    	width=1.\linewidth
    }
    {
    \figref{powerVsDarkVsFOV400} is repeated
    while varying the receive optical bandpass
    filter bandwidth $W_e$ and keeping $T_s$ fixed.
    The minimum is \ile{W_e{=}10\ \text{MHz}}
    corresponds to \ile{W_e T_s = 1}.
    In the interference-dominated regime
    the average transmit power is
    increased (in proportion to $W_e$),
    and in principle this could be compensated by
    a higher coronagraph rejection ratio.
    }

\subsubsection{Quantum efficiency}

The preceding results assume ideal
quantum efficiency \ile{\eta = 1}.
A quantum efficiency \ile{\eta < 1}
has two effects.
First it will always increase $N^S$ 
(and hence total receive aperture size)
to achieve the required average photon
rate \ile{\Lambda_A^R}.
Second, it will increase
the relative importance of dark counts
to SBR, and require an
increase in transmit average power
$P_A^T$ in the regime of large dark count rates.
This latter effect is illustrated in \figref{powerVsDarkVsEfficiency}.

\incfig
    {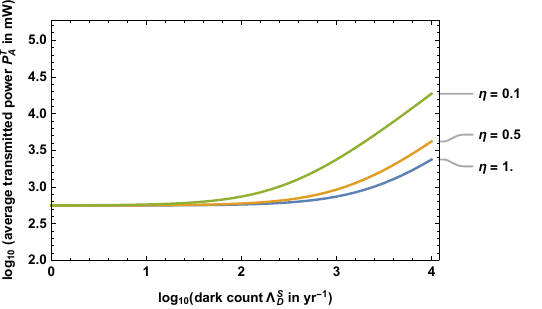}
    {
    trim=0 0 0 0,
    	clip,
    	width=1.\linewidth
    }
    {
    \figref{powerVsDarkVsFOV400} is repeated
    with all parameters held constant
    except the optical efficiency $\eta$.
    The average transmit power $P_A^T$
    has to be increased to maintain
    a fixed SBR at higher dark count rates
    $\Lambda_D^S$ since received power is reduced but
    $\Lambda_D^S$ is unaffected.
    The adverse impact of $\eta$ on $P_A^T$
    is reduced for very small
    $\Lambda_D^S$ since in that regime noise
    and interference (as well as signal) are
    reduced and thus SBR is relatively unaffected.
    }

\section{Atmospheric effects}
\label{sec:atmosphere}

For a receive aperture located on the
earth's surface,
the interaction between signal and
the earth's atmosphere at optical wavelength
introduces several significant impairments
 \citeref{1001}.
Nighttime sky irradiance is a contributor to the total background
radiation, and was incorporated into the numerical
results of  \secref{background}.
The other significant atmospheric impact  is a contribution to
outage time due to daytime solar scattering and weather,
which together reduce the rate of reliable scientific data recovery
(see \secref{reliableData} and \secref{outagemitigation}).

\subsection{Daytime sky irradiance}
\label{sec:skyRadiance}

During the daytime, sunlight scattered
through the atmosphere into the
receive aperture is a source of
background radiation that can be considered
isotropic within the coverage of a highly directive aperture.
In \tblref{background} the sky irradiance at the \leveltwo can be inferred from
Fig.8.16 in \citeref{1001}:\begin{align*}
&P_{N}^S = 3{\times}10^{-3}\  \text{W/(cm\textsuperscript{2}-sr-}\mu\text{m})
\text{ @ } \lambda_0^R{=}1 \mu\text{m}
\\
&\Lambda_{N}^S = P_{N}^S \cdot \frac{\lambda_0^R}{h c} \cdot \frac{W_e \big(\lambda_0^R\big)^2}{c}
\cdot \big(\lambda_0^R\big)^2
\text{ ph/s-Hz}
\,.
\end{align*}
The factors in the conversion from  $P_N^S$ to $\Lambda_N^S$
are (1) the power to photon rate conversion,
(2) the bandwidth expressed in terms of frequency rather than wavelength,
and (3) the $A_e \Omega_A$ product given by \eqnref{antennaThermo}.
The value of $P_{N}^S$ is about $8{\times}$ larger at \ile{\lambda_0{=}400\ \text{nm}}
and a factor of $4{\times}10^5$ smaller for a full moon as compared to daylight.

This irradiance decreases with greater angular separation
from the sun itself, and it helps that there is a large
minimum angular separation between Alpha Centauri and the
sun.\footnote{
In heliocentric coordinates the latitudinal separation is approximately
fixed at 44.7 degrees and the longitudinal separation varies
through the year.
}
Although the solar radiance is maximum at 475 nm
(corresponding to black body radiation at 5800 K),
the sky irradiance is smaller at 400 nm than at 1 $\mu$ largely due to
the smaller value of $A_e \Omega_A$.
The resulting numerical values are listed in \tblref{background}.

Unfortunately the full daylight sky irradiance in \tblref{background}
 is more than four orders of magnitude larger than the other sources of noise background.
The significance of this depends on the relative intensity of interference and dark counts,
but generally operation during daylight hours would require significantly higher probe transmit
peak power.
Thus in \tblref{receiverParameters} (as discussed in \secref{nightime}) we assume that daylight is an outage situation.
 
 \subsection{Nighttime sky radiance}
 \label{sec:nightSky}
 
 If daylight is considered an outage, the sky radiance during nighttime is a limiting factor
 on the downlink.
 At night the reflected light from the moon is an indirect source of solar background radiation
(except for a new moon or the moon below the horizon).
A full moon has a $4{\times}10^{5}$ smaller radiance than the sun.
In the absence of moonlight, the primary source of irradiance is thermal emissions
from the atmosphere, but this will be considerably smaller than cosmic zodiacal
radiation for the wavelengths of interest as long as strong emission
lines (such as OH or O\textsubscript{2}) are avoided in the choice of wavelength.
Thus, the total
night sky irradiance as a fraction $\rho_{moon}$ of the full moon value
(as determined by the moon's phase) is the approximation employed earlier.

\subsection{Turbulence}
\label{sec:turbulence}

Atmospheric turbulence causes a short-term
fluctuation in received signal power which is manifested
as a scintillation (random variation in signal strength) as well as signal incoherence
(wavefront phase aberration across
the \leveltwo area).
As the geometric size of an \leveltwo increases, 
the variation in signal power due to scintillation decreases,
while the power loss due to spatial incoherence increases  \citeref{1001}.

As the \leveltwo geometry is determined by coverage considerations,
scintillation manifests itself as a variation in erasure probability that
can only be compensated by considering high erasure probability as an
outage, 
which is mitigated using interleaving and ECC (see \secref{outagemitigation})
rather than adjustment of the \leveltwo design.

Spatial incoherence is a less severe impairment during nighttime operation,
and under some conditions may be deemed an outage.
It can in principle be compensated by adaptive optics
(the details are discussed in Sec.3.34 of \citeref{1001}).
As to whether adaptive optics will be necessary depends on the detailed
consideration of elevation, declination, and the geometric size of the
\leveltwo (rather than its area), all of which is beyond the scope of this paper.
Fortuitously, for the larger coverages necessary in the probe-swarm case
the \leveltwo is smaller and thus less susceptible to atmospheric turbulence.

\subsection{Outages}
\label{sec:outages}

Periods during which virtually all PPM frames suffer an erasure are considered
outages.
Generally the transmitter will have no knowledge
of when outages occur, but when there is
no ``fuel'' saving from avoiding transmission during outages this
has no consequences.
The outage mitigation design can assume a worst-case
outage probability $P_E$,
with the consequence that the data rate is reduced 
by a factor of $\big( 1-P_E \big)$
(see \secref{outagemitigation}).

There are twp main sources of atmospheric-origin outages are daylight and weather.
Weather, including water vapor, clouds, and storms
attenuate or block reception at random times.
The statistics of weather outages is highly site-dependent.
The receiver is aware of either of these conditions and can enforce
(and label) outages by arbitrarily enforcing 100\% erasures during these periods.

For any point on earth, nighttime lasts for half the hours averaged over one year
(actually slightly less due to refraction of sunlight near the horizon).
Near the Antarctic Circle
nightime persists for 48\% of the total hours in a year \citeref{1002}.
Thus treating daylight as an outage would reduce the scientific data rate by 52\%
at this location.

The siting of the receiver will introduce another source of
outages if its latitude results in the target star disappearing
below the horizon.  In the case of Proxima Centauri, continuous view is possible
if the receiver is sited at a sufficiently southerly
latitude\footnote{The declination of Proxima Centauri is -62.67 degrees,
so a continuous view is achieved by choosing a receiver site with
a latitude more southerly than $(90-62.67)=27.33$ degrees south.
However, to avoid significant degradations in link quality when
Proxima Centauri is low on the horizon due to increased atmospheric
degradation effects, the receiver site should be more southerly
than approximately 35 degrees south.}.

A cosmic source of outages will be flares of the target star, but these
are short-lived and will be insignificant in terms of
outage probability.

\section{Other challenging design issues}

BPPM addresses principally the technology
limitations of optical bandpass filters and detectors.
Other issues relating to the design of the physical layer
arise that impose profound technological challenges.

\subsection{Probe motion effects}
\label{sec:motion}

A low-mass probe is assumed to be traveling
(relative to the receiver frame) at relativistic speed,
and the receiver is also in motion due to earth rotation
and orbit.
When referred to baseband, due to the short optical
wavelength the resulting Doppler shifts 
must be accounted for in the design (see \secref{redShift}).

\subsubsection{Uncertainty in probe velocity}
\label{sec:redshift}

At a speed of \ile{u = 0.2 c} the combined
Doppler shift (about 20\%) and effect of time dilation (about 2\%) is about 22\%.
Since the sources of noise and interference
do not experience this shift,
the shifted wavelength determines
the atmospheric effects and interference from the
target star.
The transmit wavelength can be adjusted
toward a shorter wavelength to compensate
for the Doppler shift.

Since the swarm of probes will encompass
slightly different speed perturbations,
the result is relatively large frequency offsets
affecting WDM (see \secref{multiplexing2})
and TDM (since a common receive
wavelength is assumed).
If we allow a perturbation $\text d u$ in speed $u$,
the resulting perturbation $\text d \nu_R$
in received frequency $\nu_R$ is
\begin{equation}
    \label{eq:freqSensitivity}
   \frac{\text d \nu_R}{\nu_R} 
   = - \frac{u/c}{1 - (u/c)^2} \cdot \frac{\text d u}{u}
   = - 0.208 \cdot \frac{\text d u}{u}
\end{equation}
for \ile{u = 0.2 c} (see \secref{uncertainty}).
Thus there is a Doppler shift of $\pm 1.6$ THz for each $\pm 1$\% variation
in probe velocity at \ile{\lambda_0{=}400\text{ nm}}.

Achieving a precision in probe velocity that limits the
Doppler shift to the order of ${\sim}W_e$ seems unlikely.
If probe speed cannot be controlled tightly enough,
it will be necessary to compensate for variations in
probe speed by a coordinated configuration of transmit wavelength
following launch.
There are at least two possible options:
\begin{itemize}
\item
Since the transmission of scientific data follows encounter,
transmit wavelength configuration can 
follow encounter and precede downlink operation.
Autonomous configuration of transmit wavelength could be based 
on a sufficiently precise measurement of time to reach the target star,
combined with knowledge of target star
position.
All probes in
a swarm will be affected equally by
imprecision in knowledge of this position.
\item
If there is an earth-to-probe communication uplink 
for a short period following launch (see \secref{probeconfig}),
the received wavelength can be measured at the receiver
and used to configure the transmit wavelength.
\end{itemize}

\subsubsection{Earth motion and Doppler}
\label{sec:earthMotion}

The maximum speed of the earth relative to a heliocentric
frame due to its orbit about the sun (or revolution about its axis) is \ile{30\ \text{km/s}}
(or \ile{1670\ \text{km/h}} at the equator).
The Doppler shift is thus bounded by \ile{\pm 30\ \text{GHz}}
(or \ile{\pm 464\ \text{MHz}}) at \ile{\lambda = 1\ \mu\text{m}}.
While this Doppler can be very significant,
it is essentially the same for all probes in a swarm
and thus does not appreciably affect their relative wavelengths.
Accurate positioning of the receiver's optical bandpass filtering
does require correction based on accurate knowledge
of this motion, and any remaining uncertainty should be
accounted for by an increase in bandwidth $W_e$ with the
resulting increase in cosmic sources of background radiation
(see \secref{bandpasstradeoff}).

\subsubsection{Gravitational red shift}

During flyby of the target star, the stronger gravitational
field will result in incremental red shift of the probe's signal.
While this effect can be significant for a close-in encounter, it is largely avoided with post-encounter 
commencement of downlink transmission.
In any case downlink operation during encounter is unlikely due to
 conflicting demands on scientific instrument electrical power and
probe attitude adjustment to accommodate scientific observations.
While gravitational effects are likely insignificant in communications downlink design,
they offer a scientific opportunity for sensitive measurement of gravitational potentials
(see \secref{gravitationalEffects}).

\subsubsection{Influence on data rate}
\label{sec:timedilation}

Due to the increasing propagation delay and time dilation in the
transmitter clock relative to the receiver, the data rate
observed by the receiver is smaller by about 22\% than the
data rate generated at the transmitter.
Due to the methodology adopted in \secref{latencyVolume},
which concentrates on the earth's rest frame,
the data volumes and latency plotted in \figref{latencyVsVolume}
account for this.

\newpage
\subsubsection{Relativistic aberration}

The high speed of the transmitter moving away from the receiver
will have a relativistic effect on the transmit beam
and affect the link 
budget of \eqnref{linkbudget}.
This small incremental effect is not accounted for in this paper.

\subsection{Multiplexing options}
\label{sec:multiplexing2}

Multiplexing is the function of separating the
concurrent signals originating from different probes.
Whether there is a single shared receiver or multiple
receivers, this function is challenging due to
environmental factors like extremely low received
power and frequency shifts due to probe motion.

These challenges are magnified as the number $J_p$ of
concurrent transmitting probes increases, yielding
strong motivation to minimize $J_p$.
Available measures include, for each probe, limiting transmission during
transit and limiting the downlink operating time.
We find in \secref{masstradeoffs} that
(for a specific set of scaling laws) the 
optimum downlink time ranges from 2-9 years.
For weekly probe launches this results in \ile{J_p \sim 100 \text{ to } 500}.

There are some alternative multiplexing approaches.
The multiplexing method affects many other
aspects of the downlink design, so it should
be established early.

\subsubsection{Spatial-division multiplexing (SDM)}
\label{sec:SDM}

SDM takes advantages of angular differences between
probe trajectories to separate the signals from the probes.
For probes launched at different times of the year,
the parallax variation
combined with a drift in the launch targets 
due to proper motion of the target star will spatially
separate trajectories save for short-lived periods of
exact coincidence
(see \secref{coverage}).
This opens up the possibility of choosing a receive aperture coverage
to attenuate all probe signals save one of interest,
with any coincidences treated as an outage (see \secref{outagemitigation}).
Proper motion of the target star is particularly helpful for spatial separation, as it
opens up the possibility of combining SDM with another scheme for separating
the signals from probes with nearby launch dates.

An extreme case is to use a separate receive aperture for each downlink,
each with a very small coverage solid angle $\Omega_A$
(as implicitly assumed in \citeref{1015}).
Although extravagant in terms of ground-system cost, an important
benefit is the reduction in the required transmitter power-area product \ile{\xi_A^T = P_A^T A_e^T}
if $N^S$ is chosen to minimize this value.
However, this smaller $\xi_A^T$ may come at the expense of a considerably larger
total effective area $N^S A_e^S$ as seen in \figref{totalEffectiveArea}.

An intermediate case would be a more sophisticated single-aperture design that employs
time-dependent beamforming to separate probe signals.
In effect each \leveltwo would constitute a multiple-pixel receive optical system, with one pixel 
dedicated to each probe.

\newpage
\subsubsection{Wavelength-division multiplexing (WDM)}
\label{sec:WDM}

In WDM each probe is
assigned a different wavelength at the receiver, with
an optical bandpass filter
and dedicated optical detector servicing each probe.
WDM requires precise transmitter wavelength control,
which is a challenge complicated by Doppler shifts (see
\secref{redshift} and \secref{earthMotion})
and implementation of a serial bank of optical bandpass filters
to separate signal from different probes
(see \secref{bandpassFilter}).

\subsubsection{Time-division multiplexing (TDM)}
\label{sec:TDM}

In TDM probe transmissions are assumed to not overlap one
another in time at the receiver.
Ideally those transmissions are at a common received
wavelength, allowing for a single optical bandpass filter
and detector.
TDM requires precise transmitter knowledge of
time of arrival of its signal at earth,
which requires in turn a very precise clock and precise knowledge of its own
trajectory (see \secref{timing}).

\subsubsection{Random-access multiplexing}
\label{sec:randomAccess}

This is a variation on TDM in which transmissions have
a low duty cycle and are randomized so that the probability
of overlapping receptions in the receiver is small
 \citeref{890}.
This may be natural in combination with BPPM,
which trades higher peak power for low duty cycle (see \secref{BPPM}).
Collisions at the receiver could be treated in a similar way to outages (see \secref{outagemitigation}),
although they would complicate the ECC by causing frame errors rather than erasures.

\subsubsection{Code-division multiplexing (CDM)}
\label{sec:CDM}

In CDM probes
are assigned mutually orthogonal spreading sequences, which
allows their signals to be separated (by cross-correlation
with the spreading sequences) in spite of both wavelength- and time-overlap.
It is useful to think of CDM as maintaining orthogonality using a different
orthogonal basis for signals, where the basis functions do not align with either time or frequency
(see \secref{FSK}).

For equivalent per-probe data rates $\rate$,
to first order WDM, TDM, and CDM all expand the \emph{total} optical
bandwidth by a factor of $J_p$.
In the case of TDM this is because each probe has to
transmit at rate $J_p \rate$ during its assigned time
slot, and for CDM the spreading sequence expands the
bandwidth of each probe's signal by a factor of $J_p$.
In practice the bandwidth is larger after accounting for
guard bands (WDM) and guard times (TDM) due to imprecise
knowledge of wavelength and time at the probes.
As illustrated by BPPM, any form of TDM will also increase the peak
transmitted power requirement commensurate with $J_p$.

\subsection{Clock accuracy and stability}
\label{sec:clocks}

Particularly at high photon efficiencies, data representation by accurate timing
and/or accurate wavelength will be essential.
This represents a significant challenge in a system distributed over interstellar distances
where relativistic speeds are involved (see \secref{motion}),
and one in which atomic-clock frequency and time standards are not likely to be practical
in a low-mass probe transmitter.
Along with the generation of high peak-power pulses (see \secref{laser}),
the resulting short- and longer-term clock inaccuracies
will limit the timing accuracy (as represented by slot time $T_s$ in BPPM)
that can be obtained, and thereby will limit the photon efficiency.

There are two related issues of clock synchronization: between probe and receiver,
and among \leveltwo clocks which provide timestamps attached to photon-detection events.
While \leveltwo{s} co-located on a common site can share a common clock source with
atomic accuracy and stability, the probe clock will have relaxed crystal-oscillator accuracy and stability
and will be affected by uncertainties in probe speed (see \secref{redshift}).
It will be necessary for the receiver to estimate and track the transmit timing.
Fortunately this need not be real-time, but rather can be delegated to a post-processing
phase based on the entire record of photon-detection events gathered over the entire mission.
The probe can assist in the process by transmitting deterministic sequences of pulses
interleaved with the random encoded scientific data.
The accuracy of this tracking will be limited by the accuracy with which individual photon detections are time-stamped,
and in particular the relative phases of a common clock appearance at each of the \leveltwo{s}.

That relative offset is largely
determined by temperature variations, and is on the order
of 3 to 10 ns when using the Global Positioning System (GPS)
for time-transfer, with larger absolute diurnal and seasonal variations \citeref{1020}.
It may be possible to track these absolute variations in the post processing, similarly to the probe clock,
although the latter is much easier due to its reliance on aggregate
(as opposed to per-\leveltwo) photon counts.
Greater accuracy and stability may be feasible through a sub-terrainian temperature-controlled optical distribution of clocks
not relying on GPS.
An absolute accuracy taking into account all sources of variation on the order of 1\% of the slot time $T_s$ will be
required, placing a lower bound on $T_s$ (and hence dark count rejection).
While longer-term stability of the probe clock is less critical, shorter term instabilities will be difficult to track
at the receiver due to the low photon-detection rate (even in the aggregate).

\subsection{Probe navigation}
\label{sec:timing}

Relative to an (approximate) inertial frame defined
by the Solar System and target star,
the probe needs to be self-aware of the time and position
for both the probe itself and for earth,
for a couple of purposes.
First the scientific instrumentation and
    interpretation of its data requires
    knowledge of the probe
    trajectory, hopefully with an
    accuracy less than a fraction of
    an AU.
 Second, accurate pointing of the transmit
    aperture so that the maximum signal
    power reaches the terrestrial receive
    aperture requires accurate
    knowledge of the angle and distance
    of the earth at the 
    later time of arrival
    (see \secref{attitudeControl}).

For these purposes
the probe trajectory is
parameterized by the probe elapsed time,
the observed probe speed, and the observed
angle of the sun's and the target star's position.
There at least a couple of sources of imprecision in these
estimates.
First, after 30 yr the variation in elapsed time measured by a probe clock is 16 min
    for each one part in $10^{-6}$ variation in oscillator frequency.\footnote{
Relative to terrestrial events the time elapsed
on the probe is about 2\% shorter than on earth,
but this relativistic effect is known accurately and readily compensated.}
Second, at 4 ly and a nominal speed of $0.2 c$ the variation in propagation time is $\pm 72$ days for each $\pm 1\%$ 
variation in probe speed.
Thus a speed variation is generally a much bigger issue in accurate pointing
toward an orbiting earth than is clock accuracy.
Thus a Xtal oscillator without temperature compensation
aboard the probe may suffice for onboard estimate
of elapsed time, and whatever method is used for speed measurement and
compensation for Doppler shifts (see \secref{redshift})
also applies to estimating propagation time.

\subsection{Feedback Doppler compensation}
\label{sec:probeconfig}

The multiplexing techniques described in \secref{multiplexing2}
require control of the relative wavelength differences among
probes in a swarm to an accuracy on the order of the signal optical bandwidth $W_e$.
This is especially true of TDM, since rejection of background radiation
puts a premium on minimizing the bandwidth of a receive optical system
that is shared among probe downlinks.
Since $W_e$ is relatively small, this strongly suggests the use of
feedback to control the relative received wavelengths across probe downlinks.

Either the speed of each probe or the transmit wavelength in each
probe can be configured using feedback.
In both cases a downlink monochromatic beacon can be transmitted
during the launch or for a short period following the launch, with its observed
Doppler shift on the ground providing an accurate estimate of probe speed.
Accurately controlling the final probe speed would require an adjustment to the
time window for the directed energy beam.
This may also be necessary (or at least helpful) for navigation purposes.
Accurately configuring the transmit wavelength to compensate for
the actual probe speed following launch would require a telemetry uplink.
The latter approach has the disadvantage that the probe requires a communications
receiver, although its implications would be minimized by the large
available transmit power and the small amount of configuration data required.

\newpage
\subsection{Probe attitude control}
\label{sec:attitudeControl}

Attitude control of the probe is required for the scientific
mission (orientation of instruments) and for communication
(aiming the transmit beam back to its terrestrial communication receiver).
The pointing accuracy required for the transmission downlink
is undoubtedly the most critical requirement.
A three-axis closed-loop correction by photon thrusters
driven by the probe electrical power source
has been proposed \citeref{811}.

\subsubsection{Probe pointing accuracy}
\label{sec:pointing}

Assumptions as to receive aperture size
are predicated on precise aiming of the probe's
transmit beam to achieve the maximum power transfer.
This is a major issue in solar system probes, and
will be here as well.
Pointing can be based on probe attitude control
or beam angle configuration relative to the probe,
or some combination of the two methods.

The angular spreading of the beam is a fraction
of the Airy disk, and is thus proportional to $\lambda/d$
where $d$ is the radius of a circular aperture.
Since the target must fall within this beam, indeed near the center, the pointing
accuracy is not penalized by the great distance to the earth.
The short wavelength $\lambda^R$ in \tblref{receiverParameters} reduces the
permitted pointing error, while the relative small
size of the transmit aperture is helpful.
For example, for the parameters in \tblref{receiverParameters}
this ratio \ile{\lambda/d{=}0.8\ \text{arcsec}} is much more stringent than
\ile{58.\ \text{arcmin}} for New Horizons (see \tblref{interstellaradio})
due to the shorter wavelength.

A significant issue is what reference can be used for pointing.
Obvious candidates are the sun and the target star 
(which will fall in nearly the same direction).
If the beam were to illuminate the entire solar system,
centering it on or near the sun would suffice.
However, for the aperture in \tblref{receiverParameters}
the Airy radius becomes 3.27 au at 4.24 ly distance and \ile{\lambda_R = 1\ \mu\text{m}}
decreasing to 1.2 au at \ile{\lambda_R = 400\ \text{nm}}.
Thus, more specific aim at the position of earth is necessary.
One tradeoff would be to reduce $A_e^T$ and increase $P_A^T$, keeping
$\xi_A^T$ constant, thereby reducing the required pointing accuracy
at the expense of transmit power.
Conversely, a larger transmit aperture
(such as that assumed in \citeref{1015})
 would increase the required
pointing accuracy accordingly.

It appears likely that the beam must be aimed at the moving position of earth within its orbit
at the future date/time of signal arrival, rather than the solar system with the sun near its center.
For this purpose a second axis of reference, such as another nearby star, would be necessary.
Acquiring the direction of a pair of guide stars requires image sensing with an attitude control loop.
Generally the resolution of this sensing has to be finer than the transmit beam width,
implying a sensing aperture larger than the transmit aperture.
This suggests sharing a single aperture on the probe
between transmission and sensing, with a larger portion of
that aperture used for guide star tracking than for transmission.

\subsubsection{Attitude control operation}
\label{sec:attitudeOp}

Fortunately a low-mass probe has no moving mechanical parts or
fluids that contribute to attitude misalignment, 
so there are only 
three identifiable sources of attitude misalignment:
motion of electrons within the probe electronics,
the momentum recoil from the transmit beam, and
(likely most significantly) collisions with interstellar particles.
This suggests that
attitude adjustment and transmission should \emph{not}
occur concurrently, for several reasons:
\begin{itemize}
\item
If attitude misalignment is infrequent
the duty cycle of attitude adjustment may be low.
\item
The sharing of a common aperture
between sensing and transmission is a mass-saving
measure that is easier if the two functions are not
concurrent.
\item
Each function can utilize the entire available
electrical power, maximizing the data rate and
minimizing the time consumed by attitude adjustment.
\item
During periods of misalignment the data transmission
is likely to be unreliable so
this becomes an outage condition, whether the
probe is transmitting or not.
\item
Any effect of transmission beam momentum interfering
with attitude adjustment is minimized.
If thrusters and communications utilize the same
electrical source, then downlink transmission exerts an
equivalent recoil force on the probe.
If transmission occurs only during periods when
the attitude is within a tight tolerance, since the
probe trajectory is nearly rectilinear the only significant effect
is a tiny increase in probe speed.\footnote{
The trajectory curvature due to the target star
gravitation will introduce a small momentum component
transverse to the trajectory that needs to be
accounted for.}
\item
Although the suspension of transmission during periods of
attitude adjustment slightly reduces the average data rate,
the greater electrical energy available during each transmission
can compensate for this with a higher nominal data rate.
\end{itemize}

\subsection{Coronagraph function}
\label{sec:coronagraph}

The coronagraph function provides extra attenuation
for interference originating at the target star
(see \secref{ceffect}).
Each receive \leveltwo shared over all probe downlinks
in a swarm will have a coverage that includes probes
 just completing their target star encounter, and thus will have limited
or no attenuation of the target star radiation.
A requirement on the order of \ile{F_c{=}10^{-2}} to $10^{-4}$ follows (see \secref{ceffect}),
necessitating coronagraph-specific measures other than limiting the coverage angle.

There are various approaches, some of which are:
\begin{itemize}
\item
Delay download transmission following encounter long enough
for proper motion of the star to separate probe trajectories from the star \citeref{1015}.
\item
Arrange for an \leveltwo coverage to be limited to a single probe at a time
(as assumed in \citeref{1015}).
This is accomplished by duplicating {\leveltwo}s.
Or with some forms of TDM, the coverage of a single \leveltwo can be switched
to the single probe whose signal is currently being received.
\item
For a common \leveltwo implemented as a phased array of \levelthree{s}, 
implement multiple beams directed at the different probes.\footnote{
Such an approach is being explored at
millimeter wavelengths for application to
5G cellular \citeref{823}.
}
To avoid a reduction in SBR this would have to be
accomplished
with high quantum efficiency
and without amplification.
\item
Use techniques similar to those being pursued in the
direct imaging of exoplanets, which also requires
rejection of target-star radiation.
These techniques include
coronagraphs and interferometers
\citeref{825} and external
occulters (also known as starshades) \citeref{826}.
For example, a notch (high rejection) permanently 
located at the target star location might be
added to the \leveltwo by adjusting the phase/magnitude
in the combiner, thus creating a spatial filtering
function.
\end{itemize}
The phase and magnitude in a phased-array combiner
has to be implemented extremely accurately.
Phases also have to be dynamically adjustable to account
for parallax and the proper motion of the star (see \secref{coverage}).

A promising direction draws on advances
in direct exoplanet imaging.
However, some differences inherent to the downlink design
challenge should be noted:
\begin{itemize}
\item
For a ground-based observation platform,
dynamic pointing of the aperture due to atmospheric refraction
and turbulence are issues.
\item
The optical bandwidth is narrow and the
wavelength can be chosen to minimize the
target star interference (see \secref{bnumerical}).
\item
A single probe downlink
will typically operate for years (see \secref{masstradeoffs})
and a swarm of probes correspondingly longer.
\end{itemize}

\section{Critical technologies}

The tradeoffs quantified in \secref{background}
identify three technologies whose
stringent requirements play a significant
role in success or failure.
These are the transmit light source,
the receiver
optical bandpass filtering, and
the photon-counting optical detectors.
It is clear that the current state of source and detector technologies is not consistent with
the low-mass probe downlink application.
Fortunately they can be expected to advance in the application timeframe (see \secref{timeline}),
but this introduces considerable uncertainty.
Fortunately we have not identified any limiting physical principles that impede future advances.
Also, the configurable duty cycle $\delta$ in BPPM beneficially offers a tradeoff among
all three technologies, permitting some flexibility in their relative advancement
(see \secref{BPPM}).

\subsection{Transmit light source}
\label{sec:laser}

In an optical direct-detection communication system
employing PPM and BPPM,
high peak powers are fundamental to achieving
high photon efficiency \citeref{1019}.
This is reflected in the theoretical limit
of \eqnref{logPAR}, which predicts a logarithmic
relationship between the achievable BPP with reliable data recovery and PAR.
High PAR is achieved by reducing slot time $T_s$ and increasing peak power $P_P^T$
to keep the pulse energy $P_P^T T_s$ constant.
This relationship of PAR and BPP is discussed
in practical terms in \secref{highBPP2}.
However this straightforward approach results in multiple-kilowatt peak powers
(see \figref{powerVsSlotVsDarkPeak}).
Optical amplification of semiconductor
laser outputs can attain Watt-level peak powers, and
Q-switched lasers can attain kilowatt-levels
but suffer from low electrical-to-optical conversion efficiency
\citeref{889}.

\subsubsection{Design parameter adjustment}

The required peak power can be reduced by several adjustments in design parameters
singly or in combination, although these unfortunately all trigger other issues.
Reducing the coverage solid angle $\Omega_A$ benefits the peak power because of the
higher sensitivity of a large \leveltwo (see \figref{powerVsDarkVsFOV400}), 
but the total receive \levelone
area is also increased (see \figref{totalEffectiveArea}).
Increasing the transmit aperture area $A_e^T$ is effective if the power-area $P_P^T A_e^T$ is fixed.
Reducing the dark count rate results in lower peak power (see \figref{powerVsSlotVsDarkPeak}),
but the effect is limited by the continued presence of moonlight interference.
Smaller peak power can be achieved by reducing the photon efficiency BPP
and hence lower PAR (see \figref{PapVsAreaSwarm}), 
but this requires very large total receive aperture area to compensate for the reduced efficiency.

\subsubsection{Pulse compression}
\label{sec:pulsecompression}

Optical interference can be used in place of emitting high powers directly from an optical source.
A straightforward example of this is the ganging of multiple semiconductor lasers in the transmitter with an optical combiner.
Unfortunately this approach is expected to be inconsistent with the low-mass objective.

Pulse compression technology is often used to generate narrow high-power optical pulses.
It relies on the observation that for two pulse shapes $p_1 (t)$ and $p_2 (t)$ with the same bandwidth $W$,
$p_2 (t)$ can be obtained from $p_1 (t)$ by a linear filter with transfer function \ile{F(f) = P_2 (f) / P_1 (f)},
where $P(f)$ is the Fourier transform of $p(t)$.
For example, a pulse with constant envelope \ile{\big| p_1 (t) \big| = A} and approximate
time duration \ile{T \gg 1/W} can be converted to a pulse $p_2$ with approximate time-duration $1/W$ and
envelope \ile{\big| p_2 (t) \big| = A \sqrt{W T}} if energy is conserved.\footnote{
A  function bandlimited to $W$ cannot be strictly time-limited, and a function  time-limited to $T$ cannot be
strictly bandlimited.
However these conditions can be approximated if \ile{W T{\gg}1} \citeref{1013}.
The narrow pulse $p_2 (t)$ violates this condition and will be only roughly confined to time duration $1/W$.
}
This process converts a longer lower-amplitude pulse into a narrower high-amplitude pulse
with power gain equal to $W T$ (the bandwidth-time product).

Pulse compression has to be performed in the optical domain,
typically by creating targeted interference with arrangements of
diffraction gratings \citeref{1006} or prisms \citeref{1005}.
This apparatus achieves wavelength-dependent
group delays to align shorter- and longer-frequency components in time.
A \emph{structured receiver} approach has been proposed that uses more complex
long-duration waveforms with matched filtering \citeref{888}.
Due to the linearity of Maxwell's equations
the pulse-compression filtering could be performed in the receiver, reducing the mass burden on the probe.
However, while pulse compression
works well in generating picosecond pulses, it would be difficult to generate
\ile{10-100\text{ ns}} pulses as required in this application.
This is because light travels distance of one meter in \ile{{\sim}3\text{ ns}} in a vacuum, and 
such an optical apparatus would necessarily be physically large.
Of course shorter pulses could be generated, but the required reduction in pulse duration
and attendant increase in peak power would be counter-productive.

\subsubsection{Modulation code layer}
\label{sec:FSK}

The need for short pulses can be attributed to the choice of PPM
for the modulation code layer, and it is intriguing to consider alternatives.
One approach is to extend the code-division multiplexing idea described in \secref{CDM}
to the modulation coding layer for a single probe downlink.

Consider a set of $L$ waveforms \ile{\{ p_i (t) , 1 \le i \le L \}},
each bandlimited to \ile{W\text{ Hz}} and confined to time interval \ile{0 \le t \le T} with the
property that
\begin{equation*}
\int_{- \infty}^{\infty} p_i (t) p_j^* (t) \,\text{d} t = 
\begin{cases}
A\, & i=j \\
0\, & i \ne j
\end{cases}
\end{equation*}
This requirement can be approximately satisfied whenever \ile{L \le W T},
with increasing accuracy as \ile{W T \to \infty} \citeref{1013}.

The modulation coding layer can transmit one of the $L$ waveforms,
and at the receiver a parallel set of filters (one matched to each possible waveform)
can estimate which waveform was transmitted \citeref{1018}.
A PPM modulation code is a special case of this scheme with \ile{L = M \sim W_e T},
where $M$ is the number of slots in a PPM frame,
and frequency-shift keying (FSK) is another.
FSK and other options could intrinsically allow the light source to generate
a waveform with constant power with duration $T$
(rather than duration $T/M$ as in PPM).
However, there is a major disadvantage, and this is the need for $L$
optical detectors, one for each matched filter, multiplying the total dark count rate accordingly.
PPM has the unique property that a single optical photon-counting detector
(with its dark counts) is needed.

Intermediate cases are possible.
For example if \ile{M = I \cdot J} then each candidate transmitted pulse could be confined to
one of $I$ time slots (each with duration $T/I$) and one of $J$ frequencies.
Only \ile{J < M} optical detectors would be needed.

\subsection{Optical bandpass filtering}
\label{sec:bandpassFilter}

Bandpass filtering in the optical domain 
is a critical system component with
several requirements that are challenging
to achieve individually and collectively.
Numerical results in \secref{slotpeak} suggest
a BPPM slot time \ile{T_s{\sim}0.1{-}1\ \mu\text{s}},
which implies \ile{W_e{\sim}1{-}10\ \text{MHz}}.
An
optical bandwidth $W_e$ larger than necessary results in
larger cosmic source of background (see \secref{bandpasstradeoff}).
At \ile{\lambda{=}1\ \mu\text{m}}, \ile{W_e{=}1\ \text{MHz}} results in
a Q-factor (ratio of center frequency to
bandwidth) of $3 {\times} 10^{8}$.
Such Q's are being approached with
recent technology.\footnote{
For example
cavity-based high-Q optical bandpass filters with
\ile{Q = 8 {\times} 10^7} are reported in
\citereftwo{812}{813}.
}

There are other requirements on the bandpass filtering.
If WDM is used to separate the probe signals (see \secref{WDM})
we actually need a bank of bandpass filters with nearly adjacent passbands.
To avoid signal attenuation, this filter bank has to be serial rather than parallel,
with each filter splitting off the optical signal in one band while providing
low insertion loss outside this designated passband.
Further, the location of these passbands have to be agile to adjust to
changing Doppler shift due to the earth's motion (see \secref{earthMotion})
and possibly also to adjust for variations in the probe transmit wavelengths (see \secref{redshift}).

\subsection{Optical single-photon detectors}
\label{sec:opticalDetector}

The numerical results
point to the need for optical detectors with
dark count rates that are extremely low
(see \secref{darkcounts}).
Superconducting detectors,
which can successfully detect individual photons
and have intrinsically low dark count rates,
will be necessary.

Dark-count rates of
\ile{\Lambda_D{=}10^{-4}\ \text{Hz}}
have been reported for a superconducting nanowire detector \citeref{1010}.
This is two orders of magnitude higher than the objective of \tblref{receiverParameters},
which translates to \ile{\Lambda_D^S{=}10^{-6}\ \text{Hz}}.
In \citeref{1010}
the black body radiation from the connecting optics is reported
as the major source of dark counts, and this source is substantively
different than intrinsic dark counts in the detector itself since the
dark count rate is proportional to optical bandwidth in the former
and independent of bandwidth in the latter.
Low dark count rates were obtained
by inserting a cold \ile{100\ \text{GHz}} bandpass
filter between optics and the nanowire.
However, the signal optical bandwidth in \tblref{receiverParameters}
is four orders of magnitude lower than this, opening an opportunity
to further reduce the optics blackbody radiation.

In fact there is an opportunity to locate all the receiver optical bandpass filtering
at the output of the optics and input of the optical detector.
In this case the dark count rate originating as blackbody radiation in the optics would be
comparable between PPM and FSK, since the total bandwidth 
$W_e$ for the two modulation schemes are nominally equal (see \secref{FSK}).
This example illustrates the intimate relationship between the design of the
modulation coding and physical constraints and characteristics.

Superconducting detectors are intrinsically free of dark counts,
but material impurities
and external sources of radiation (radioactive decay and cosmic rays)
may be an issue.
Some other measures can be taken to minimize the dark-count rate, 
including the following ideas:
\begin{itemize}
\item
Cryogenic cooling of the optics and other photonic elements.
\item
Better shielding of optics and detectors  from external radiation.
\item
Many intrinsic dark counts may be manifested by out-of-band (higher-energy)
photons,
and these events may be recognizable in the electrical pulse amplitude
at the detector output.
\item
Share a single optical detector over multiple {\leveltwo}s, reducing the
per-\leveltwo dark count rate accordingly
(see \secref{detectorShare}).
For example multiple fibers might be attached to a large-area detector.
These fibers may need to be optically isolated, or this may be unnecessary
at the extremely low photon detection rates expected.
\item
It is also feasible to operate with \ile{\SBR{<}1}, although the theoretical photon efficiency
falls off rapidly as SBR decreases in this regime.
While this would allow for greater dark count rates, all else equal, it would have other
negative consequences.
\end{itemize}

\section{Data reliability}
\label{sec:highBPP2}

Scientific data, especially after aggressive compression,
must be recovered with very high reliability to maintain the
integrity of scientific conclusions and outcomes.
Achieving this high reliability in the face of impairments like noise and
interference, atmospheric turbulence, and outages is
a challenge.
In the interest of a smaller receive aperture we seek high photon efficiency
BPP (see \secref{reducingPower}), and this magnifies the challenge.
Even after the aforementioned impairments are tamed, signal shot noise
(a feature of the laws of quantum mechanics) places theoretical
limits on the BPP that can be achieved consistent with reliable data recovery.
Here we discuss this issue from an intuitive perspective intended
to be accessible to those not versed in communication and information theory, with
technical details relegated to Appendices.

\subsection{Reliability is the hard part}
\label{sec:reliabilityHard}

There is no theoretical limit on photon efficiency
if reliability is not demanded.
The challenge is in \emph{actually achieving} high reliability
in the recovery of scientific data in spite of  high photon efficiency.
This is easily illustrated using the encoding scheme of PPM, one frame
of which was illustrated in \figref{BPPM}c.
The reliability challenge comes about because even
with $K_s^R$ \emph{average} detected photons for
each occupied slot (one out of $M$), the actual
number of photons is \emph{random}.
This is a manifestation of the signal shot noise.

Assuming no spurious photon detections from background radiation
(this would be another source of randomness to overcome),
let the number of detected photons $Y$ in time
interval $\timeslot$ be a random variable,
which quantum mechanics predicts has a Poisson distribution.
If the slot in question is the one containing non-zero average power,
and the average number of detected photons in this slot is  \ile{E[Y] = K_s^R}, then
\begin{equation}
\label{eq:poisson}
\prob{Y = k} = \frac{(K_s^R)^k e^{-K_s^R}}{k!}
\,.
\end{equation}
An erasure occurs in a PPM frame when no photons are detected in
all $M$ slots, and this erasure has to be mitigated by ECC.
Based on \eqnref{poisson},
this event \ile{Y = 0}  occurs with probability
\ile{\prob{Y=0} = e^{- K_s^R}}.

Since one PPM frame communicates $\log_2 M$ bits using
an average of $K_s^R$ detected photons, the photon efficiency
is
\begin{equation}
\label{eq:nakedBPP}
\BPP = \frac{\log_2 M}{K_s^R} = \frac{m}{K_s^R}
\,.
\end{equation}
Obviously \ile{\BPP \to \infty} as \ile{K_s^R \to 0}, so
achieving a large BPP is not an issue.
The problem arises when we consider reliability,
since a side effect of reducing $K_s^R$ is \ile{\prob{Y=0} \to 1} as \ile{K_s^R \to 0},
and thus a large and increasing fraction of PPM frames are erased.
Even \ile{K_s^R = 1} in the absence of ECC would result in unacceptable reliability,
since in that case an erasure occurs in 37\% of
PPM frames on average, which is not suitable for scientific data.\footnote{
In the physics literature one occasionally sees an analysis
of PPM with \ile{Y \equiv 1}, claiming a deterministic
number of detected photons and a large resulting
value of BPP (see \citeref{891} for an example).
This analysis is not valid because it neglects
the stochastic nature of quantum photon detection,
which at \ile{K_s^R = 1} results in an unacceptably large erasure probability.
}
A more acceptable erasure probability
of \ile{e^{-K_s^R}{\approx}10^{-7}} can be obtained
with \ile{K_s^R{=}16}, but this would result in
\ile{\BPP{<}1} for practical values of $m$.

Conceptually we can achieve high BPP (albeit in an impractical way) by manipulating $m$.
For any value of \ile{K_s^R}, even a larger value like \ile{K_s^R{=}16}, \eqnref{nakedBPP} indicates that
any arbitrarily large BPP can be achieved by choosing $m$ sufficiently large,
because this conveys an increasing number of bits at the
expenditure of a fixed number $K_s^R$ average photon detections.
However, this quickly becomes impractical due to a vanishing small slot duration $T_s$.
For example, if we choose the typical values of \ile{\BPP{=}10} and \ile{K_s^R{=}16},
then we infer from  \eqnref{nakedBPP}  that \ile{m{=}160} is required.
To achieve a data rate of \ile{\rate_0{=}1\ \text{b/s}} with conventional PPM we must
map 160 data bits into a PPM frame with duration \ile{160\ \text{seconds}}, requiring a slot time of
\begin{equation}
\label{eq:nakedM}
T_s = \frac{160\ \text{s}}{2^{160}}
\approx 10^{-34}\ \text{ps}
\,,
\end{equation}
which is, needless to say, far beyond the capability of our electronics.

Although impractical,
improving BPP by increasing $m$
does uncover an important principle.
Namely, a way to achieve high photon efficiency 
coincident with high reliability is to increase the PAR (and consequently the bandwidth) of the signal.
This is suggested by the theoretical limit of \eqnref{logPAR} as well.
This principle applies at both radio and optical wavelengths.\footnote{
For the radio case, where the channel model is quite different, this principle is described and quantified in \citeref{708}.
}
We now illustrate how efficiency and reliability can be achieved in a practical way
by greatly moderating the bandwidth requirement.

\subsection{Modulation and coding architecture}

The theoretical bound of \eqnref{logPAR} suggests that
\ile{\PAR{\sim}2^{10}} is necessary to achieve \ile{\BPP{\sim}10\text{ b/ph}},
a value dramatically lower than
would ever be achievable by
using a `raw' PPM frame to communicate
our data as in \eqnref{nakedBPP}.
A practical architecture achieving this lower PAR is shown
in \figref{layers}.
It is organized into layers, with each layer split between
coordinated functionality in the transmitter and receiver.
This architecture, in conjunction with specific choices
for the modulation code and error-correction code (ECC)
is not only theoretically solid, but
\ile{\BPP{=}13} has been achieved in bench testing \citeref{834}.

The physical layer
involves all the physical elements of the downlink, including optical source and
detector and transmit and receive apertures.
Its input is an intensity-vs-time waveform, and its output is a sequence
of photon detection events.

\incfig
	{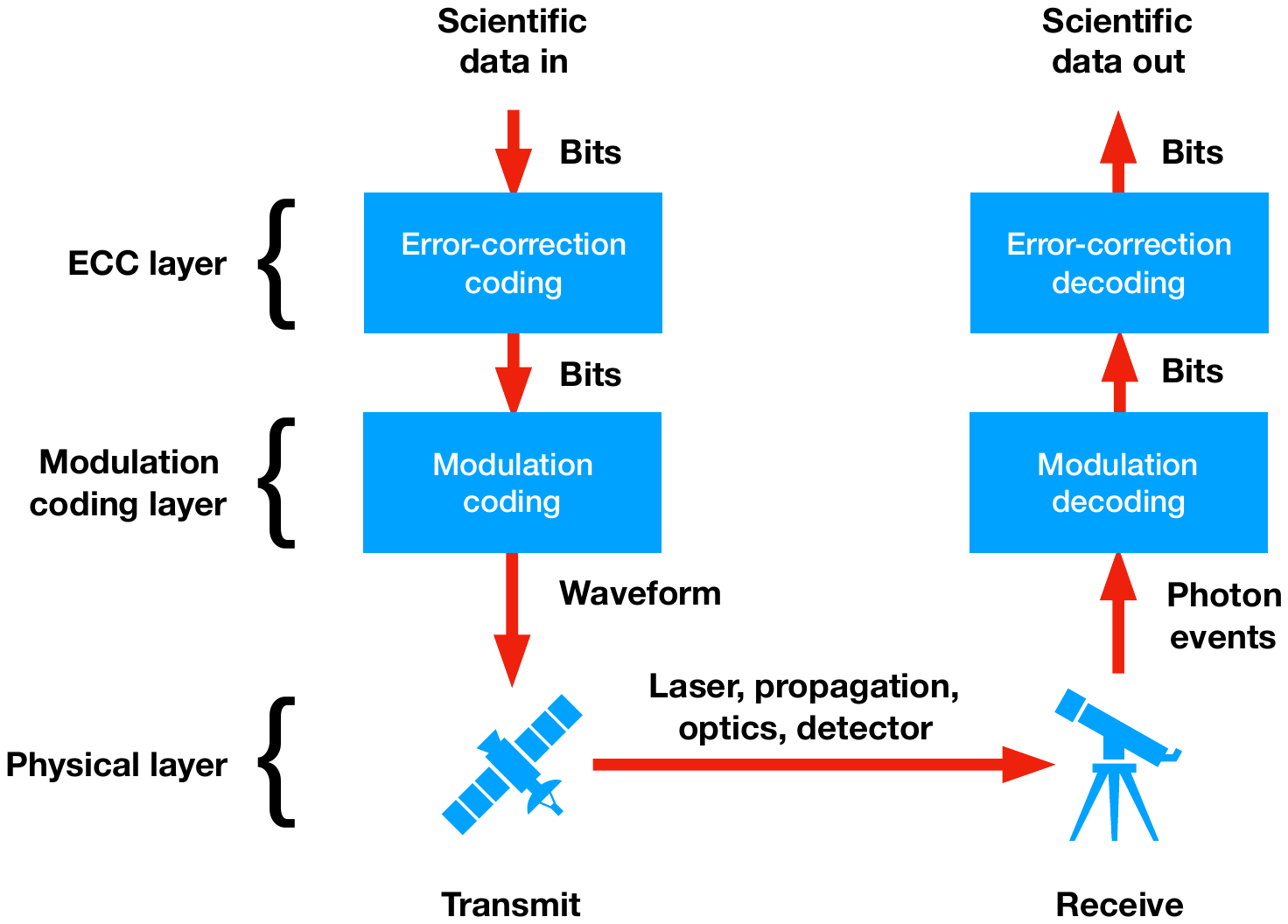}
    {
    trim=40 40 30 30,
    clip,
    width=.98\linewidth
    }
    {
    Coordinated transmit-receive communications architecture.
    Functionality is divided into three layers, with
    each layer divided into a transmit and receive component.
    Logically each layer in the transmitter is coordinated with its counterpart
    in the receiver.
    }

The role of the modulation coding layer is to represent 
discrete data by a continuous-time intensity waveform at the transmitter,
and interpret the resulting photon events observed in the receiver
in terms of the transmit data (but with poor reliability).
For our purposes our modulation code is described by BPPM
in \figref{BPPM}, which consists of a sequence of PPM frames,
each frame representing $\log_2 M$ bits of data.
This simple scheme, operating in conjunction with the ECC layer,
allows us to achieve high
photon efficiency BPP with high reliability in data recovery.

\newcommand{\timeword}{T_c}
\newcommand{\intensity}{\Lambda}

\subsection{ECC layer}
\label{sec:ECC}

In contrast to the approach taken in \secref{reliabilityHard},
the way to achieve a large BPP is to operate the
modulation code layer with very poor reliability;
that is, choose a small value for $K_s^R$ (see \eqnref{PPM_BPP}),
which is a photon-starvation mode.
The value \ile{K_s^R{=}0.2} chosen in \tblref{receiverParameters}
results in a frame erasure probability \ile{e^{-0.2}{=}0.82},
so approximately 82\% of frames are lost to erasures.
Although this choice may seem arbitrary, it is further justified in \secref{BPPvsReliability}.

The role of the ECC
layer is to reconstruct a 
highly reliable replica of the scientific data.
In order to dramatically improve the reliability, the
ECC coding layer in the transmitter
adds redundancy to the scientific data.
In the receiver the ECC decoding layer makes use of
this added redundancy (the 
precise structure of which is known to the receiver) 
to reconstruct a replica of the scientific
data with dramatically improved reliability.

The principle behind the ECC layer can be described abstractly as follows.
For \ile{\BPP{=}10.9\text{ bits/ph}} in \tblref{receiverParameters}, each scientific data bit recovery is
based on an average of 0.092 photon detection events.
A way (the only way)
to achieve reliability in photon-starvation mode is to recover multiple data bits based on a commensurate large number of photon detections,
exploiting the law of large numbers to yield less random variability in the number of photon detection events.
This can still be accomplished in the context of a PPM modulation coding layer by representing the data bits
by pulses with duration $T_s$.
Thus \ile{\BPP{=}10.9} might actually be achieved by exchanging an average of 100 detected photons for 1090 bits.
At a data rate of \ile{\rate{=}1\text{ b/s}} this implies an average of 100 photon detection events within 1090 sec (18 minutes),
or (almost) equivalently 100 detected pulses on average.
There must be  \ile{2^{1090}({\sim}10^{328}}) distinguishable patterns of pulses,
one pattern for each possible combination of 1090 bits
(compare this to the \ile{{\sim}10^{80}} atoms in the visible universe).
The error-correction decoder examines the random pattern of actual detected pulses,
and infers the most likely pattern, and hence the most likely combination of 1090 bits that were represented
in the transmitter.

\subsubsection{Small codebook example}

More concretely, ECC in the context of a PPM modulation coding layer
associates \ile{k > 1} bits with a group of \ile{l > 1} PPM frames,
each having \ile{M = 2^m} slots.
The set of $2^k$ frame groupings (one for each set of $k$ inputs bits) is
called a \emph{codebook}.
Since these $l$ PPM frames could in principle represent as many as $l m$ bits,
not all the possibilities are included in the codebook as long as \ile{k < l m}.
This is what we mean by \emph{redundancy}, which can result in dramatically 
improved reliability by circumventing erasures that have occurred.

This concrete description of the ECC layer can be illustrated as in \figref{ECCexample} for three very simple examples.
In \figref{ECCexample}(a) \ile{k=1} bits (2 codewords) are represented in a codebook of \ile{l = 1} PPM frame with \ile{M = 2} slots,
in (b) \ile{k=2} bits (4 codewords) are represented in a codebook of \ile{l = 2} PPM frames with \ile{M = 4} slots,
and in (c) \ile{k=4} bits (16 codewords) are represented in a codebook of \ile{l = 4} PPM frames with \ile{M = 4} slots.
The redundancy in the three cases is zero for (a) and 0.5 for (b) and (c).
If the transmit energy per PPM frame remains fixed, 
the average detected photons per frame has the same value $K_s^R$ for all cases.
Thus, the photon efficiency has the same value (\ile{\BPP = 1/K_s^R}) for all three cases, and thus the probability of
a single erasure remains the same.

\incfig
	{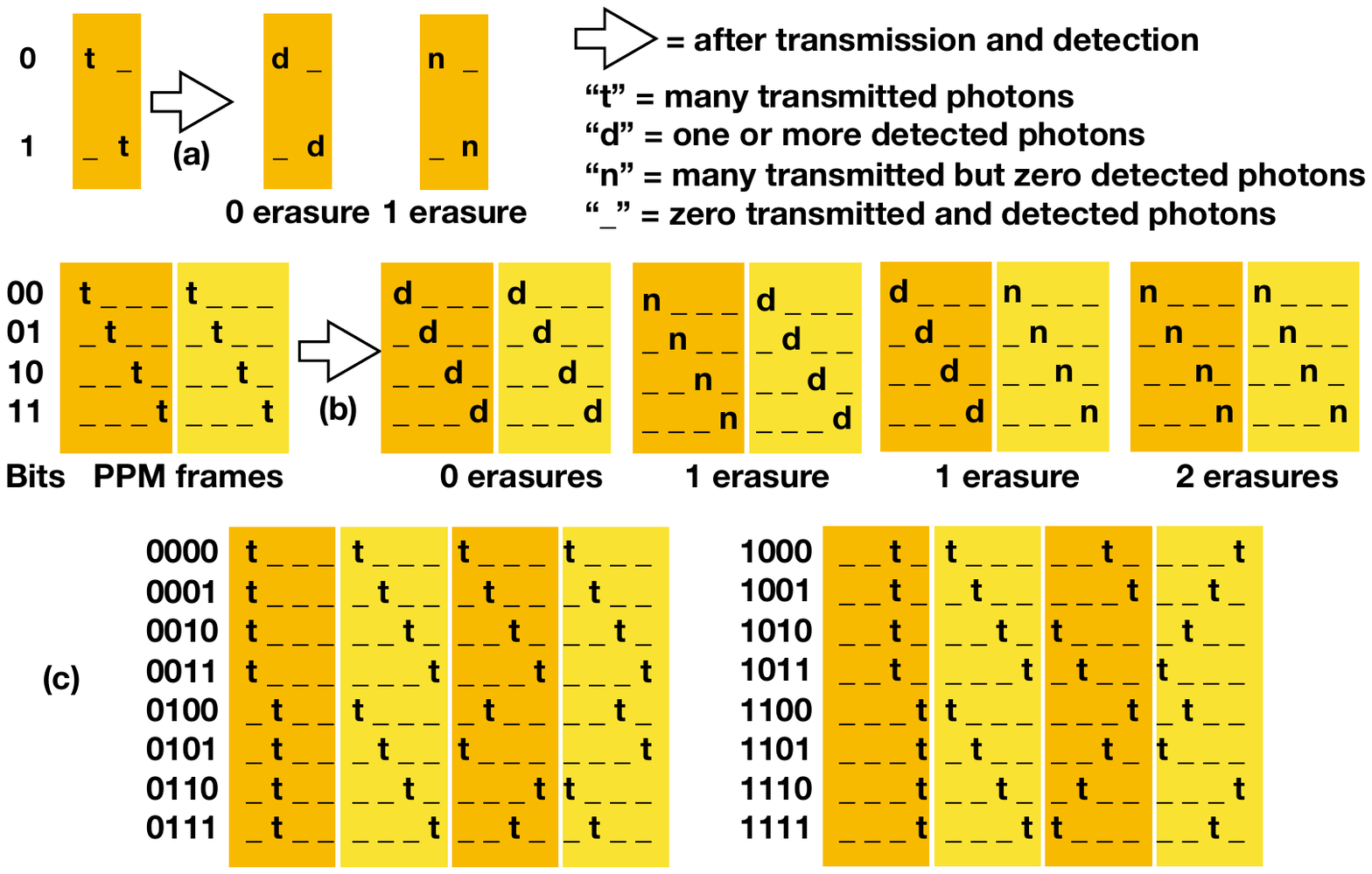}
	{
	trim=0 70 0 30,
	clip,
	width=1\linewidth
	}
	{
	An example of the benefits of error-correction coding (ECC)
	to improve the data-recovery reliability
	in conjunction with pulse-position modulation (PPM).
	(a) One scientific data bit is mapped into a single PPM
	frame with \ile{M=2} slots.
	The effect of an erasure is shown.
	(b) Two data bits are mapped into a pair of PPM frames,
	each with \ile{M=4} slots.
	The three possible erasure scenarios are shown.
	(c) Four data bits are mapped into four PPM frames, each
	with \ile{M=4} slots.
	Erasure scenarios are not shown.
	While the photon efficiency is the same in (a), (b), and (c),
	mapping a larger number of data bits into a larger number of frames results in improved
	immunity to erasures.
}

In \figref{ECCexample}(a) and (b) all possible erasure events are enumerated (for brevity this is omitted in (c)).
In (b) the input data can be inferred even with a single erasure, and data is lost only if there are two erasures.
In (c) each pair of codewords agree in at most one PPM frame but disagree in at least three, so
	one or two erasures (out of a possible four) result in no data loss.
We can easily calculate the probability of these events, with the result that for \ile{K_s = 12}
(\ile{\BPP = 0.083\text{ b/ph}}) the probability of data loss is $6.1{\times}10^{-6}$ for (a),
$3.8{\times}10^{-11}$ for (b), and $9.3{\times}10^{-16}$ for (c).
The reliability improves dramatically.
Increasing the size of the codebook 
can be beneficial, as in this example, because without
any change in BPP the data recovery becomes less susceptible to frame erasures.

\subsubsection{Theoretical constraint}
\label{sec:theoreticalPPMvsPhotonCounting}

The photon efficiency for the codebooks displayed in \figref{ECCexample} are nowhere near the
\ile{\BPP{=}10.9\text{ b/ph}} that we are seeking.
One of the central results of information theory is that under certain conditions
the existence of a sequence of ever-larger codebooks is guaranteed,
with the desirable property that the probability of data loss decreases to zero.
In particular, for the
parameters BPP,  $K_s^R$ and $M$ 
the condition is (see \secref{PPMcapacity})
\begin{equation}
\label{eq:PPM_BPP}
\BPP <  \frac{1 - e^{- K_s^R}}{K_s^R} \cdot \log_2 M
< \log_2 M
\,.
\end{equation}
This is consistent with theoretical bound on \eqnref{logPAR}
presented earlier.\footnote{
The value of BPP assumed in \tblref{designEquations} and in the numerical
calculations of \secref{background} assume equality in \eqnref{PPM_BPP}.
This is recognizably optimistic,
but something close to this should be achievable in practice
with an appropriate design of ECC accompanied by large processing resources
in the receiver.
}
This confirms the significance of photon-starvation mode,
because BPP in \eqnref{PPM_BPP} approaches \eqnref{logPAR} as \ile{K_s^R \to 0}
and thus we must choose a very small value of $K_s^R$ to achieve a BPP near the
theoretical maximum for the chosen $m$.

\subsubsection{Role of redundancy}

We are assured by \eqnref{PPM_BPP} that photon starvation
isn't detrimental to data reliability, as long as ECC overcomes the
inherent unreliability at the modulation coding layer.
The explanation for this is the inherent redundancy.
The average rate of photon detections is \ile{\rate / \BPP}, and
with $K_s^R$ average photons per PPM frame, the rate of PPM frames is
\ile{\rate / K_s^R \cdot \BPP}.
Thus the ``raw'' bit rate at the input to the modulation coding layer is
\begin{equation*}
\frac{m \cdot \rate}{K_s^R \cdot \BPP} = \frac{12 \cdot 1.}{0.2 \cdot 10.9}
= 5.5\ \text{b/s}
\,.
\end{equation*}
Thus, 4.5 b/s of this ``raw'' bit rate is redundancy, and 1 b/s is scientific data
(82 \% redundancy and 18\% scientific data).
According to \eqnref{PPM_BPP},
this redundancy offers a sufficient opportunity to overcome the frequent erasures
in PPM frames suffered in the modulation coding layer.

While the theory backing up \eqnref{PPM_BPP} strongly suggests that
 a very large ECC codebook is required to obtain high BPP, 
 we have to fall back on best practices
 in actually choosing such a codebook.
 For a high SBR, high reliability can be obtained \citeref{834} using Reed-Solomon coding \citeref{876},
which is well suited to overcoming frequent erasures.\footnote{
For a laboratory
demonstration of high BPP in reliable optical communications, see
\citeref{834}.
There are also tutorial papers 
\citeref{172} 
and textbooks
\citereftwo{171}{658} that expand on the theoretical basis of ECC.
}
At lower values of SBR more advanced and modern coding techniques have to be adopted,
a topic beyond the scope of this paper.

\subsubsection{Quantum limit}
\label{sec:BPPvsReliability}

The question arises as to the degree to which the architecture of \figref{layers}
and the specific choice of PPM for the modulation coding limits performance.
Theoretical limits on photon efficiency BPP with reliable data recovery 
are identified in the Appendices:
\begin{itemize}
    \item
    Any physical layer that makes use of electromagnetic
    transmission is governed by a fundamental
    quantum limit on reliable data recovery
    (see \secref{holevo}).
    This limit is expressed in terms of the photon density PPD
    (the average number of photons per dimension).
    There has been some exploration of technologies that may be
    able to approach this limit \citeref{877}.
    \item
    Any physical layer that modulates optical power
    in the transmitter and performs direct detection
    (photon counting) in the receiver
    has, at high SBR, a theoretical limit on BPP
    with reliable data recovery
    defined by \eqnref{logPAR}
    (see \secref{photonCountingCapacity}).
    \item
    The limit of \eqnref{PPM_BPP} comes
    arbitrarily close to the photon-counting limit as \ile{K_s \to 0}
    (see \secref{PPMcapacity}).
\end{itemize}

The shortfall of PPM (and hence photon-counting more generally)
to the quantum limit is plotted in
\figref{BPPvsPPD} (this applies to BPPM equally well).
The quantum limit is expressed in terms of PPD, the
average photons per dimension, so the BPP for PPM is plotted against
the same metric for different values of $m$.
The specific values of \{PPD,BPP\} for \ile{K_s = 0.2} are
 identified by dots, which fall near the maximum BPP achievable,
 thus justifying this assumption in \tblref{receiverParameters} and the
 previous numerical results.

\incfig
	{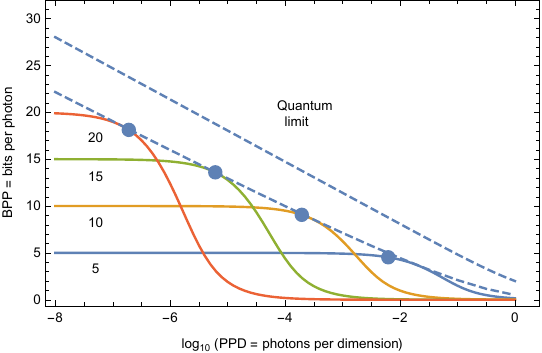}
	{trim=0 0 0 0,width=1.0\linewidth}
	{
	A plot of the photon efficiency BPP at the theoretical limit
	of reliable data recovery (see \secref{PPMcapacity}).
	BPP is plotted against
	\ile{\log_{10} (\text{PPD})},
	where PPD is the
	photon density (average photons per dimension),
	For PPM we have \ile{\text{PPD} = K_s / M},
	and PPD is manipulated by varying $K_s$.
	The different curves are for
	the values \ile{m \in \{ 5,10,15,20\}}, where the number
	of slots per PPM frame is \ile{M = 2^m}.
	The bottom dashed curve is the largest BPP 
	that can be achieved by PPM (by optimizing
	the value of $m$ for each PPD)
	and the upper dashed curve is the quantum limit
	on reliable data recovery.
	The dots correspond to \ile{K_s = 0.2}, the
	nominal value chosen in \tblref{receiverParameters},
	which is nearly BPP-maximizing for the $m$-range of interest.
	This displays the gap between PPM and the
	quantum limit, which shrinks for larger PPD.
	This gap captures the maximum increase in BPP that may be
	feasible with future technologies that achieve a more sophisticated
	manipulation of quantum states.
	}

As $m$ increases in \figref{BPPvsPPD}, BPP increases
as predicted in \eqnref{logPAR}, since for PPM we have
\ile{\PAR = M}. 
The largest BPP achievable by PPM is plotted as the
lower dashed curve.
Our nominal value \ile{K_s = 0.2}
approximates this optimum choice.
If a larger BPP is desired, it is preferable to increase $m$ rather
than further reduce $K_s$.

\subsection{Outage mitigation}
\label{sec:outagemitigation}

Outages are characterized by erasures which are not due to signal shot noise, but rather
environmental factors experienced by a terrestrial-based receiver such as sunlight and weather.
The data loss due to outages will inevitably reduce the rate at which scientific data is communicated
reliably immediately following encounter to \ile{\rate_a < \rate_0}.
A simplistic approach to overcoming outages would be to repeat the transmission of
scientific data multiple times.
We want $\rate_a$ to be as large as feasible, and
fortunately there are far more efficient outage mitigation techniques based on a modification to the
modulation coding and ECC layers in \figref{layers}.
We can demonstrate theoretical upper limits on $\rate_a$ for a particular statistical model of
fully random outages, modify the modulation coding layer to approximate this model,
and choose ECC codebooks which come close to these theoretical limits.

\subsubsection{Outage characteristics}

Outages manifested as frame erasures
(rather than frame errors) are simpler to deal with
(see \secref{BPPM} for the distinction),
and since the receiver is aware of daylight and weather conditions
it can enforce outages-as-erasures
by simply ignoring the received signal during periods of questionable SBR.
In this case the receiver will experience daylight and weather
outages as contiguous bursts of PPM frame erasures.
This distinguishes outage erasures from shot-noise erasures,
which are temporally completely random.

\subsubsection{Modification to modulation coding layer}

In principle the increase in erasure frequency due to outages can be overcome
by choosing a more powerful form of ECC.
In practice, however, it is very difficult for ECC to deal effectively with
temporally highly correlated erasures.
Fortunately this challenge can be overcome by adding
interleaving to the modulation coding
layer as shown in  \figref{erasureChannel}a.
The interleaver in the transmitter scrambles the order of the bits in the input bit stream,
and the de-interleaver in the receiver reverses
that operation to recover the original ordering.\footnote{
Operating the interleaving and de-interleaving at the bit-level simplifies the
following analytical argument.
In practice the interleaving may be performed at the PPM frame level.
}
The modulation-layer decoder outputs a stream of symbols with three possible values
$\{0,1,\text{E}\}$, where 'E' indicates that this bit is unknown because it experienced an erasure
(it was one of $m$ bits in an erased PPM frame).
Whether this erasure was due to shot noise (with probability $\text e^{-K_s}$) or outages 
(with probability $\pW$ for weather and $\pD$ for daylight) cannot be inferred by the PPM decoder,
but nevertheless the knowledge of which bits have suffered an erasure is helpful in the subsequent
error-correction decoding.

The interleaver utilizes a deterministic but pseudo-random algorithm with the goal of destroying
the correlation of the erasures and make them appear to occur randomly and independently
(like a sequence of flips of a biased coin).
This requires a `depth' of interleaving that is larger than the correlation scale
of the frame erasures, which implies longer than the longest outage period 
(the longest credible weather-outage event).
Nominally the number of erasures per error-correction codeword
then obeys a binomial probability distribution which is identical for all codewords, 
yielding a more effective and efficient ECC.

\incfig
	{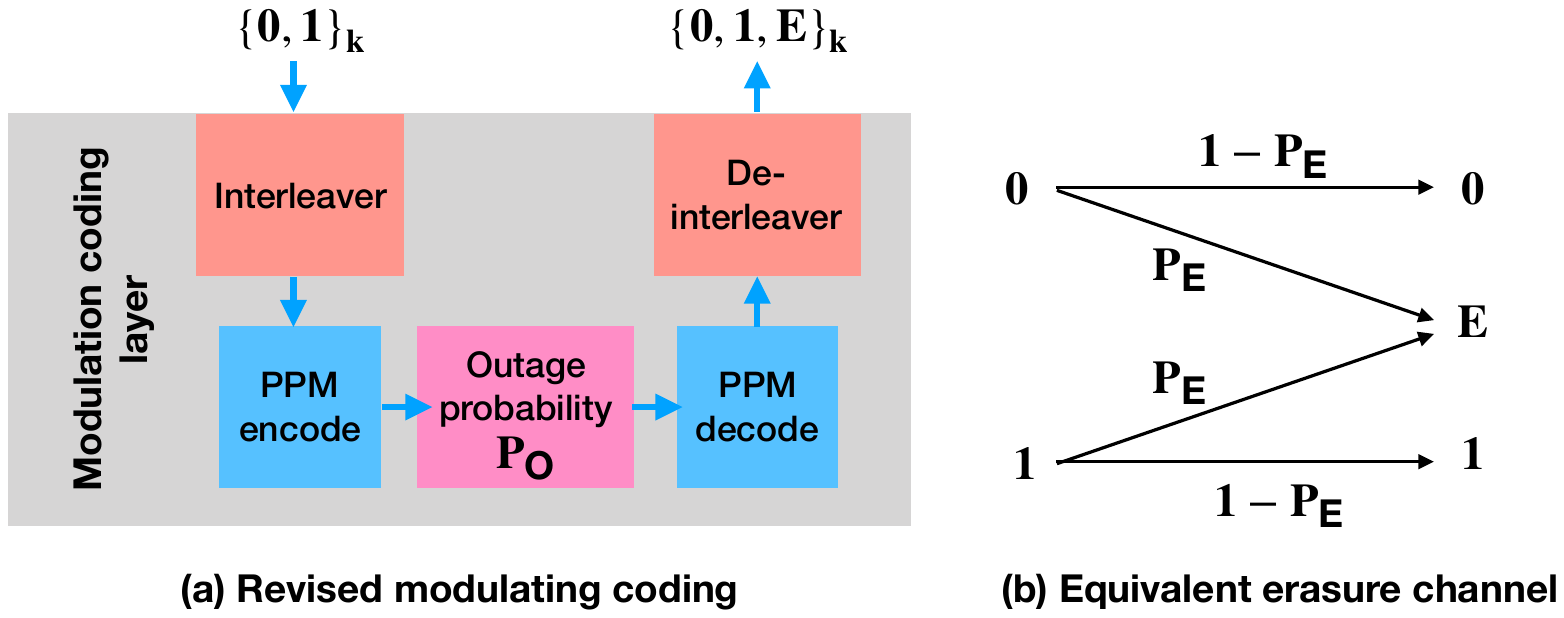}
	{
	trim=0 270 0 0,
	clip,
	width=1.0\linewidth
	}
	{
	Modification to the PPM modulation code layer which converts it
	to a binary erasure channel.
	(a) Bit-level interleaving
	and coordinated de-interleaving 
	convert grouped erasures `E' (due to shot noise, daylight, weather, etc.) into 
	an equal number of pseudo-randomly distributed erasures.
	The PPM coding/decoding introduces erasures due to shot noise,
	while the transmission introduces additional erasures due to outages with
	probability $\pW$.
	(b) Single bit erasure channel with erasure probability $\pE$,
	with transition probabilities $\pE$ and $1-\pE$.
	Due to interleaving we assume that successive channel 
	uses are statistically independent.
	}

\subsubsection{Modification to ECC layer}

It is argued in \secref{outageBPP} that the detected photons per frame
$K_s^R$ and rate of PPM frames $T_I^{-1}$ should not be changed with the addition of outages.
Rather,  the reliability of data recovery is theoretically preserved in the presence of outages
if the input scientific data rate is reduced from its nominal $\rate_0$ by a factor of
 \ile{\big( 1-\pW \big) \big( 1-\pD \big)}.
That is, the data rate is reduced on average by the fraction of PPM frames that are correctly decoded,
which is intuitively the best we could hope for.

Since $K_s$ should not be changed in response to outages, neither should
 $P_P^T A_e^T T_s$, $A_e^S$ or $N^S$.
Assuming that $T_I$ is also not changed, neither is the average transmit power $P_A^T$.
The sole effect of a reduction in data rate to $\rate_a < \rate_0$ is an increase in the redundancy in the
ECC in order to maintain fixed data-recovery reliability in the face of the added
erasures due to outages.
Since the data rate is lowered without a reduction in average power, 
an unsurprising side effect is to reduce the photon efficiency BPP
by the same factor.
Fortunately this has no practical impact for a continuous electrical
power source not based on fuel consumption.

If rate $\rate_a$ is unacceptably small, it can always be increased by choosing a larger $\rate_0$.
This of course will have implications throughout the design such as
increased average power, increased duty cycle with an attendant increase in the impact of dark counts, etc.

\section{Conclusions}

The design of a communication downlink from
low-mass interstellar probes is extremely challenging.
While the downlink violates no laws of physics
and appears to be feasible in theory, 
the challenge arises when the characteristics of
available technologies are taken into account.
Innovation and invention in both system concepts
and in the constituent technologies will be needed.

The primary areas of necessary technology development
identified in this paper are:
\begin{itemize}
\item
Low-mass light sources achieving high peak powers
possibly combined with optical pulse compression technology in either
transmitter or receiver.
\item
Low-mass transmit aperture and coordinated attitude adjustment
and pointing with sufficient accuracy.
\item
A feasible multi-probe multiplexing scheme taking into
account inaccuracies in the knowledge of time and speed and distance,
including Doppler shift.
\item
Approaches to a near-earth uplink for configuration of transmit wavelength.
\item
Optical \leveltwo possibly incorporating
adaptive optics to counter atmospheric turbulence,
achieving uniform sensitivity over a relatively large non-circular coverage angle,
and with sufficient coronagraph rejection of target star radiation.
\item
Optical bandpass filters with sufficiently narrow bandwidth,
and with agility and configuration capability to accommodate multi-probe
multiplexing and uncertainties in Doppler shift.
\item
Receive optics and optical detectors with very low dark count rates, with possibly
detector sharing over multiple \leveltwo{s}.
\item
Low-mass and low-power processing to realize the requisite
compression and error-correction coding on board the probe.
\end{itemize}
This paper can serve as a roadmap to
further investigation and research,
and the ultimate outcome depends on the success
of those efforts.

\section*{Acknowledgements}

PML gratefully acknowledges funding from NASA NIAC
NNX15AL91G and NASA NIAC NNX16AL32G for the NASA Starlight program
and the
NASA California Space Grant NASA NNX10AT93H,
a generous gift from the Emmett and Gladys
W. Technology Fund, as well as support
from the Breakthrough Foundation for its Breakthrough StarShot
program.
More details on the NASA Starlight program can be found
at \url{www.deepspace.ucsb.edu/Starlight}.

\appendix

\twocolumngrid

\section{Relativistic effects}
\label{sec:dopplerModel}

Since a low-mass probe travels at relativistic speed,
relativistic effects enter.

\subsection{Transmitter and receiver motion}
\label{sec:redShift}

Assume an inertial frame $S$ 
(one approximating
a heliocentric coordinate system is convenient).
Relative to the trajectory of a photon from probe to receiver,
the probe and receiver have a longitudinal component of velocity
and a small transverse component (which is neglected).
Let the longitudinal components for the probe and
a receiver relative to $S$ be $u_p$
and $u_r$ respectively.
The relationship between
transmitted wavelength $\lambda_T$
(observed in the rest frame of the transmitter)
and received wavelength $\lambda_R$
(observed in the rest frame of the receiver)
is\footnote{
This relation, including the product law, is derived in \citeref{806}.
Note however that Eq. (17) of that reference is incorrect except at
\ile{\rho = \pm 1}, with the corrected value given by \eqnref{xiVsu2}.
}
\begin{align}
\label{eq:xiVsu1}
&\xi =
\frac{\lambda_R}{\lambda_T} = \frac{\nu_T}{\nu_R} =
 e(u_p,-1) \cdot e(u_r,+1)
  \\
\label{eq:xiVsu2}
 &e(u,\rho) =  \frac{1- \rho u/c}{\sqrt{1 - u^2/c^2}}
    \,.
\end{align}
Thus $\xi$ is the product of four factors, each of which models
a distinct physical effect.
The two numerators account for Doppler shifts due to
changing propagation delay, while the two denominators
model the relativistic time dilations
of the transmit and receive clocks.
Nearly desired $\lambda_R$ can be achieved by adjustment
of $\lambda_T$ to compensate for the first factor, while
smaller variations in $u_p$ and $u_r$ have to be accounted
for in other ways (see \secref{redshift} and \secref{earthMotion}).

\subsubsection{Probe speed}

The systematic red shift due to the nominal probe speed
 \ile{u_p = 0.2 c} results in \ile{\xi = 1.22} (or a 22\% shift).
 This should be compensated by an adjustment in $\lambda_T$.
Since the energy of each photon is \ile{E = h c/\lambda_R},
the energy of each photon at the receiver is lower than at the probe.
Energy seen by an observer in $S$ is conserved, including
this photon energy loss and kinetic energy gains for the
probe
(the recoil effect) and receiver
(photon absorption)  \citeref{872}.
The probe transmitter thus acts as a photon
engine which continually increases $u_p$ by a tiny amount.
From the perspective of the receive signal, the motion affects
wavelength and not
photon count, and thus has
no effect on the data rate $\rate$
obtainable by a direct-detection receiver.

\subsection{Uncertainty in probe speed}
\label{sec:uncertainty}

Assuming that \ile{u_r = 0},
based on \eqnref{xiVsu1},
any uncertainty in probe speed $u_p$ relates directly
to an uncertainty in received frequency $\nu_R$.
Differentiating the relation \ile{\xi \nu_R = \nu_T}
with respect to $u_p$,
we get
\begin{equation}
\label{eq:dNuR}
   \xi \cdot \frac{\text d \nu_R}{\text d u_p} 
    + \nu_R \cdot \frac{\text d \xi}{\text d u_p}
    = 0
    \,.
\end{equation}
Eq.\eqnref{freqSensitivity} follows from
substituting \eqnref{xiVsu2} in \eqnref{dNuR}.

\subsection{Gravitational potential effects}
\label{sec:gravitationalEffects}

There are small gravitational effects on the optical communications photons due to the 
gravitation potential in the target stellar system as well as our solar system. 
Gravitational effects on the probe cause it to speed up as it approaches the target
and slow down after encounter.
The photons leaving the spacecraft before it passes the target undergo a gravitation redshift as it climbs out of the target system potential back to the Earth and a gravitational blueshift as they enter our solar system and are detected at the Earth or nearby (lunar base etc). 
For photons emitted after the target encounter there will additional effects that need to be included due to the infall and exiting of the photons as the pass the vicinity of the target system.

This potential modifies the photon energy and hence the frequency in proportion to the integrated difference in 
gravitational potential divided by \ile{c^2} or one-half the ratio of the Schwarzschild radius divided by distance to object (star, planet etc). 
With an extremely narrow linewidth (bandwidth) of the transmit laser gravitational potential may be detectable,
offering an opportunity for extremely sensitive measurements of gravitational potential.
For example, with a transmit laser linewidth of 
10 KHz at 300 THz a fractional frequency offset \ile {3{\cdot}10^{-11}} may be observable.
This is of course modified by the modulation and dependent on the laser technology. 
For comparison the Schwarzschild radius of our Sun is about \ile{3\ \text{km}} and at a distance of 
\ile{1\ \text{AU}} (\ile{1.5{\cdot}10^{8}}\ \text{km}) this yields a gravitational shift relative to infinity of \ile{1{\cdot}10^{-8}}. 
While small compared to the relevant Doppler shifts due to velocity effects, it is still $300\times$ larger than the fractional laser linewidth. 
Even the graviational potential of the Earth is relevant
(Schwarzschild radius \ile{9\ \text{mm}} with physical radius of
\ile{6.4{\cdot}10^6\ \text{m}} yielding shift \ile {8{\cdot}10^{-10}}).

\section{Antennas}
\label{sec:antennaTheory}

Certain fundamental principles of antennas
or apertures
for electromagnetic communication illuminate both
opportunities and constraints on our downlink.
Our interest here is only in theory that applies
to an \leveltwo, which is where the coverage
and the SBR are determined.
Our \leveltwo is assumed to be a diffraction-limited
optical aperture with a single optical
detector, which is essentially a
single-pixel optical telescope.
Our receive aperture as a whole is \emph{not} diffraction-limited
(it has multiple detectors, and radiation
adds incoherently), so the following 
principles do not apply.
However, the coverage and SBR determined
at the \leveltwo level are preserved in the
scale-out to $N$ {\leveltwo}s (see \secref{scaleout}).

\subsection{Principles}
\label{sec:principles}

As a consequence of Maxwell's model of electromagnetism,
the \emph{reciprocity theorem} states that when one
antenna is used for transmission and another for
reception, their roles can be reversed without
any change to the outcome 
(see Appendix D of \citeref{870}).
It is often easier to analyze an antenna in transmit
mode, even if it is to be used in receive mode, or
\emph{vice versa}.

Suppose a plane wave at wavelength $\lambda$ and equivalent
frequency $\nu$
has specific intensity $\mathcal I_\nu$ as a function
of frequency $\nu$, defined as the flux (power per cross-sectional
area) per frequency interval \citeref{822}.
The corresponding intensity in 
frequency interval $\text d \nu$ is $\mathcal I_\nu \cdot \text d \nu$.
Since the bandwidths of interest are very small for interstellar
communication (to eliminate most noise and interference),
it is convenient to use $\nu$ and recognize that any variation
in $\lambda$ can usually be neglected.
Thus we assume a fixed wavelength \ile{\lambda = \lambda_R}, and simultaneously
(and somewhat inconsistently) allow $\nu$ to vary over a bandwidth $W_e$.

For astronomical sources, which can often be considered as isotropic emitters
(at least over a finite range of directions of interest) 
it is often more convenient to use the
specific brightness $\mathcal B_\nu$, which is the power per
unit solid angle per frequency interval.
At great distances a spherically expanding wave (as from
a point source) can be considered a plane wave at the antenna
(as long as the difference between a plane and a sphere is a small fraction of
a wavelength).
In that case, for a spherical wave at a distance $D$ from
the source, the conversion 
from specific brightness to specific intensity is \ile{\mathcal I_\nu = \mathcal B_\nu/D^2}.

\subsection{Effective area and gain}

When a plane wave with specific intensity $\mathcal I_\nu$ 
impinges on an antenna, we can measure
the specific total power $P_\nu$ delivered to an impedance-matched termination (at radio wavelengths) or an ideal
optical detector (at optical wavelengths).
The \emph{effective area} of the antenna is defined as
\ile{A_e = P_\nu/\mathcal I_\nu}, which has the units of area.
$A_e$
usually depends strongly on the angle of incidence,
and any dependence on $\nu$ can often be neglected over a small
bandwidth $W_e$.

Often we
are interested in the maximum value of $A_e$
for the antenna's preferred direction (its ``pointing''
or ``main beam'').
For an aperture with geometric area $A_g$,
in general \ile{A_e \le A_g} with
equality when the aperture is diffraction-limited
and the direction of propagation of the plane wave
is aligned with the pointing.

An isotropic radiator does not favor one direction over another.
Its specific brightness
$\mathcal B_\nu$  is constant in all directons.
For a lossless isotropic radiator and lossless
(free-space) propagation, by conservation of
energy
\ile{4 \pi \mathcal B_\nu = P_\nu}.

The radiation pattern of an antenna of interest is conveniently specified
by the ratio of its specific brightness in a particular direction to that
of an isotropic radiator with the same power input.
This value \ile{G = 4 \pi \mathcal B_\nu/P_\nu} is the
\emph{directivity} or \emph{gain} of the antenna in that direction.\footnote{
Directivity refers to an ideal antenna, while
gain takes any losses in the antenna structure
into account. Here we assume an ideal lossless
antenna so there is no distinction.}
For a lossless diffraction-limited antenna,
$P_\nu$ equals $\mathcal B_\nu$ 
integrated over all directions, so
\begin{equation}
\label{eq:consOfPower}
P_\nu = \iint \mathcal B_\nu \cdot \text d \Omega
\quad\text{or}\quad
\frac{1}{4 \pi}
\iint G \cdot \text d \Omega = 1 \,.
\end{equation}
In words, the gain of any antenna averaged over all solid angles is unity,
and identical to the gain of an isotropic radiator in any direction.

It follows from the laws of thermodynamics \citeref{870} that
for some particular direction, for an antenna with gain $G$
and effective area $A_e$ in that direction,
\begin{equation}
\label{eq:antennaThermo}
    G = \frac{4 \pi A_e}{\lambda_R^2}
    \,
\end{equation}
Thus the gain of an antenna (a mostly transmission characteristic)
is proportional to effective area (a mostly reception characteristic)
expressed in wavelengths.
It follows that for an isotropic radiator
\ile{A_e = \lambda_R^2 / 4 \pi} in all directions.

\subsection{Highly directive antennas}
\label{sec:directiveAntenna}

For optical communications and astronomy we are interested in
highly directive antennas.
Consider an ideal directive antenna that in transmit mode concentrates
all its flux uniformly across solid angle $\Omega_A$
(called its main beam),
and there is zero flux outside
this solid angle of coverage.
Such an antenna is diffraction-limited
when the phase shift from optical source
to free-space is fixed for all angles within
$\Omega_A$.
Such an antenna used
in receive mode has a fixed effective area $A_e$ for any direction within the main beam.
Note that the main beam area need not be circular for this to hold (this is important
in our \leveltwo application).

Then based on \eqnref{consOfPower} the gain of this antenna is
\ile{G = 4 \pi / \Omega_A}, and
substituting into \eqnref{antennaThermo}
\begin{equation}
\label{eq:areaSolidAngle}
A_e \Omega_A = \lambda_R^2
\,.
\end{equation}
The solid angle covered by this ideal main beam
is inversely proportional to effective area, the latter expressed in
units of wavelength.
As expected, directive antennas with a larger
effective area are more directive.
This idealized result can be approximated for a practical directive
antenna, which introduces sidelobes due to diffraction.

\subsection{\Leveltwo response to different sources}
\label{sec:apertureResponse}

The specific power $P_\nu$ delivered to a matched termination
or absorbent surface
by a lossless diffraction-limited aperture
is now evaluated for the types of radiation sources of interest
in interstellar communication.

\subsubsection{Interference}
\label{sec:interferenceResponse}

Interference from the target star can be approximated as 
an isotropic point source with constant specific brightness $\mathcal B_\nu$ at all angles.
To a receive antenna at great distance $D_0$ from the target star this will appear as a plane wave
with specific intensity \ile{\mathcal I_\nu =\mathcal B_\nu /D^2}
and thus
\begin{equation}
\label{eq:interferencePower}
    P_\nu = A_e \mathcal I_\nu =  A_e \cdot \frac{\mathcal B_\nu}{D^2}
\end{equation}
for a receive antenna with effective area $A_{e}$ in the
direction of the interferer.

\subsubsection{Probe transmitter}
\label{sec:signalResponse}

For communications, the optical bandwidth $W_e$ will always be
at least as large as the signal bandwidth.
Thus it is more appropriate to leave out the `specific' in
intensity, brightness, and power, and model the total
power over all frequencies.
At distance $D$ the average receive power $P_A^S$ in a lossless
antenna (\leveltwo) with effective area $A_e^S$ is
\begin{equation}
\label{eq:Friis}
    P_A^S = \frac{P_A^T}{4 \pi} 
    \cdot \frac{ 4 \pi A_e^T}{\lambda_T^2}
    \cdot \frac{1}{D^2}
    \cdot A_e^S =
    \frac{A_e^T A_e^S P_A^T}{\lambda_T^2 D^2}
\end{equation}
for transmit antenna with average transmit power $P_A^T$ and effective area $A_e^T$, 
confirming \eqnref{FriisPower}.
The four product terms in \eqnref{Friis} are (1) the brightness
of the transmitter if its radiation were isotropic, (2) the
transmit antenna gain, (3) the translation
from brightness to intensity
at the receiver, and (4) the receive aperture equivalent area
(which converts intensity to power).
This is the well-known Friis transmission equation \citeref{871}.

Although the transmit and receive wavelengths
$\lambda_T$ and $\lambda_R$ differ
substantially due to Doppler shift (see \secref{redShift}),
$P_R$ in \eqnref{Friis} is not dependent on $\lambda_R$.

\subsubsection{Noise}
\label{sec:noiseResponse}

Finally consider noise due to unresolved sources of radiation.
From the perspective of the receive antenna, we can consider
this to be an isotropic source of radiation.
In practice the radiation is never fully isotropic,
because it is limited in extent (like Zodiacal radiation in a distant
solar system) or is not fully visible to a terrestrial antenna 
(due to occlusion by the ground).
However, if the antenna is highly directional, and its
coverage coincides with radiation that is well approximated as
uniform with direction, then the isotropic model should be accurate.

Isotropic radiation has constant specific brightness $\mathcal B_\nu$.
For an antenna with gain $G$ as a function of direction,
the received power is
\begin{equation}
\label{eq:noisePower}
P_\nu = \iint G \mathcal B_\nu \cdot \text d \Omega
= \mathcal B_\nu \cdot \iint G \cdot \text d \Omega
= 4 \pi \mathcal B_\nu
\end{equation}
making use of \eqnref{consOfPower}.
The received specific power is independent of the antenna effective area.
This is why we can meaningfully speak of a CMB `noise temperature'
at microwave frequencies
without referring to $A_e^S$.

A couple of special cases are of interest.
First, if the isotropic specific brightness $\mathcal B_\nu$ 
were actually generated by
an isotropic radiator with power $P_\nu$,
then we know that \ile{G \equiv 1} and 
\ile{B_\nu = P_\nu / 4 \pi} consistent with \eqnref{noisePower}.
Second, if a highly directive antenna receives isotropic radiation,
then \eqnref{noisePower} becomes
\begin{equation}
P_\nu = \iint_{\Omega_A} \frac{4 \pi}{\Omega_A} \cdot \mathcal B_\nu \cdot \text d \Omega
= 4 \pi \mathcal B_\nu
\,.
\end{equation}
In this case $P_\nu$ is independent of 
$\Omega_A$ (and hence effective area $A_e$)
because the total noise power due to the size of the main beam and
the gain within that main beam precisely offset.

An alternative interpretation of this last observation is that the
received power is proportional to $A_e^S$ 
(the conversion factor of input flux to power at the detector) 
and also $\Omega_A$ (the fraction of total isotropic power reaching the detector)
(see Eq. (8.76) of \citeref{1001}).
This product $A_e^S \Omega_A$ is constant 
for a diffraction-limited highly directive antenna (based on \eqnref{areaSolidAngle}).

\section{Data volume vs. latency}
\label{sec:latencyVolume}

Let $D_0$ be the distance
to the target star and \ile{D>D_0} the distance to
the probe.
The total latency
(elapsed time since launch) for data transmitted from
distance $D$ is
\begin{equation}
\label{eq:latency}
T_L = D \left(\frac{1}{u_0} + \frac{1}{c} \right)
\end{equation}
for fixed probe speed $u_0$.
The terms in \eqnref{latency} are respectively
the elapsed time $t$ for the probe to
reach distance $D$, and a
photon emitted at distance $D$ to return to
to the receiver,
both measured relative to the rest frame of the receiver.

The appropriate data rate $\rate$ for purposes
of data volume is the transmitted data rate \eqnref{rateVsDistance}.
The total data volume transmitted
between distance $D_0$ and $D$ is
\begin{equation}
\label{eq:volume}
\volume = \int_{D_0/u}^{D / u}
\rate (\massratio, u_0 t) \, \text{d}t =
\frac{\rate_0 D_0}{u_0} \cdot \left( 1 - \frac{D_0}{D} \right)
\,.
\end{equation}
This entire $\volume$ is decoded at the receiver
during time duration $T_L$.
Eliminating $D$ from \eqnref{latency} and \eqnref{volume}, the latency can be expressed directly in terms of the volume,
\begin{equation}
\label{eq:latencyDecomposition}
T_L = \frac{D_0}{u_0} + \frac{D_0 \volume}{D_0 \rate_0 - u_0 \volume} + \frac{D_0^2 \rate_0}{c ( D_0 \rate_0 - u_0 \volume )}
\end{equation}
where the three terms are the transit time, the transmission time, and the return propagation time
all expressed in terms of an earth-based clock.
After substituting for $\rate_0$ and $u_0$ in terms of $\massratio$
(see \secref{masstradeoffs}),
this latency can be minimized over the choice of $\massratio$.

For any $\rate_0$, $\volume$ approaches a finite
asymptote\\
\ile{\volume \to \rate_0 D_0/u_0}
as \ile{D \to \infty}.
This ultimate limit on volume with infinite latency is due to the decreasing
data rate with distance.
Larger $\rate_0$ and a smaller $u_0$ help to increase $\volume$,
the latter effect due to the slower $D$-related falloff in $\rate$.

There will also be a small reduction in available electrical power during transmission
(due for example the half life of a radioactive source),
and this is not taken into account in \eqnref{volume}.

\section{Quantum limit on data reliability}
\label{sec:holevo}

The actual design of a concrete ECC for the photon-counting channel
will not be pursued here, because ECC design is itself an
extensive and sophisticated project.
Nevertheless it is important to determine how large a BPP consistent with reliable data
recovery might be expected, and also quantify how larger 
BPP relates to larger PAR and bandwidth $W_e$ because those two parameters
interact strongly with other aspects of the downlink design.

\subsection{Channel model}

A given physical configuration is not generally associated
with a theoretical limit on the data rate $\rate$, because
we must also specify other details of the physical layer
such as aperture sizes and detection method before such a
limit can be established.
Any theoretical limit is associated with a specific \emph{statistical channel model},
which specifies, for each possible input
(typically a continuous-time or discrete time signal),
a probability distribution of the output.
Here we find it useful to address three such models:
the quantum limit (see \secref{holevocapacity}),
direct-detection of light (see \secref{photonCountingCapacity}),
and direct-detection in conjunction with
a PPM modulation layer (see \secref{PPMcapacity}).

\subsection{Channel capacity}

Communication theory offers a way to establish
a theoretical limit on the photon efficiency BPP
which can be achieved consistent with reliable data recovery
through the concept of \emph{channel capacity} $\capacity$.
Channel capacity can be calculated 
(using statistical techniques beyond
our scope) for any specific channel model.
A channel model can be discrete-time or continuous time,
with any combination of discrete or continuous inputs or
outputs.
Channel capacity does \emph{not} place a limit on the
data rates that can be achieved (see \secref{reliabilityHard}).
What it does reveal is a theoretical limit on what 
data rate $\rate$ can
be achieved with arbitrarily high reliability, applying to the specific
channel model in question.
The associated photon efficiency is easily inferred
from \ile{\BPP{=}\capacity/\avgpwr}.
In particular, capacity reveals that:
\begin{itemize}
    \item
    For any \ile{\rate < \capacity}, 
    a modulation code and ECC is guaranteed to exist
    that can achieve data rate $R$ in conjunction with
    any arbitrarily stringent reliability requirement.
    Channel capacity does not reveal any concrete way
    to achieve \ile{\rate \to \capacity}, as that is an
    entirely separate issue.\footnote{
    Based on decades of research it is generally
    possible to narrow the gap between $\rate$ and $\capacity$, although
    attempting to do so may be impractical
    (due to complexity, storage and
    processing overhead, PAR, bandwidth, etc.)}
    \item
    The existence of capacity does not preclude 
    \ile{\rate \ge \capacity} for some concrete implementation, for arbitrarily large $\rate$.
    However it asserts that high
    reliability cannot be achieved at such data rates.
    Generally the highest achievable reliability
    will deteriorate as $\rate$ increases.
\end{itemize}
Capacity is very useful as a tool in the preliminary
design of a communication link because it represents an
ideal to shoot for; that is, what BPP is possible
for reliable data transmission.
Also it tells us, for any concrete implementation,
how much further improvement may be possible with the
application of additional effort and resources.

\subsection{Holevo capacity}
\label{sec:holevocapacity}

For optical communications,
the most general channel model assumes that the transmitter
and receiver can manipulate an arbitrarily large
number of quantum states directly and simultaneously.
The resulting \emph{Holevo capacity} establishes
a technology-independent theoretical limit on the 
data rate that can be achieved with reliable data recovery \citeref{863}.
Since we cannot avoid the randomness inherent in
quantum mechanics in any physically meaningful way,
Holevo capacity represents an ultimate and theoretical limit
to aspire to, and reveals what future progress may be
feasible with new technologies not available to us today.
Recent research is making some inroads in
approaching the Holevo limit \citeref{888}.

PPM makes use of a repeated transmission of PPM frames,
each composed of
$M$ timeslots of duration $T_s$.
Thus it uses temporal modes only
(avoiding the use of wavelength, spatial, and
polarization modes to convey information).
For a fair comparison with
Holevo capacity we also limit the
Holevo capacity to temporal quantum modes
making use of timeslots with the same duration $T_s$.
Then for average power $\avgpwr$ the average
photon count per timeslot is
\ile{\PPD = T_s \avgpwr}.
The photon efficiency BPP at the Holevo quantum limit
is then \citeref{800}
\begin{equation}
\label{eq:HolevoBPP}
\BPP =
(1 + 1/\PPD) 
\cdot \log_2 (\PPD +1) - 
\log_2 (\PPD)
\,.
\end{equation}
Of course this must be greater than 
either the BPP achievable by a direct-detection
receiver (see \secref{photonCountingCapacity}) 
or PPM
(see \secref{PPMcapacity}).
This applies in the absence of background radiation,
which suffices for our purposes.

\section{Theoretical limits on direct detection}
\label{sec:photonCountingCapacity}

A practical channel model for today's technology
is modulation of optical intensity in the transmitter
and direct detection of photons in the receiver.

 \incfig
	{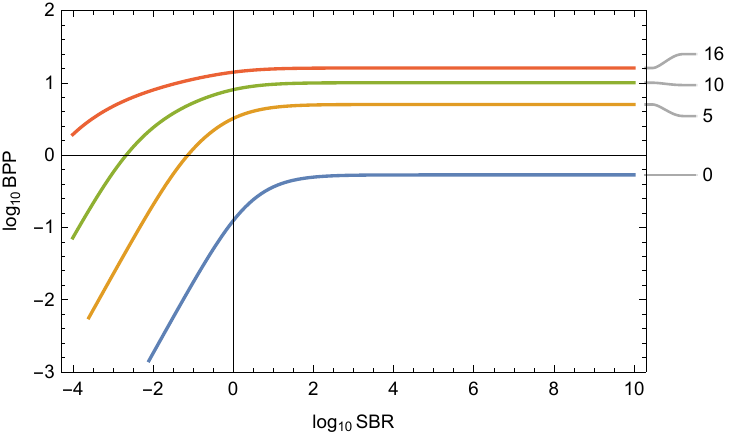}
    {width=.97\linewidth}
    {
    For a continuous-time photon-counting channel, a
    log-log plot of BPP vs SBR with the different
    curves corresponding to different values of PAR.
    The curves are labeled
    by \ile{\log_{2} \PAR}, so for example
    16 corresponds to \ile{\PAR = 2^{16} = 65,536}.
    The axis \ile{\log_{10} \BPP = 0} corresponds to a photon efficiency of 1 b/ph.
    The axis \ile{\log_{10} \SBR = 0} corresponds to
    equal average signal and background power.
    }

\subsection{Constraints}
\label{sec:constraints}

If we allow infinite transmit power, any data rate $\rate$ can be achieved reliably.
To get practically meaningful results, we must place constraints on the received power.
Assume that the intensity (average photons per second) over a finite interval $T_c$ is $\intensity_n (t)$,
which we can think of as a codeword.
These are the peak power $\peakpwr$ and average  power $\avgpwr$ constraints
\begin{align*}
\label{eq:ccconstraints}
&0 \le \Lambda_n (t) \le \peakpwr
\,,\;
0 \le t \le \timeword
\\
&\frac{1}{\timeword} \int_{0}^{\timeword} \Lambda_n (t) \cdot \diff t = \avgpwr
\,,\;
1 \le n \le N_c
\,.
\end{align*}
We make two simplifying assumptions which
have no effect on the capacity $\capacity$.
First, each and every codeword is constrained to have the
same energy $\timeword \avgpwr$.
Also, every codeword has an
ON-OFF character (\ile{\Lambda_n (t) = \peakpwr}
or \ile{\Lambda_n (t) = 0} for \ile{0 \le t \le T_c}).
Photon rates \ile{\{ \avgpwr, \peakpwr \}}
are referenced to the detector output,
taking account of everything that happens in the
physical layer (including quantum efficiency).
Also assume that there is a fixed background rate $\backpwr$,
and define \ile{\PAR = \peakpwr/\avgpwr} and \ile{\SBR = \avgpwr/\backpwr}.

 \subsection{Capacity result}
 
 The direct-detection channel is modeled by photon detection
 events governed by Poisson arrival statistics.
 This is the most random arrival process, in which
 inter-arrival times are statistically independent and
 obey an exponential distribution.
 The capacity has been determined for this statistical
 model under the peak and average power constraints
 described in \secref{constraints}.
Define two parameters
\begin{displaymath}
s = \frac{1}{\PAR \cdot \SBR}
\,,\quad
q = \min \left\{ \frac{1}{\PAR},\, \frac{(1+s)^{1+s}}{e s^s} - s \right\}
\,,
\end{displaymath}
and then \citeref{799}
\begin{equation}
\label{eq:fullcap}
\frac{\capacity}{\avgpwr} = \BPP = \PAR \cdot \log_2 
\frac
{(1+s)^{q (1+s)} s^{(1-q) s}}
{(q+s)^{q+s}}
\,.
\end{equation}
In the limit as \ile{\SBR \to \infty} (or equivalently \ile{s \to 0}),
\eqnref{fullcap} simplifies to
(yielding \eqnref{logPAR} for \ile{\PAR > e})
\begin{align}
\label{eq:highSBRcap}
&\BPP \to - \PAR \cdot q \log_2 q
\\
\notag
&q \to \min \left\{ 1/\PAR ,\, 1/e \right\}
\,.
\end{align}

\subsection{BPP is unbounded}
\label{sec:unbounded}

Notably the capacity $\capacity$ is unbounded
even for finite average power,
\begin{equation}
\BPP \rightarrow \infty
\;\text{as}\;
\PAR \to \infty
\;\text{for any SBR} < \infty
\,.
\end{equation}
This confirms that a PAR constraint is necessary to prevent
infinite BPP, and more importantly indicates
that increasing $\peakpwr$ increases BPP and $\capacity$
even as $\avgpwr$ is held constant.
The uncertainty principle will intervene
at very high PAR \citereftwo{798}{836}
to render our channel model physically unrealizable.

\subsection{Effect of background radiation}

A log-log plot of BPP vs SBR
is shown in  \figref{bitsPerPhotonVsABlargeSBR}
over a wide range (14 orders of magnitude) of SBR.
The value of BPP is nearly constant with the value \eqnref{logPAR}
for large SBR.
In the region of low SBR, the highest achievable BPP
falls off rapidly.
This is clearly a region to be avoided by choosing a
sufficiently large transmit power $P_A^T$.
Notably even in the regime of low SBR, the background
can still be overcome by choosing a sufficiently large PAR.

\section{Theoretical limits on pulse-position modulation}
\label{sec:PPMcapacity}

Constraining the implementation to
the concrete architecture of \figref{layers}
must necessarily reduce the capacity
relative to the direct detection model of \secref{photonCountingCapacity},
which does not constrain the modulation code in any way.
We can ascertain the adverse effect of choosing a particular modulation
code by associating a new bits-in to bits-out channel model
with the
modulation code-in to modulation code-out.
In the case of PPM this new channel model
has bits-in and bits-plus-erasures-out,
with an erasure probability \ile{e^{-K_s}}.
The  resulting theoretical limit on reliable data recovery
is well known to be \eqnref{PPM_BPP}
\citeref{798}.

One distinction between PPM and the assumptions
behind the Holevo capacity
is that the $K_s$ average photons are
all isolated in a single timeslot per frame rather than
allowed to fall
in the available dimensions in any pattern.
The other distinction is the direct manipulation of
quantum states allowed in Holevo, which
lacks a practical realization in today's technology.

\subsection{Effect of outages}
\label{sec:outageBPP}

The idea of using a de-interleaver to randomize the temporal statistics of outages
was introduced in \figref{erasureChannel}a, and for theoretical convenience assumes that erasures occur
at the bit level (although in practice they may well occur at the PPM frame level).
An ideal \emph{erasure channel} is pictured in \figref{erasureChannel}b.
It models a statistical relationship between input bits  $\{0,1\}$
and output symbols  $\{0,1,\text{E}\}$ in terms of a set of transition
probabilities, where $\pE$ is the probability of a single bit-level erasure
(which equals the probability of a PPM frame erasure).
Further, it is assumed that the individual symbols in \figref{erasureChannel}b
are statistically independent, a condition that is strongly violated with PPM frame erasures
or outages, but that a well-designed
interleaver can approximate.

The capacity $C_u$ of the ideal erasure channel of \figref{erasureChannel}b is well known to be
\ile{C_u = \big(1 - \pE \big)} per channel use, or
\begin{equation*}
C_u = \big( 1 - \pE \big) = \big(1 - \text e^{-K_s} \big) \cdot \big(1-\pW \big) \cdot \big(1-\pD \big)
\end{equation*}
where $K_s$  is the average detected photons per PPM frame.
This assumes that the three sources of erasures
(shot noise, weather, and daylight) are statistically independent.
Weather outages and daylight may be correlated, in which case $C_u$
can be adjusted accordingly.

The rate of channel uses in  \figref{erasureChannel}b as a model for 
\figref{erasureChannel}a is $m/T_I$, and thus the capacity of the modulation
coding layer is $m C_u /T_I$.
Relative to the case of no outages, this capacity is reduced by a factor of
\ile{\big(1-\pW \big) \big(1-\pD \big)}, and thus the data rate $\rate_0$
has to be reduced by the same factor to ensure that reliable data recovery remains possible.
Since the average photon detection rate is $K_s/T_I$, the upper bound on BPP is changed to
\begin{equation}
\label{eq:BPPwOutages}
\BPP < m \cdot \big(1- \pW\big) \cdot \big(1-\pD \big)  \cdot \frac{1- \text e^{-K_s} }{K_s}
\,.
\end{equation}
This is consistent with \eqnref{PPM_BPP} for the case \ile{\pW =\pD = 0}.
As in the outage-free case, 
for a fixed value of $m$ the greatest photon efficiency is obtained as \ile{K_s \to 0},
and there is no motivation to modify $K_s$ to account for outages.

The foregoing is a theoretical limit on $C_u$ and BPP.
For any specific choice of a class of error-correction codes (such as the Reed-Solomon codes),
it is appropriate to minimize the error rate by the choice of $K_s$ and $T_I$
taking into account the characteristics of this particular code.
In this context there may be some dependence of $K_s$ on $\pW$ and $\pD$.


\bibliographystyle{aasjournal}

 \end{document}